\documentclass[a4paper,english,12pt]{book}
\pdfoutput=1
\usepackage{fancyhdr}
\usepackage{babel}
\usepackage{a4}
\usepackage{amsfonts}
\usepackage{amsthm}
\usepackage{eucal}
\usepackage[intlimits]{amsmath}
\usepackage{epsfig}
\usepackage{graphicx}
\usepackage{lscape}
\usepackage{longtable}
\usepackage{rotating}
\usepackage{supertabular}
\usepackage[noprefix,intoc,refpage]{nomencl}
\usepackage[percent]{overpic}
\usepackage{tocbibind}
\usepackage{caption}
\usepackage{epstopdf}
\usepackage[printonlyused]{acronym}
\usepackage[toc]{appendix}

\usepackage{datetime}
\usepackage{multibib}
\newcites{own}{Publications}
\usepackage{subcaption} 
\usepackage{algorithmic}
\usepackage{algorithm}
\DeclareMathOperator*{\argmin}{arg\,min}
\DeclareMathOperator*{\argmax}{arg\,max}
\usepackage{enumitem}
\usepackage{multirow}
\usepackage{cite}
\usepackage{bm}
\usepackage{pdfpages}
\usepackage[hyphens]{url}
\usepackage{color}
\usepackage[colorlinks=true,citecolor=blue]{hyperref}
\hypersetup
{
    pdfauthor={Hauke Holtkamp},
    pdfsubject={PhD Thesis},
    pdftitle={Enhancing the Energy Efficiency of Radio Base Stations},
    pdfkeywords={energy efficiency, radio base station, radio resource management, RRM, MIMO, OFDM, OFDMA, power model, power consumption, earth, interference avoidance}
}

\makeglossary

\begingroup%
\ifx\MakeISMBox\undefined%
\gdef\MakeISMBox#1#2#3{%
  \begin{overpic}[width=#1]{#2}%
    \put(98,2){\makebox(0,0)[br]{#3}}%
  \end{overpic}}%
\fi\endgroup%

\begingroup%
\ifx\color\undefined%
\gdef\color[#1]#2{}%
\fi\endgroup%

\begingroup%
\ifx\href\undefined%
\gdef\href#1#2{#2}%
\fi\endgroup%

\newcommand{\nomunit}[1]{%
}

\renewcommand{\chaptermark}[1]{%
 \markboth{\MakeUppercase{%
 \chaptername} \ \thechapter.%
 \ #1}{}}

\newcommand{\ie}{\emph{i.e.}} 
\newcommand{\eg}{\emph{e.g.}}

\newcommand{\ea}{\emph{et al.}}

\newcommand{\coo}{$\mathrm{CO_2}$} 

\newcommand{\B}{\boldsymbol}

\begin{document}
\newboolean{edthesis}
\setboolean{edthesis}{false}

\newcommand{\dir}{preface}

\ifthenelse{\boolean{edthesis}}
 {
   \begin{prefacepart}
   \pagenumbering{roman} 
}
{
  \frontmatter
}

\thispagestyle{empty}

\begin{minipage}{\textwidth}
\end{minipage}
\begin{center}
\vspace{2cm}
{ \Huge Enhancing the Energy Efficiency \\of Radio Base Stations
  \par
  \vspace{0.5cm} 
{\Large Hauke Andreas Holtkamp \par}
}
\vfill
\centerline{\includegraphics[width=0.35\textwidth]{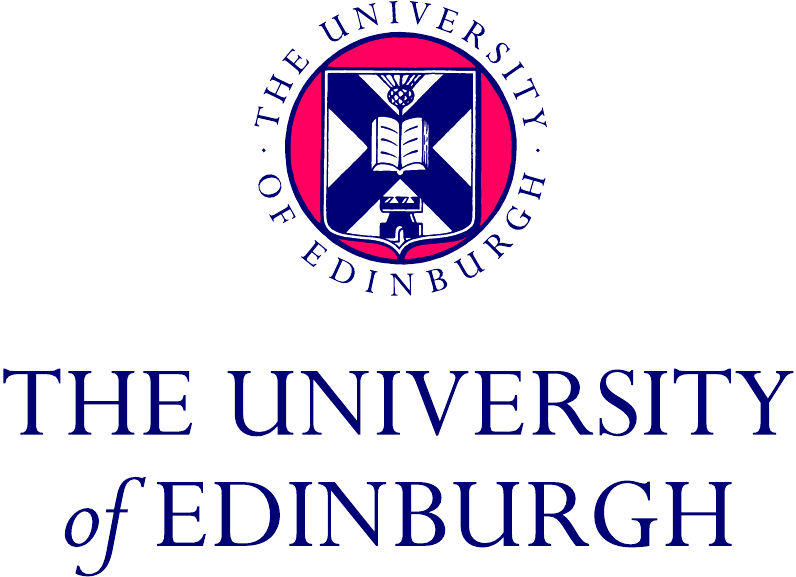}}
\vspace{0.5cm}
Thesis submitted in fulfilment of\\
the requirements for the degree of\\ 
Doctor of Philosophy\\ 
to the\\
University of Edinburgh --- 2013
\end{center}

\newpage
\thispagestyle{empty}

\chapter*{Declaration}
\vspace*{2\baselineskip}
I declare that this thesis has been composed 
solely by myself and that it has not been submitted, either in whole or
in part, in any previous application for a degree.
Except where otherwise acknowledged, the work presented is entirely my
own.
\vspace{6\baselineskip}\\
\begin{flushright}
\hspace*{\fill}
Hauke Andreas Holtkamp
\newline
\newdateformat{mydate}{\monthname[\THEMONTH] \THEYEAR}
\mydate\today
\end{flushright}

\cleardoublepage

\chapter{Abstract}
\markboth{\MakeUppercase{Abstract}}{\MakeUppercase{Abstract}}
This thesis is concerned with the energy efficiency of cellular networks. It studies the dominant power consumer in future cellular networks, the \ac{LTE} radio \ac{BS}, and proposes mechanisms that enhance the \ac{BS} energy efficiency by reducing its power consumption under target rate constraints. These mechanisms trade spare capacity for power saving.

First, the thesis describes how much power individual components of a \ac{BS} consume and what parameters affect this consumption based on third party experimental data. These individual models are joined into a component power model for an entire \ac{BS}. The component model is an essential step in analysis but is too complex for many applications. It is therefore abstracted into a much simpler parameterized model to reduce its complexity. The parameterized model is further simplified into an affine model which can be applied in power minimization.

Second, \ac{PC} and \ac{DTX} are identified as promising power-saving \ac{RRM} mechanisms and applied to multi-user downlink transmission. \ac{PC} reduces the power consumption of the \ac{PA} and is found to be most effective at high traffic loads. DTX mostly reduces the power consumption of the \ac{BB} unit while interrupting transmission and is better applied in low traffic loads. Joint optimization of these two techniques is found to enable additional power-saving at medium traffic loads and to be a convex problem which can be solved efficiently. The convex problem is extended to provide a comprehensive power-saving \ac{OFDMA} frame resource scheduler. The proposed scheduler is shown to reduce power consumption by 25-40\% in computer simulations, depending on the traffic load.

Finally, the thesis investigates the influence of interference on power consumption in a network of multiple power-saving \acp{BS}. It discusses three popular alternative distributed uncoordinated methods which align \ac{DTX} mode between neighboring \acp{BS}. To address drawbacks of these three, a fourth memory-based \ac{DTX} alignment method is proposed. It decreases power consumption by up to 40\% and retransmission probability by around 20\%, depending on the traffic load.

\chapter{Lay Summary}

This thesis is about reducing the power consumption of mobile phone networks. Base Stations, the antennas on rooftops everywhere, currently require a large amount of electricity which is expensive. This thesis studies base stations and proposes mechanisms which reduce their power consumption while providing an unchanged connection quality. This is achieved by reducing the base station power when only few people use their mobile phones.

First, the thesis describes how much power the individual electronic elements of a BS consume and what parameters this consumption depends on. These individual models are joined into a component power model for an entire base station. With this power model, one can study a base station in theory without having an expensive laboratory. The component model is very complex and is therefore further simplified into a so-called parameterized model.

Second, two techniques are studied which reduce the power consumption of base stations when it is transmitting data to mobile phones. One is Power Control (PC), which reduces the power at the antenna when it is not needed, because users are standing near the base stations. The other is Discontinuous Transmission (DTX). DTX turns off the base station for a very short time, which is not noticeable for mobile phones. Using the parameterized power model, PC is found to work very well when many people use their phones, for example, at daytime. DTX is more effective, when few people use their phones, for instance at night-time. It is found that we can mathematically combine both techniques to save even more power. Using a mathematical technique called convex optimisation, this can be done extremely fast. Using this technique, the power consumption could be reduced by 25-40\% in computer simulations, depending on how many people use their phones at one time.

Finally, the thesis investigates how much power an entire network consumes when all base stations are transmitting. When all base stations transmit at the same time, the network becomes noisy. It is better for base stations to take turns. The thesis discusses three popular alternative methods which try to organize when the different base stations in the network transmit. To address disadvantages of these three, a fourth method is proposed which provides a trade-off between user satisfaction and power-saving. A small decrease in network quality could reduce power consumption by 20\%, depending on how many people use their phone at one time.

\chapter*{Acknowledgements}
\markboth{\MakeUppercase{Acknowledgements}}{\MakeUppercase{Acknowledgements}}

This document could not have been produced without help. I owe deep gratitude to my academic supervisor, Professor Harald Haas of the University of Edinburgh, for believing in me and providing me with guidance and support. This work is in its current shape only through his teachings, insistence and attention to detail. Furthermore, I am grateful to the thesis examiners, Dr James R. Hopgood of the University of Edinburgh, and Dr Emad Alsusa of the University of Manchester. 

Extended thanks goes to DOCOMO Euro-Labs GmbH in Munich. There, Dr. Gunther Auer was the best industry supervisor any PhD student could wish for. He pulled me back on track when I drifted. Guido Dietl helped me through extended technical discussions. The entire DOCOMO team over the years were my friends, sharpened my technical understanding and provided support when it was needed. Thanks, 
Emmanuel,
Iwamura-san,
Marwa,
Matthias,
Patrick,
Petra,
Samer,
Serkan,
Shinji,
Toshi,
and Zubin.

At the University of Edinburgh, 
Bogomil,
Harald,
Nikola,
and Stefan
have helped me to get through the university jungle and lead the way to a PhD. Go Jacobs/IUB alumni!

My friends have provided me with the indispensable social support. Most of all I am thankful to my parents without whom I would not be where I am today.

{\setlength{\parskip}{0ex plus 0.5ex minus 0.2ex}

\ifthenelse{\boolean{edthesis}}
 {
  \begin{singlespace}
 }{}

\tableofcontents
\listoftables
\listoffigures
\chapter{List of Acronyms}
\begin{acronym}[WINNER]

\acro{3GPP}[3GPP]{3rd Generation Partnership Project} 
\acro{AA}[AA]{Antenna Adaptation}
\acro{AC}[AC]{Alternating Current}
\acro{ACLR}[ACLR]{Adjacent Carrier Leakage Ratio}  
\acro{ADC}[ADC]{Analog-to-Digital Converter}  
\acro{ASIC}[ASIC]{Application-Specific Integrated Circuit}  
\acro{ASIP}[ASIP]{Application-Specific Instruction-set Processor}  
\acro{BA}[BA]{Bandwidth Adaptation} 
\acro{BB}[BB]{Baseband} 
\acro{BS}[BS]{Base Station}
\acro{CDM}[CDM]{Code Division Multiplexing}  
\acro{CDMA}[CDMA]{Code Division Multiple Access}  
\acro{CMOS}[CMOS]{Complementary metal–oxide–semiconductor}  
\acro{CSI}[CSI]{Channel State Information}
\acro{DAC}[DAC]{Digital-to-Analog Converter}  
\acro{DC}[DC]{Direct Current}
\acro{DTX}[DTX]{Discontinuous Transmission}
\acro{EARTH}[EARTH]{Energy Aware Radio and neTwork tecHnologies}
\acro{FDD}[FDD]{Frequency Division Duplex}
\acro{FDM}[FDM]{Frequency Division Multiplexing}  
\acro{FDMA}[FDMA]{Frequency Division Multiple Access}
\acro{FFT}[FFT]{Fast Fourier Transform}  
\acro{FPGA}[FPGA]{Field-Programmable Gate Array}  
\acro{GOPS}[GOPS]{Giga Operations Per Second}  
\acro{GSM}[GSM]{Global System for Mobile communications} 
\acro{HetNet}[HetNet]{Heterogeneous Network}
\acro{ICT}[ICT]{Information and Communication Technologies}  
\acro{IQ}[IQ]{In-phase/Quadrature} 
\acro{LNA}[LNA]{Low-Noise Amplifier}  
\acro{LTE}[LTE]{Long Term Evolution}
\acro{M2M}[M2M]{Machine-to-Machine} 
\acro{MCS}[MCS]{Modulation and Coding Scheme} 
\acro{MIMO}[MIMO]{Multiple-Input Multiple-Output}
\acro{OFDM}[OFDM]{Orthogonal Frequency Division Multiplexing}
\acro{OFDMA}[OFDMA]{Orthogonal Frequency Division Multiple Access}
\acro{OPEX}[OPEX]{Operating Expenses} 
\acro{PA}[PA]{Power Amplifier}
\acro{PAPR}[PAPR]{Peak-to-Average Power Ratio} 
\acro{PC}[PC]{Power Control}
\acro{PRAIS}[PRAIS]{Power and Resource Allocation Including Sleep}
\acro{QoE}[QoE]{Quality of Experience} 
\acro{QoS}[QoS]{Quality of Service} 
\acro{RAPS}[RAPS]{Resource allocation using Antenna adaptation, Power control and Sleep modes}
\acro{RF}[RF]{Radio Frequency}
\acro{RCG}[RCG]{Rate Craving Greedy}
\acro{RRH}[RRH]{Remote Radio Head} 
\acro{RRM}[RRM]{Radio Resource Management}
\acro{SC-FDMA}[SC-FDMA]{Single-carrier FDMA} 
\acro{SIMO}[SIMO]{Single-Input Multiple-Output}
\acro{SINR}[SINR]{Signal-to-Interference-and-Noise-Ratio}
\acro{SISO}[SISO]{Single-Input Single-Output}
\acro{SNR}[SNR]{Signal-to-Noise-Ratio}
\acro{SOTA}[SotA]{State-Of-The-Art}
\acro{TDD}[TDD]{Time Division Duplex} 
\acro{TDM}[TDM]{Time Division Multiplexing}
\acro{TDMA}[TDMA]{Time Division Multiple Access}
\acro{UE}[UE]{User Equipment}
\acro{UMTS}[UMTS]{Universal Mobile Telecommunications System} 
\acro{VCO}[VCO]{Voltage-Controlled Oscillator} 
\acro{WSN}[WSN]{Wireless Sensor Network} 

\end{acronym}

\cleardoublepage
\printnomenclature[0.6in] 
\ifthenelse{\boolean{edthesis}}
 {
  \end{singlespace}
 }{}
}

\ifthenelse{\boolean{edthesis}}
 {
   \end{prefacepart}
   \pagenumbering{arabic} 
}
{
   \mainmatter
}

\acresetall 

\renewcommand{\dir}{chapter1/chapter/}
\chapter{Introduction}
\label{ch:introduction}

\section{Overview}
In the first section of this chapter, it is argued why energy efficiency in wireless networks is an important research topic and what goals this thesis sets out to achieve. The second section introduces the contributions of this thesis towards achieving these goals. The third section outlines the thesis structure. In the last section, the \ac{EARTH} project is briefly introduced which provided laboratory results for the contents of Chapter~\ref{ch:device}.

\section{Thesis Context}

\begin{figure}
\centering
\includegraphics[width=0.8\textwidth]{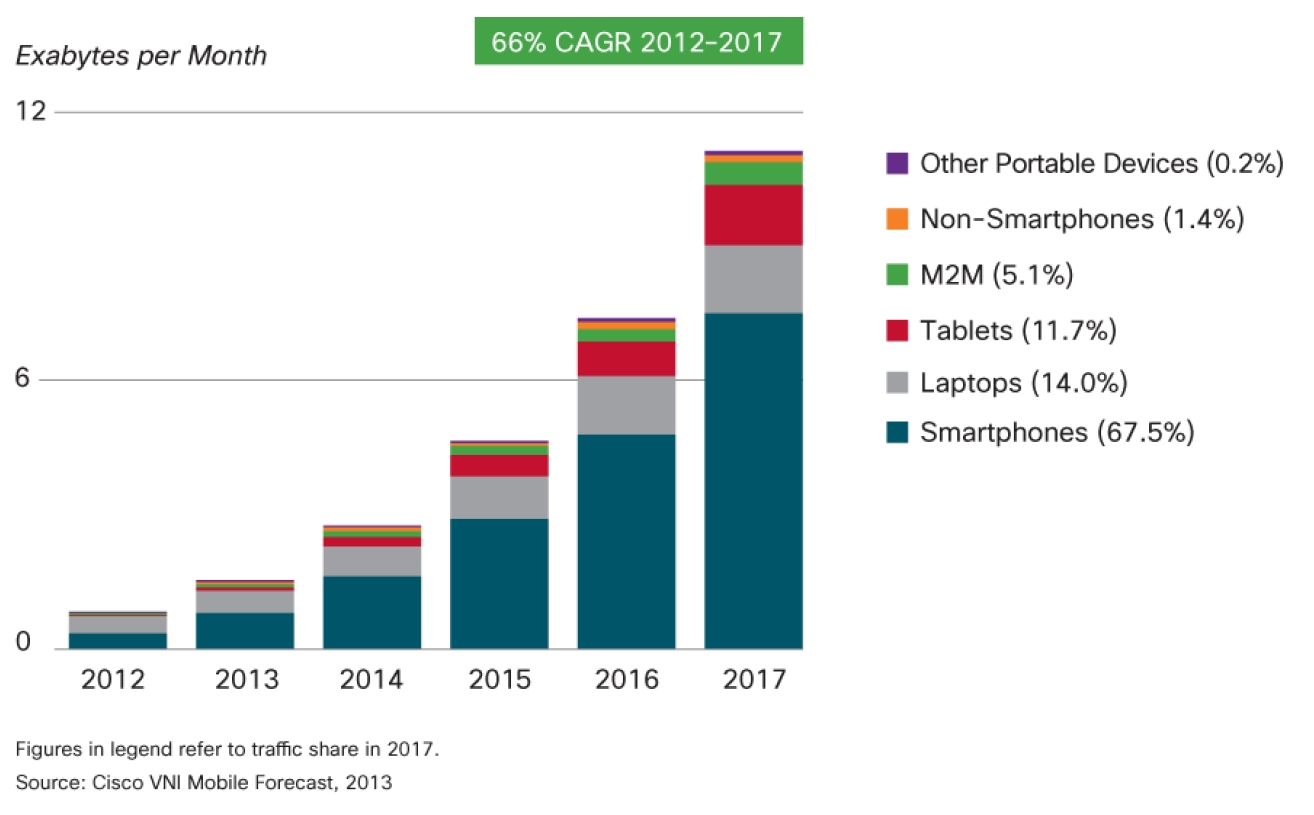}
\caption[Mobile traffic forecast 2012-2017]{Mobile traffic forecast 2012-2017 with a \acl{CAGR} of 66\%~\cite[p.\,7]{c1301}.}
\label{fig:trafficforecast}
\end{figure}

Since the emergence of mobile communications, both the number and the density of mobile devices have constantly increased~\cite{bzm0101}. They are predicted to rise further, fueled by innovations such as \ac{M2M} communication and the `Internet of Things'~\cite{wtjhj1101,book:bddrl0501}. It is predicted that a trillion devices will be connected to the Internet by the end of 2013~\cite{w1001}, with a growing share of these on mobile networks. Fig.~\ref{fig:trafficforecast} shows a forecast of the resulting mobile traffic until 2017. To serve this growing traffic load, the capacity, size and density of the infrastructure network are continually upgraded by network operators. 

\acrodef{CAGR}[CAGR]{Compound Annual Growth Rate}

While for the past 20~years network capacity, reliability and deployment were the main concerns during these upgrades, new factors like rising energy prices, increasing consumer attention to the emission of \coo, the deployment of \acp{BS} in remote off-grid locations, and disaster recovery have come into play. To respond to these new factors, the energy efficiency of the current and future mobile networks needs to be enhanced. Delivery of bits through the cellular network has to become cheaper. \coo~emissions have to be reduced. \acp{BS} in remote locations must become self-sustaining. And maintaining connectivity after disasters is critical as the dependence of people and services on mobile communication in a disaster aftermath grows. 

Lowering the power consumption of the \acp{BS} in cellular networks addresses all of these problems. It reduces the network operational cost and the required \coo~emission. Installing grid-independent \acp{BS} becomes more feasible when less power is needed for their operation. Furthermore, connectivity can be provided for a longer period of time in case of disastrous power interruptions.

However, the causes of power consumption in \acp{BS} have not been studied in depth. Whether and by how much this power consumption can be lowered through operational modifications is an open problem. The answer to this problem affects the design of future transmission schemes and hardware. Once the most effective mechanisms for reducing power consumption are understood, they can be used to operate \acp{BS} with lower power consumption and, thus, with higher efficiency. Radio resource schedulers need to be developed which enhance the efficiency of operation in both individual \acp{BS} and a network of \acp{BS}.

As such, this thesis sets out to achieve the following goals:
\begin{itemize}
 \item Identify how the power consumption of a \ac{BS} is comprised in terms of hardware components. Evaluate what operating parameters have the strongest effect on the \ac{BS} power consumption. Model this consumption in equations such that it can be reused and modified.
 \item Identify promising operational techniques to reduce a \ac{BS}'s power consumption without affecting the \ac{QoS}. Generate resource schedulers which employ such techniques.
 \item Understand how power consumption can be reduced in a network of multiple \acp{BS}. Provide solutions to address power consumption on the network level.
\end{itemize}


\section{Thesis Contributions}
This thesis contributes towards the enhancement of energy efficiency in wireless networks by reducing the power consumption of radio \ac{BS} through \ac{RRM} without affecting the \ac{QoS}. In particular, it provides the following contributions:
\begin{itemize}
 \item First, the power consumption of \ac{SOTA} \ac{LTE} \acp{BS} is studied and presented in detail. The number of antennas, the transmission power and the lowered consumption through \ac{DTX} are derived as the most relevant operating parameters. A general, practical power model for cellular \acp{BS} is constructed.
 \item Second, the attainable power savings of power control, \ac{DTX} and \ac{AA} are quantified. The three techniques are jointly applied to a new energy efficient and spectrum efficient \ac{OFDMA} scheduler.
 \item Third, the alignment of \ac{DTX} time slots between neighbouring \acp{BS} is identified as a multi-cell power-saving mechanism. Three alternative alignment schemes are studied with regard to their power consumption. The findings lead to a novel time slot alignment scheme which is shown to overcome the limitations of the other techniques.
\end{itemize}

\section{Thesis Structure}
Chapter~\ref{ch:motivation} first addresses the issue why this work is concerned with the supply power consumption of cellular \acp{BS} in particular. Next, it provides an overview of Green Radio research alongside a review of relevant literature. Finally, a technical background for Chapters~\ref{ch:device}, \ref{ch:cell}, and \ref{ch:network} is provided about the topics \ac{LTE}, multi-carrier technology, multi-antenna transmission, convex optimization, and computer simulation techniques.

Chapter~\ref{ch:device} begins with an analysis of the power consumption of a \ac{SOTA} \ac{BS} in order to identify opportunities for power saving. Each subcomponent is individually inspected and described with regard to its power consumption. A comprehensive \ac{BS} power model is established as the sum of each subcomponent's power consumption. Relevant and promising saving mechanisms are identified. A parameterized power model is proposed which encompasses these saving mechanisms and abstracts architectural details in favor of simplicity. Finally, an affine power model is derived, which is used through the remainder of this thesis. The affine mapping this model provides between transmission power and supply power is advantageous for its application in optimization.

Chapter~\ref{ch:cell} applies the knowledge on \ac{BS} power consumption from Chapter~\ref{ch:device} by proposing power-saving \ac{RRM} mechanisms. On the basis of the affine power model, \ac{PC} and \ac{DTX} are studied. The potential of their joint optimization is identified. The power savings achieved by this method over the \ac{SOTA} are quantified in simulation. The joint optimization of \ac{PC}, \ac{DTX} and \ac{AA} is applied in a comprehensive scheduler for power-saving in \ac{OFDMA} downlink transmission within a cell.

Chapter~\ref{ch:network} adds the consideration of intercell interference. When multiple \acp{BS} are considered, the use of sleep modes in each \ac{BS} results in significant fluctuations of interference. Constructive alignment of interference is proposed to exploit this effect and can be employed for a decrease of power consumption. To address multi-cell power saving resource allocation, \emph{distributed \ac{DTX} with memory} is proposed and compared to alternative alignment mechanisms. Network simulations are used to estimate achievable savings.

Finally, conclusions of the above research, a discussion of limitations and an outlook for future research is provided in Chapter~\ref{ch:conclusion}.

Regarding the format of this document: There are two non-overlapping bibliographies at the end of this document. One contains publications by the author of this thesis. References to these publications are made with alphabetic indices, such as `[a]' or `[b]'. References to other literature are written with numeric indices such as `[1]' or `[22]'. 

\section{The \acs{EARTH} Project}
\label{earthproject}

\acrodef{FP}[FP]{Framework Programme}
\acrodef{WINNER}[WINNER]{Wireless world INitiative NEw Radio}
The research presented in Chapters~\ref{ch:device} and \ref{ch:cell} of this thesis has received funding and experimental data from the \acf{EARTH} project. Initiated by the European Union's \ac{FP} 7, the \ac{EARTH} project aligned cooperation between 15~industry and academic institutions towards the common goal of driving up energy efficiency in current and future cellular networks. The project lasted from January 2010 to June 2012. It was founded to work on 
\begin{itemize}
\item deployment strategies,
\item network architectures, 
\item network management,
\item adaptation to load variations with time,
\item innovative component designs with energy efficient adaptive operating points,
\item and new radio and network resource management protocols for multi-cell cooperative networking.
\end{itemize}


The most prominent outcomes of the project were
\begin{itemize}
 \item the cellular network life cycle analysis~\cite{fmbf1001},
 \item the energy efficiency evaluation framework~\cite{aggsoisgd1101,std:earthd23}, 
 \item the \ac{BS} power model~\cite{ddgfahwsr1201},
 \item hardware implementations of \acp{PA} with improved dynamics~\cite{gfwem1101} and a sleep mode enabled small cell~\cite{std:earthd43},
 \item and numerous individual techniques which reduce the power consumption of cellular networks such as the ones presented in Chapter~\ref{ch:cell}~\cite{std:earthd33}.
\end{itemize}


All project deliverables are available on the project website for reference~\cite{01b}.

\acresetall 

\renewcommand{\dir}{chapter2/chapter/}
\chapter{Motivation and Background}
\label{ch:motivation}

\section{Overview}
This chapter first outlines the motivation why reducing the power consumption of cellular \acp{BS} constitutes a large steps towards energy efficient wireless networks and how this reduction can be achieved. Second, it discusses the difficulties of quantifying energy efficiency and provides an overview of Green Radio research topics in literature. Finally, technical concepts which are applied in Chapters~\ref{ch:cell} and \ref{ch:network} are briefly introduced for the unfamiliar reader.

The work in Section~\ref{sect:eebs} has been previously published by the author of this thesis in~\cite{aghimfbhz1001}. The technical background in Section~\ref{background} is sourced from literature as referenced.


\section{Energy Efficient Base Stations}
\label{sect:eebs}

\begin{figure}
\centering
\includegraphics[width=0.6\textwidth]{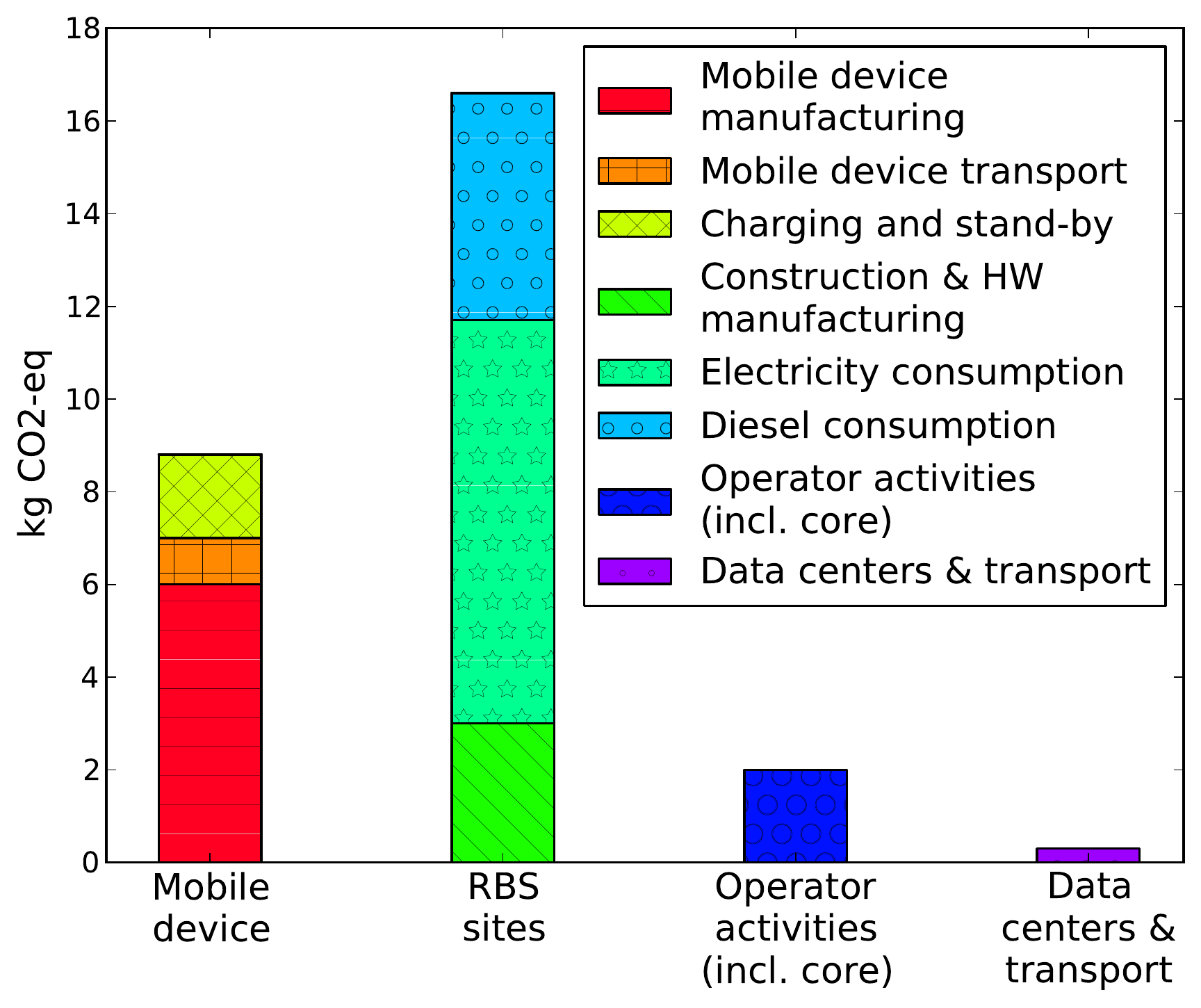}
\caption[The carbon footprint for an average subscriber in 2007]{The carbon footprint (\coo-equivalent emissions, see Section~\ref{sect:quantEE}) for an average subscriber in 2007~\cite{mmlfl1001}, with a focus on radio base stations (RBS).} 
\label{fig:4}
\end{figure}

Life cycle analyses of mobile networks and their equipment have shown that the access infrastructure (\acp{BS}, data centers, controllers) generates significantly more \coo~than the connected mobile devices~\cite{fmbf1001}. They also reveal that mobile devices cause the majority of \coo~emissions during manufacturing due to their battery-optimized operation and short life times. In contrast, \acp{BS} tend to be less power optimized with longer life times leading to the majority of \coo~emitted during operation. This is illustrated in Fig.~\ref{fig:4} by the detailed carbon footprint of the average mobile network subscriber in 2007, the most recent data available. It shows that only 20\% of the \coo~emissions over the life of a mobile device were caused during operation. On the contrary, 82\% of \ac{BS} emissions are owed to operation, either as diesel or electricity consumption.

While the power consumption of mobile devices has always been a topic of research and development due to the constraints imposed by battery operation, \acp{BS} were usually installed in urban locations with good connections to the power grid providing little incentive for efforts in energy efficiency. However, as briefly mentioned in Chapter~\ref{ch:introduction}, the energy efficiency of cellular \acp{BS} is receiving more attention due to several factors~\cite{ls0801}:
\begin{itemize}
 \item Rising energy prices~\cite{ahjlss1001} combined with decreasing profit margins~\cite{fz0801} require operators to optimize their operational expenses by decreasing power bills.
 \item In countries with incomplete or unreliable power grids, \acp{BS} are operated with grid-independent power sources like diesel generators or renewable energy sources~\cite{fz0801}. The costly deliveries of diesel and size of renewable energy sources like wind engines or solar panels make low consumption of the \ac{BS} desirable~\cite{nnr1001}. 
 \item In the face of global warming, customers have developed an understanding and sensibility for power consumption. Companies aim to apply \emph{green} labels to their products~\cite{hhakvghtk1101,s1201}.
 \item Disasters like the 2011 Tohoku earthquake and tsunami have shown that the installed battery backups were insufficient for long power grid interruptions~\cite{o1201}. After the disaster, \acp{BS} (and, thus, most communication) were unavailable for days until repair units had reconnected them to the power grid. Efforts in disaster recovery include low-power emergency modes and a generally lower power consumption to enable \acp{BS} to operate longer on battery power~\cite{r1101}.
\end{itemize}
For these reasons, reducing the operating power consumption of \ac{LTE} \acp{BS} is the goal of the work presented in this thesis. Recent analyses provided information on how this goal could be achieved. In particular, it was found that while the cell traffic load fluctuates significantly, \acp{BS} are mostly unable to adapt to these fluctuations.

\begin{figure}
\centering
\includegraphics[width=0.7\textwidth]{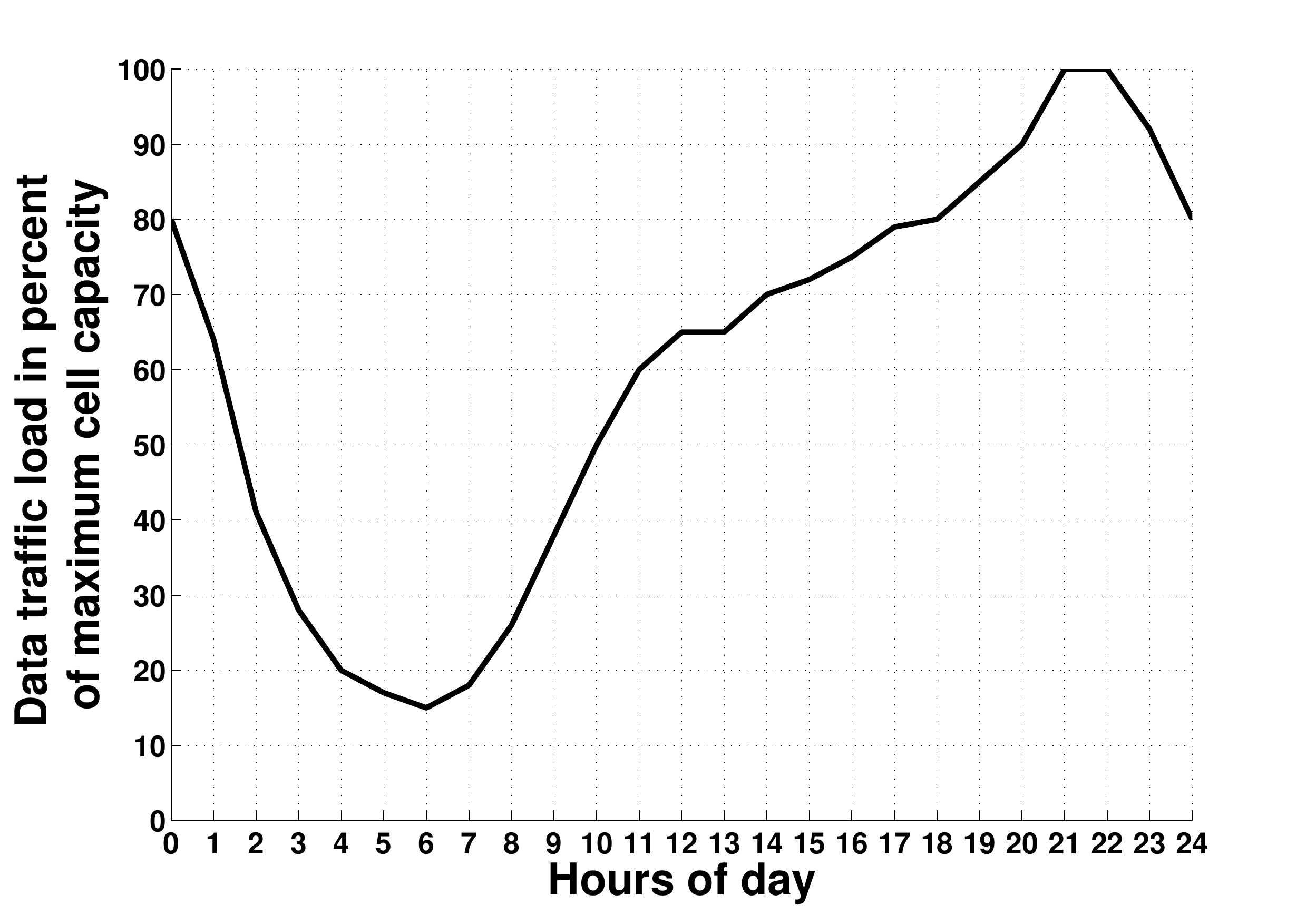}
\caption[European daily traffic pattern]{European average wireless daily traffic in 2010~\cite{aggsoigdb1101}.}
\label{fig:dailyload}
\end{figure}

\acp{BS} have been designed and deployed to provide peak capacities to minimize outages. However, in practical scenarios, traffic is rarely at its peak. While \acp{BS} may be designed for crowded midday downtown situations, there are strong variations over time and location with regard to the density of mobile traffic. For example, see the typical mobile traffic in Europe over the course of a day in Fig.~\ref{fig:dailyload}. It varies over a range of more than 80\% and has an average of only 60\% load~\cite{aggsoigdb1101}. Yet, although load varies so strongly, the power consumption of cellular \acp{BS} was found to vary as little as 2\% over the course of a day~\cite{arfb1001}. \ac{BS} power consumption is thus not yet sufficiently adaptive to the traffic demand. Hence, the times of day when the traffic load is below peak provide room for exploitation. During such times, the capacity of a \ac{BS} can be  reduced to achieve lower power consumption without affecting the user satisfaction or \ac{QoS}. How this could be achieved is the contribution of this thesis.

Note here an important differentiation between Green Radio technology proposals being either in design or in operation. Design proposals are made to affect the architecture or hardware of a \ac{BS}, such as changing the type of \ac{PA}. Design changes may take a long time to reach the market due to specification, approval or manufacturing. In contrast, operation proposals, like turning a \ac{BS} off when it is unused, can reach the market much faster. They may be applicable to existing hardware and can be applied via software upgrades.

Although design and operation changes can be made independently, they affect each other. For example, a change in design---like the introduction of a secondary low power wake-up controller---may enable new types of operation like a cognitive wake-up functionality. In turn, the popular application of a certain type of operation, like sleep modes, may have consequences for future design decisions by providing incentive for producing sleep-mode-enabled hardware. In this thesis, this intertwining of design and operation is taken into account by first identifying the structure and potential of the hardware and then exploiting this potential in operational improvements to the \ac{BS}.

\section{Quantifying Energy Efficiency}
\label{sect:quantEE}
Enhancing the energy efficiency of communication networks has led to a field of research popularly labelled \emph{Green Radio}, which this thesis is part of. This field encompasses all efforts taken to increase the ratio of network quality over energy spent or \coo~emitted. Making `radios greener' can be achieved by either increasing network quality at unchanged energy expenditure, by providing equal quality at a lower cost, or both. Here, typical network quality indicators to be improved are capacity, fairness, latency, reliability, and range. Cost is often measured in \coo~emission, energy consumption, or power consumption. Note that energy consumption is equal to the product of the average power consumption and a certain duration. In other words, power is energy normalized over time.

When discussing Green Radio research and results, it is important to be aware of some intricacies and necessary differentiations. Energy efficiency is not easily compared when considering different network quality indicators. For example, at equal energy consumption, an increase in capacity may come at the cost of increased latency, which cannot be easily weighted against each other. Alternatively, a decrease in power consumption through reduced network quality may trigger users to change their behavior which negates the initial savings. These problems are present in many fields of research and have lead to discussions about \ac{QoS}, which tries to assess the alternative a user may be more satisfied with~\cite{akrswv0101a}. But since an industry standard for the \ac{QoS} does not yet exist, the common ground is to measure capacity as the most basic metric while stating the considered scenario assumptions such as geographic location, transmission duration, latency, \emph{etc.}~\cite{std:earthd23}.

Just as there are several ways to measure network quality, there are several alternatives to measuring the cost incurred. The simplest is to only consider \emph{transmission power} at the antenna of a wireless transmitter. This number is usually well-known as it is regulated and targeted in hardware design. A more comprehensive approach is to consider total power spent by a transmitter which includes the heat generated while producing or processing a wireless signal. This number can usually be measured at the mains supply or deduced from battery durations. It is called the \emph{supply power}. But even the supply power required by a device during operation may not display the full picture. Aside from operation, a wireless device requires production before and disposal after. These steps can consume significant energy, particularly in electronic devices. They are composed of resource acquisition and mining, transportation, design, facility construction, production, and recycling. When widening the inspection of power or energy consumption to this global level, the consideration of watts or joules is insufficient. Rather, the global cost of the steps is measured in terms of \coo~emission. Comparing \coo~emissions instead of power consumptions allows taking into account that the \coo~emitted by generating a watt of power varies depending on many factors. For instance, a \ac{BS} operated in a remote location on imported diesel may cause much higher \coo~emissions than a solar-powered \ac{BS} on an urban rooftop. As energy generation usually does not lead to the emission of pure \coo, but rather a mixture of global warming affecting gases, the effects of these actions are typically normalized to \coo-equivalent (or \coo-eq). For example, the \ac{ICT} industry is estimated to have caused about 1.5\% of global \coo-eq emissions in 2007~\cite{mmlfl1001}. From 2007 to 2020, the \coo-eq emission caused by mobile networks is predicted to grow from 0.2\% to 0.4\%~\cite{fmbf1001}. Thus, a doubling of the energy efficiency of mobile communications would allow keeping \coo-eq emissions on the 2007 level which is a common research and development target~\cite{bzb1001}.

As described above, both measuring network quality as well as measuring cost are very intricate. Therefore, the metric of watts of supply power per bit alongside clear network definitions has been chosen as the reference for this thesis. When applying techniques described here to larger scenarios, the additional definitions provide the information necessary for converting watts to \coo-eq over a regional expanse and duration of operation.

\section{Green Radio in Literature}
Historically, the early works on energy efficiency for wireless networks have been triggered by a desire to extend battery lives in cellular, ad hoc or sensor networks, for example see~\cite{yhe0201,cgb0401,cgb0501,bk0401a}. By the early 2000s, growing environmental concerns had lead to extensive life cycle analyses of cellular networks ~\cite{mob0101,wl0001}. These analyses collected all steps in a devices' life, quantified them with regard to energy expenditure and \coo~emissions and joined them with the number of units sold and installed. The resulting statistics revealed that the power consumption of mobile network infrastructure was significant and that the \coo~emissions of the \ac{ICT} industry were as high as those of international air traffic~\cite{fmbf1001,fz0801}. Through these findings, interest grew with regard to the energy efficiency of transmitters which are powered by the mains grid.

As a consequence, in 2008, the collaborative \emph{Green Radio} project was initiated in the UK which coined the term~\cite{m01}. The efforts were extended in 2009 by the European \ac{EARTH} project~\cite{gbfzis0901}. Both projects encouraged specifically the enhancement of the energy efficiency of cellular access networks by improving their architecture and operation. The works proposed since then can be distinguished by the network aspects they consider. 

For one, there are works which address the hardware of the radio access network. A general overview of such hardware improvements is provided in~\cite{fbzfgjt1001}. Specifically, the improvement of transceiver units is considered in~\cite{gfwem1101,bdgd1101}. New \ac{PA} architectures are proposed in~\cite{hkhg1001}. The \ac{EARTH} project has summarized proposed hardware improvements to \ac{LTE} \acp{BS} in~\cite{std:earthd43}.

Another group of works is the field of \acp{HetNet}. These works are concerned with the power consumption of future networks which are predicted to consist of a very large number of low power small \acp{BS}, such as micro, pico, or femto, in addition to macro \acp{BS}. The study in~\cite{kff1101} finds the optimal density of small cells to match a traffic requirement from an energy perspective. In~\cite{rff0901}, the area power consumption metric is proposed to enable comparison between different \acp{HetNet}. Badic~\ea~\cite{bolh0901} formalize the trade-off that is present between capacity and power consumption in \acp{HetNet}. In~\cite{kt1201}, it is suggested that the overlap between macro and small cells in \acp{HetNet} should be used in combination with sleep modes when the capacity increase provided by small cells is not needed. The authors in~\cite{hkd1101} argue that due to the complexity of \acp{HetNet}, they should be self-organizing and self-adapting to the traffic situation.

As a third topic, several kinds of sleep modes are proposed. Long sleep modes with duration of minutes or hours are very effective in reducing power consumption, but require a wake-up mechanism and reduce coverage. Short sleep modes in the order of microseconds to seconds, which are also called dozing, micro sleep or \ac{DTX}, do not pose these problems but promise smaller reductions of the power consumption. As a consequence, long sleep modes are proposed for situations in which coverage can be maintained through other means. For example, when network densities are sufficient such that neighbouring \acp{BS} can take over the coverage of sleeping \acp{BS}~\cite{ok1001,abh1101}. Short sleep modes can be applied independent of coverage. The authors in~\cite{fmmjg1101} propose to use \ac{DTX} when a cell is completely empty, posing a simple but very inflexible mechanism. In~\cite{sec1001} it is assumed that a \ac{BS} consists of multiple independent transmitters, some of which can be deactivated according to traffic requirements, thus increasing the adaptivity to load. With regard to an entire network of sleep capable cells, Abdallah~\ea~\cite{acc1201} study the alignment of sleep modes between neighbouring cells and conclude that sleep modes should be synchronized and orthogonal for maximum efficiency.

Related to sleep modes is the field of cell breathing, cell shrinking or cell zooming through the adjustment of the transmission power~\cite{ha1201,asb1002,nwgy1001}. By introducing flexibility with regard to the transmission power of a \ac{BS}, a network can flexibly adjust to coverage requirements by increasing or decreasing cell sizes. For example, an increase of a cell's size may be needed when neighbouring cells enter sleep mode. 

An extension of sleep modes is the topic of adapting multi-antenna transmission. Rather than using all installed antennas, a \ac{BS} could deliberately reduce its \ac{MIMO} degree to deactivate some antennas when high capacities are not required~\cite{hafm1201,sf1201,kcvh0901,xylzcx1101}. These approaches are labeled \ac{MIMO} adaptation, antenna adaptation or \ac{MIMO} muting. An alternative approach to exploit multiple antennas for energy efficiency is proposed by Wu~\ea~\cite{wsdh1201} and Stavridis~\ea~\cite{ssdhg1201}, who, through---spatial modulation---use all antennas, but never at the same time.

Resource scheduling is a diverse group of works in which different approaches are proposed. Relaxing the delay requirement can be exploited for opportunistic power saving~\cite{gs1101,cj1201}. Spreading signals over a larger bandwidth allows reducing the transmission power~\cite{a1201,vh1101}. Approaches such as game theory promise gains by providing trading mechanisms for a self-organized allocation of resources for enhanced energy efficiency~\cite{mps0701}.

In order to exploit the typically deployed overlapping radio access network generations, it is proposed in~\cite{std:earthd33} to combine a deactivation of one network technology, \eg~\ac{GSM}, with the fall-back to another such as \ac{LTE}. However, this generates issues with backward compatibility and \ac{QoS}.

To exploit spare backhaul capacity, it is suggested in~\cite{sbckk1101} to use coordinated multi-point transmission between multiple \acp{BS} while deactivating parts of each \ac{BS}. However, the benefits strongly depend on the additional power consumption caused by the backhaul link. When combined with sleep modes, it was found in~\cite{hywl1101} that this approach only reduces power consumption in the presence of high data rate users.

In~\cite{hg1101} it is proposed to switch sectorized macro \acp{BS} to omni-directional operation when the capacity requirements are low. While this promises to save two thirds of the power consumption in typical three-sector setups, it also requires the installation of additional omni-directional antennas at \ac{BS} sites.

The most far-reaching proposals for energy efficient cellular networks break with the cellular assumptions presumed in \ac{GSM}, \ac{UMTS} or \ac{LTE}. By separating data and control planes from one another, future networks could have large, long-range `umbrella' control transmitters which take care of discovery and coverage. These would overarch small data plane \acp{BS} which are only activated when they are needed. Such a network setup would provide a very high capacity at a very high degree of load flexibility. These proposals are referred to in literature as `ghost' or `phantom' cell concepts~\cite{sdd1101,tbt1301}, since the small data plane \ac{BS} are `invisible' to the \ac{UE}.

For good surveys on Green Radio topics, the reader is referred to~\cite{hhakvghtk1101,aggsoigdb1101,wvclk1201,e0201a,cpgbc1201}.

Note that detailed literature backgrounds are again provided for each individual chapter in Chapters~\ref{ch:device}, \ref{ch:cell} and \ref{ch:network}.

In this wide field of research, this thesis can be placed in the group of works dealing with resource scheduling. 


%
%
%
%
%
%









\section{Technical Background}
\label{background}
%

The following sections outline some fundamental concepts which reappear in Chapters~\ref{ch:device}, \ref{ch:cell}, and \ref{ch:network} of this thesis, namely, \ac{LTE}, multi-carrier technology, multiple antenna transmission, convex optimization, and the usage of simulations to model complex cellular systems.

\subsection{\ac{LTE}}

\begin{figure}
\centering
\includegraphics[width=0.9\textwidth]{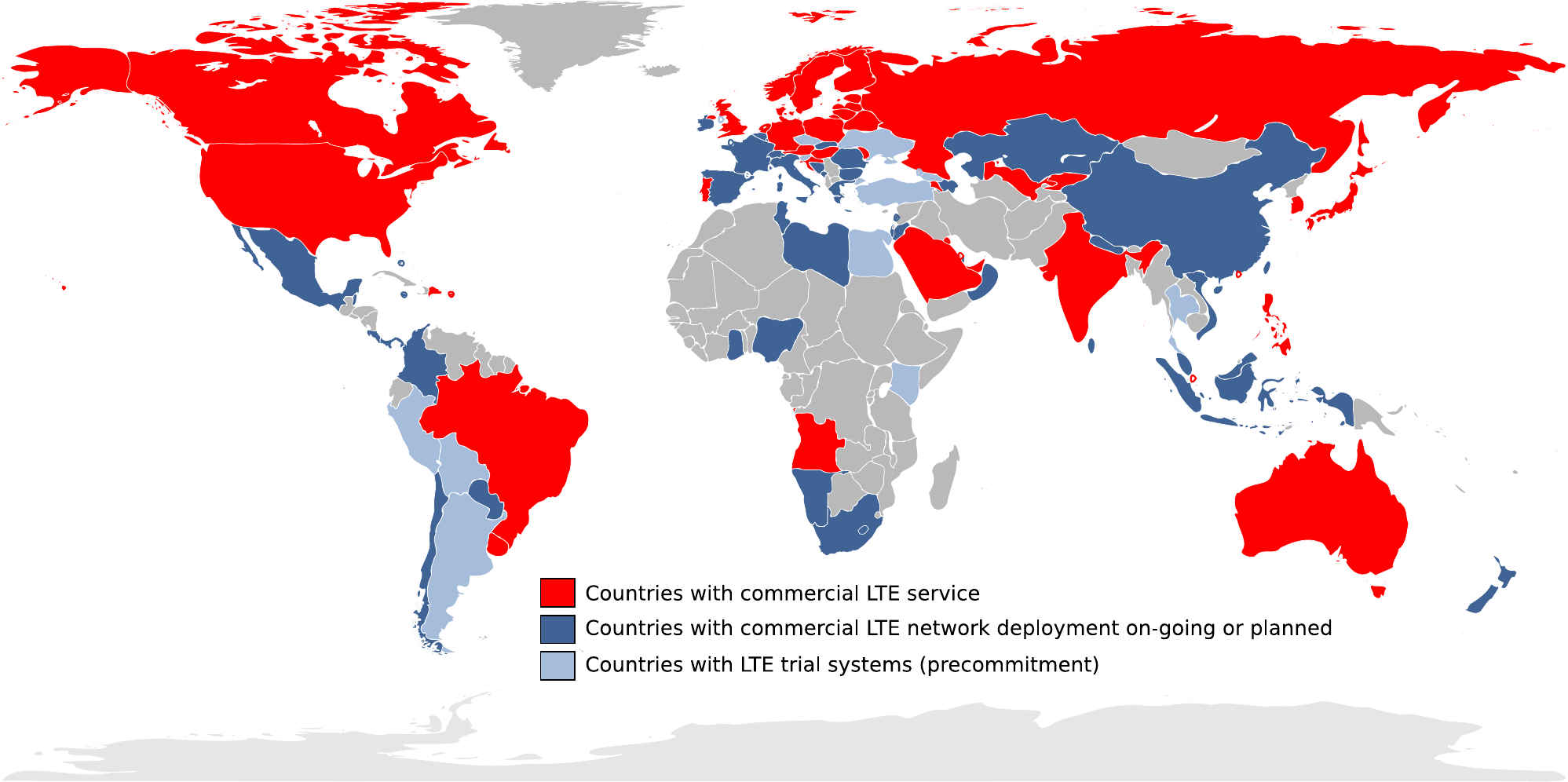}
\caption[Adoption of \ac{LTE} technology]{Adoption of \ac{LTE} technology as of May 8, 2012~\cite{tech:b}.}
\label{fig:worldmapLTE}
\end{figure}

\begin{figure}
\centering
\includegraphics[width=0.9\textwidth]{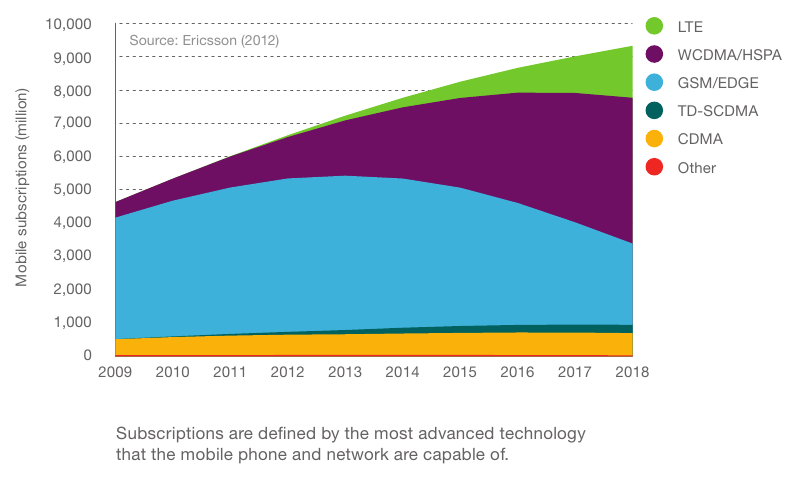}
\caption[Mobile subscriptions by technology, 2009-2018]{Mobile subscriptions by technology, 2009-2018~\cite[p.\,7]{tech:c}.}
\label{fig:subscriptions}
\end{figure}

\ac{LTE}~\cite{book:dps1101} is a wireless access standard superseding the \ac{GSM} and \ac{UMTS} for increased network capacities. It is predicted to be adopted in 93 countries by the end of 2013~\cite{tech:b}. See Fig.~\ref{fig:worldmapLTE} for an illustration on the world map. To this extent, the number of \ac{LTE} subscriptions is rising steadily as it replaces older network technologies, as shown in Fig.~\ref{fig:subscriptions}. \ac{LTE} is being continually standardized and developed by the \ac{3GPP}. It is designed to achieve the following set of goals~\cite{book:stb0901}:
\begin{itemize}
 \item faster connection establishment and shorter latency;
 \item higher user spectral efficiencies of more than 5 bps/Hz on the downlink using link bandwidths of up to 20~MHz;
 \item uniformity of service provision independent of mobile position in a cell;
 \item improved cost per bit;
 \item flexibility in spectrum usage to accommodate to national band allocations;
 \item simplified network architecture;
 \item seamless switching between different radio access technologies;
 \item reasonable power consumption for the mobile terminal.
\end{itemize}

\acrodef{SAE}[SAE]{System Architecture Evolution}
As \ac{LTE} is a wireless technology on a shared medium, it is required to handle resource sharing and interference. These two important challenges are addressed in Chapters~\ref{ch:cell} and \ref{ch:network}, respectively. Note that while \ac{LTE} also consists of non-radio aspects such as the \ac{SAE}, this thesis is only concerned with its \ac{BS} and radio access aspects. 

\subsection{Multi-carrier Technology}
\label{sect:multicarrier}
\begin{figure}
\centering
\includegraphics[width=0.6\textwidth]{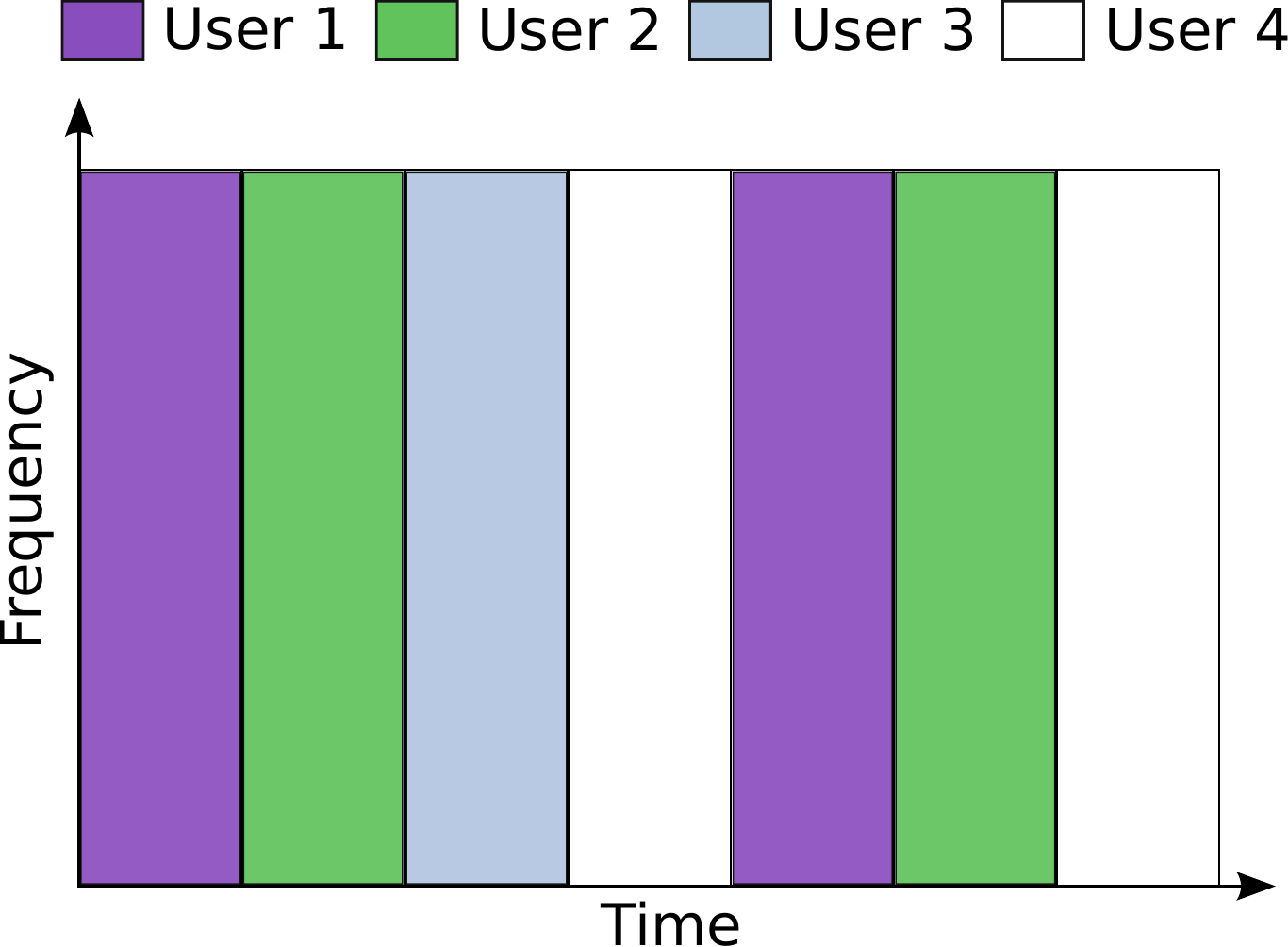}
\caption[An illustration of \acs{TDM}]{An illustration of \ac{TDM}. Different users are assigned non-overlapping shares of the transmission time spanning a wide bandwidth.}
\label{fig:TDM}
\end{figure}

\begin{figure}
\centering
\includegraphics[width=0.6\textwidth]{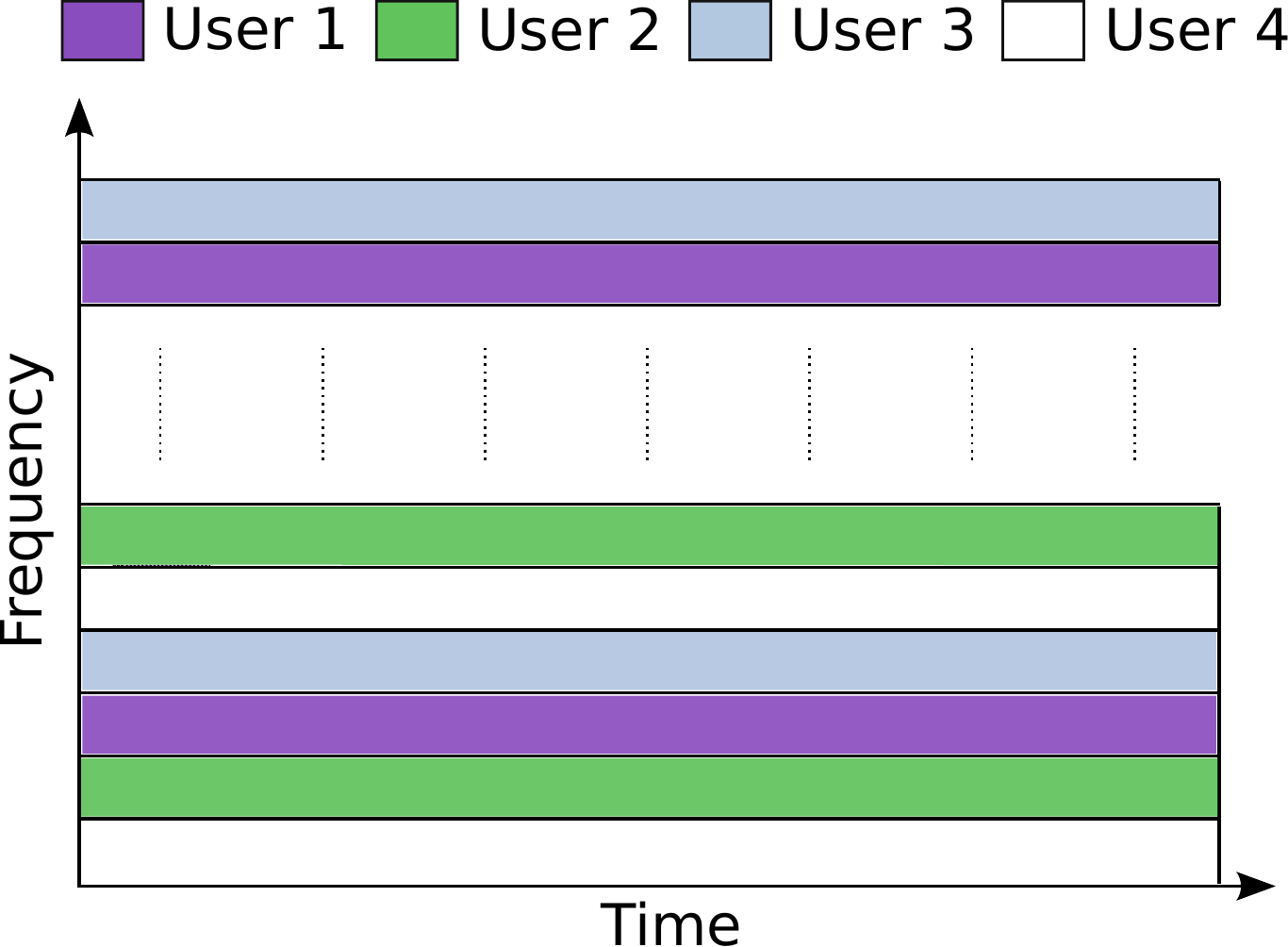}
\caption[An illustration of \acs{FDM}]{An illustration of \ac{FDM}. Different users are assigned non-overlapping shares of the transmission bandwidth over a significant duration.}
\label{fig:FDM}
\end{figure}

The wireless medium is inherently shared. Wireless transmissions occur simultaneously in resources such as the frequency spectrum, time, and location. Therefore, transmissions can negatively affect one another through interference. The technique of dividing a resource into chunks to avoid overlap between multiple transmissions is called \emph{multiplexing}. {Time} is typically shared via \ac{TDM}, {frequency} via \ac{FDM}, and location via installing \acp{BS} a significant distance apart from one another. Multiplexing allows providing multiple independent links carrying independent information streams on the wireless medium. These streams can be allocated to different users or scheduled to the same user for increased sum data rates, thus increasing the scheduling flexibility. In \ac{TDM}, the shared resource is the time domain in which only one link can be active at any point in time and all links are spread over the entire available bandwidth. This concept is illustrated in Fig.~\ref{fig:TDM} for four users. In \ac{FDM}, the shared resource is the frequency domain. Here, links only take up a portion of the available frequency spectrum. See Fig.~\ref{fig:FDM} for an illustration of \ac{FDM} with multiple carriers and four users. Both of these technologies have been applied commercially in wireless technologies for decades. For example, \ac{TDM} is the access scheme of choice in the \ac{GSM} standard~\cite{book:gg9801}, whereas \ac{FDM} has been employed in cordless phones\cite{book:g1001}. A third popular multiple access scheme, \ac{CDM}---\eg~used in the \ac{UMTS}~\cite{p9801}---is not discussed here as it is not related to this thesis.

\ac{OFDM} extends \ac{FDM} to provide a very flexible high capacity multiple access scheme. Similar in notion to \ac{FDM}, \ac{OFDM} subdivides the bandwidth available for signal transmission into a multitude of narrowband subcarriers. The channels affecting these subcarriers are sufficiently narrow in the frequency domain to be considered flat-fading~\cite{book:dps1101} which is important for link adaptation. Unlike in \ac{FDM} where frequency bands are non-overlapping, \ac{OFDM} subcarriers are packed tightly and do overlap. Yet, through selection of the right pulse shape, the signals on each subcarrier remain orthogonal and can be transmitted in parallel with little to no mutual interference.

\begin{figure}
\centering
\includegraphics[width=0.6\textwidth]{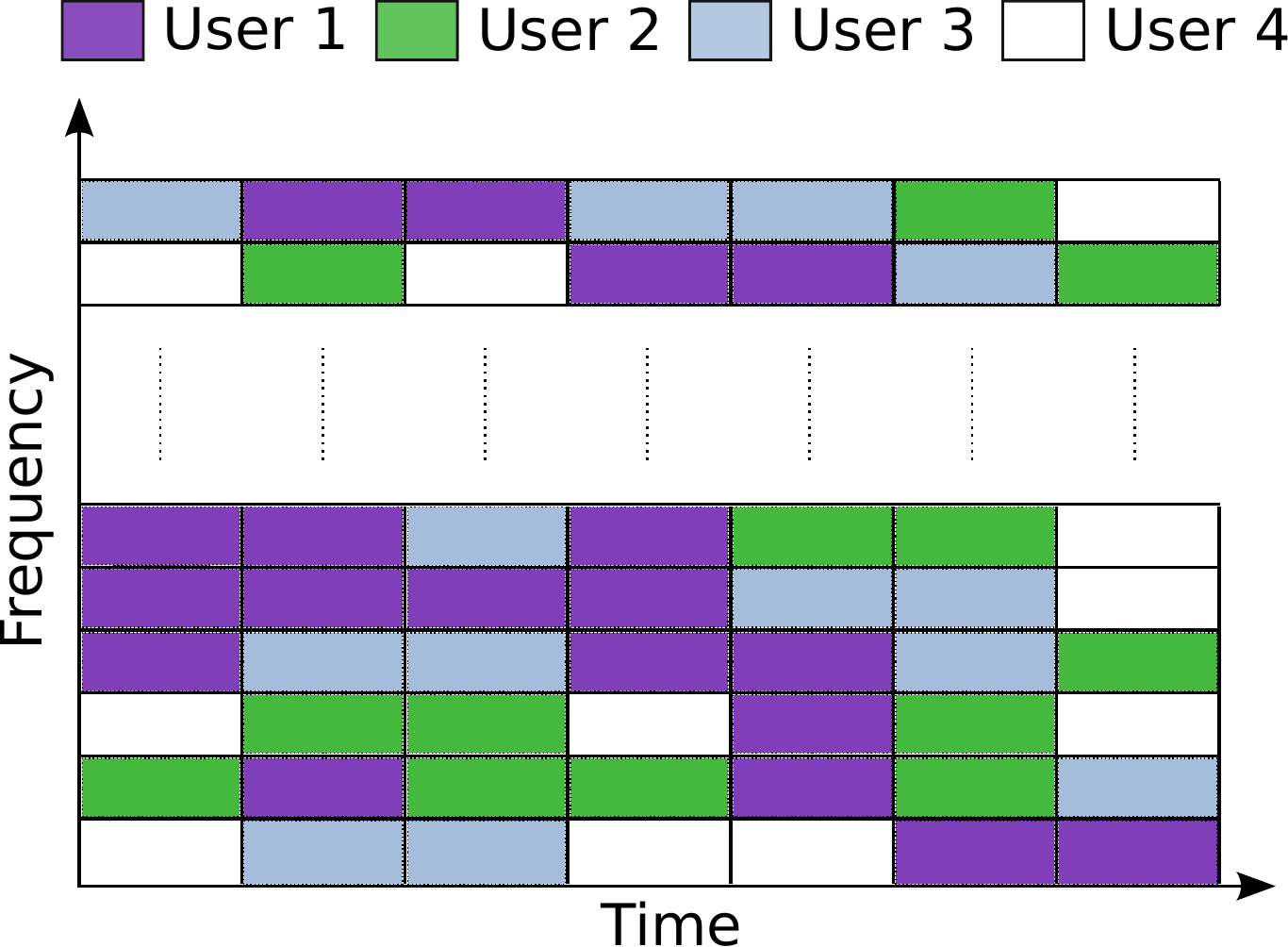}
\caption[Resource allocation in a combined \acs{OFDMA}/\acs{TDMA} system]{Example of resource allocation in a combined \ac{OFDMA}/\ac{TDMA} system constituting an \ac{OFDMA} frame~\cite{book:stb0901}.}
\label{fig:OFDMATDMA}
\end{figure}

\acrodef{QAM}[QAM]{Quadrature Amplitude Modulation}
\acrodef{QPSK}[QPSK]{Quadrature Phase Shift Keying}
In addition to this fine separation in frequency, in \ac{OFDM}, the transmission duration is divided into short slots to create an \emph{\ac{OFDM} block} with flat-fading characteristics in which one modulation symbol is transmitted (Multiple antenna technology allows the transmission of multiple symbols, as described in the following section). The modulation symbols can be from any modulation alphabet such as \ac{QPSK}, 16-\ac{QAM}, or 64-\ac{QAM}. In \ac{LTE}, such an \ac{OFDM} block is labeled a \emph{resource element}. The available bandwidth and a duration called \emph{frame} are divided into a number of resource elements. Multiple resource elements are combined to constitute \emph{resource blocks}. When \ac{OFDM} is used to grant multiple users access to the shared transmission medium, it is called \acf{OFDMA}. An illustration of a resulting \ac{OFDMA} frame consisting of a set of resource blocks scheduled to different users is shown in Fig.~\ref{fig:OFDMATDMA}. For a typical \ac{LTE} system with a  subcarrier spacing $\Delta_w = 15$\,kHz, a system bandwidth $W = 10$\,MHz, 12~subcarriers per resource block and a 10\%~guard band, the number of available resource blocks over the frequency domain is
\begin{equation}
 N = \frac{0.9 W}{12 \Delta_w}=50.
\end{equation}
\nomenclature{$W$}{Total system bandwidth in Hz}
\nomenclature{$\Delta_w$}{Subcarrier spacing in Hz}
\nomenclature{$N$}{Number of subcarriers}
\nomenclature{$T$}{Number of time slots}

Although a resource block consists of multiple subcarriers constituting resource elements, it is the smallest unit which can be individually scheduled, according to the \ac{LTE} standard~\cite{std:3gppts36321}. For this thesis, it is thus assumed that a resource block consists of one subcarrier and stretches over the duration of one time slot. A frame is comprised of $T$ time slots. For example, with $N=50$ and $T=10$, there are
\begin{equation}
 NT = 500
\end{equation}
resource blocks available for scheduling in an \ac{OFDMA} frame.

\begin{figure}
\centering
\includegraphics[width=0.9\textwidth]{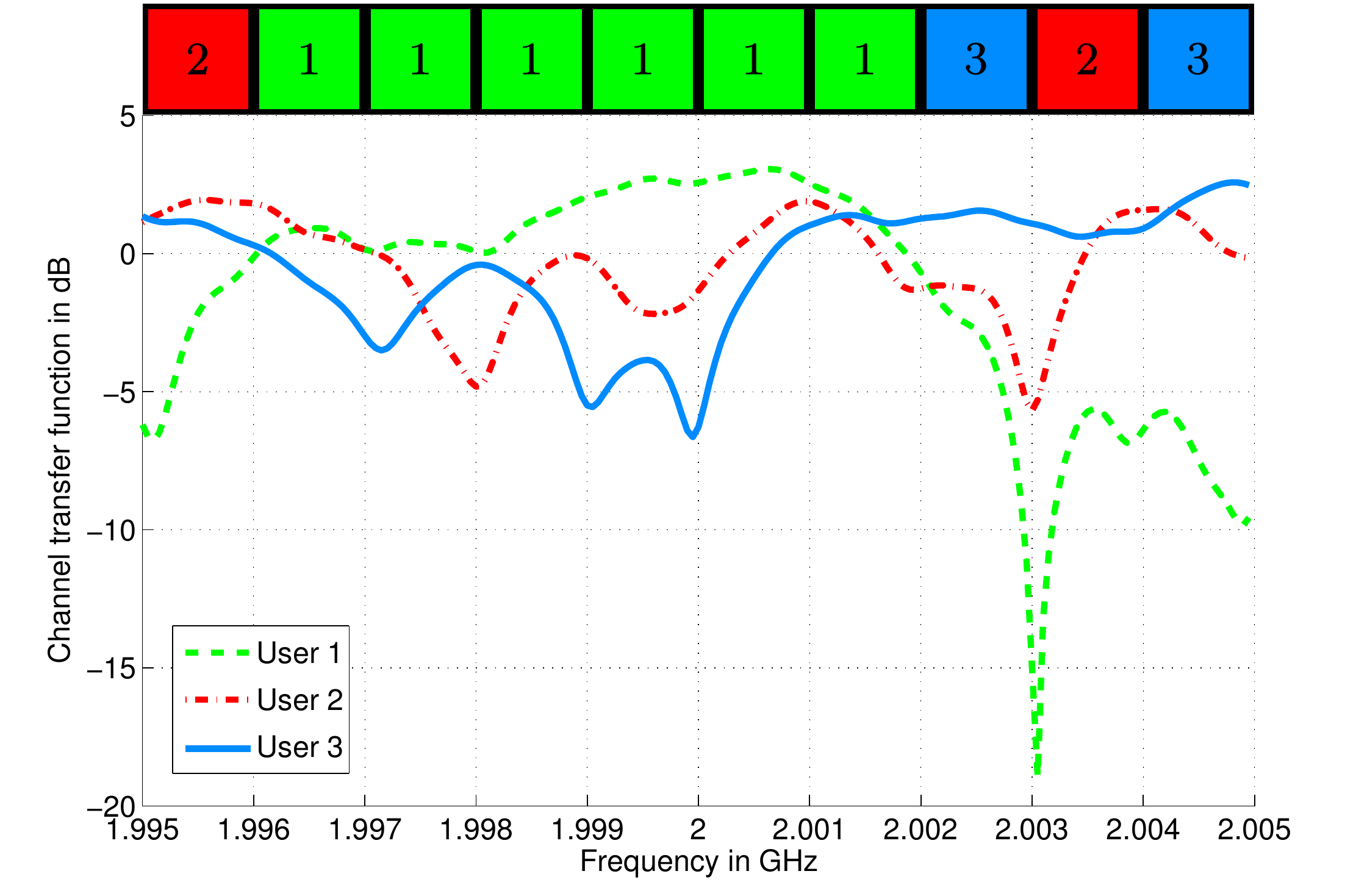}
\caption[Example of multi-user diversity]{Example of multi-user diversity scheduling. For each frequency chunk the user with the best channel is selected.}
\label{fig:multiuserdiversity}
\end{figure}

The option to schedule each resource block to any user in the system enables exploitation of an effect called \emph{multi-user diversity}. As the channels on each resource block between each user and the \ac{BS} is different, resource blocks can be scheduled such that only good channels are used. Statistically, over a certain bandwidth, number of time slots, and number of users, it is very probable that a good channel exists. This high probability results in an increase of the transmission data rate compared to a system which does not exploit this effect. An example of scheduling frequency by multiuser diversity is illustrated in Fig.~\ref{fig:multiuserdiversity}. Opportunistic scheduling between multiple users in a system is an important technique employed in Chapter~\ref{ch:cell}.

\subsection{Multiple Antenna Technology}
The transmission and reception of wireless signals with multiple antennas is a standard technique in modern communication systems~\cite{book:k0601}. Exploiting multi-path scattering, \ac{MIMO} techniques deliver significant performance enhancements in terms of data transmission rate and interference reduction. The term \emph{multi-path} refers to the arrival of transmitted signals at an intended receiver through differing angles, time delays and/or frequency shifts due to the scattering of electromagnetic waves in the environment, \ie~over multiple paths. Consequently, the received signal power fluctuates through the random superposition of the arriving multi-path signals. This random fluctuation allows the recognition and separation of the paths signals take between the multiple transmit antennas and multiple receive antennas, allowing to benefit from two important effects. First, like with multi-user diversity described in the previous section, having multiple signal paths reduces the probability of a deep fade. Hence, the reliability of the transmission is increased. The benefit of this effect is called the spatial diversity gain. Second, under suitable conditions such as rich scattering in the environment, separate data streams can be transmitted on the multiple paths. Thus, at the same time, on the same frequency spectrum, and in the same place, multiple data streams can be transmitted, increasing the data rate compared to a system with fewer antennas~\cite{book:b0701}. The benefit of this effect is called the spatial multiplexing gain.

\begin{figure}
\centering
\includegraphics[width=0.8\textwidth]{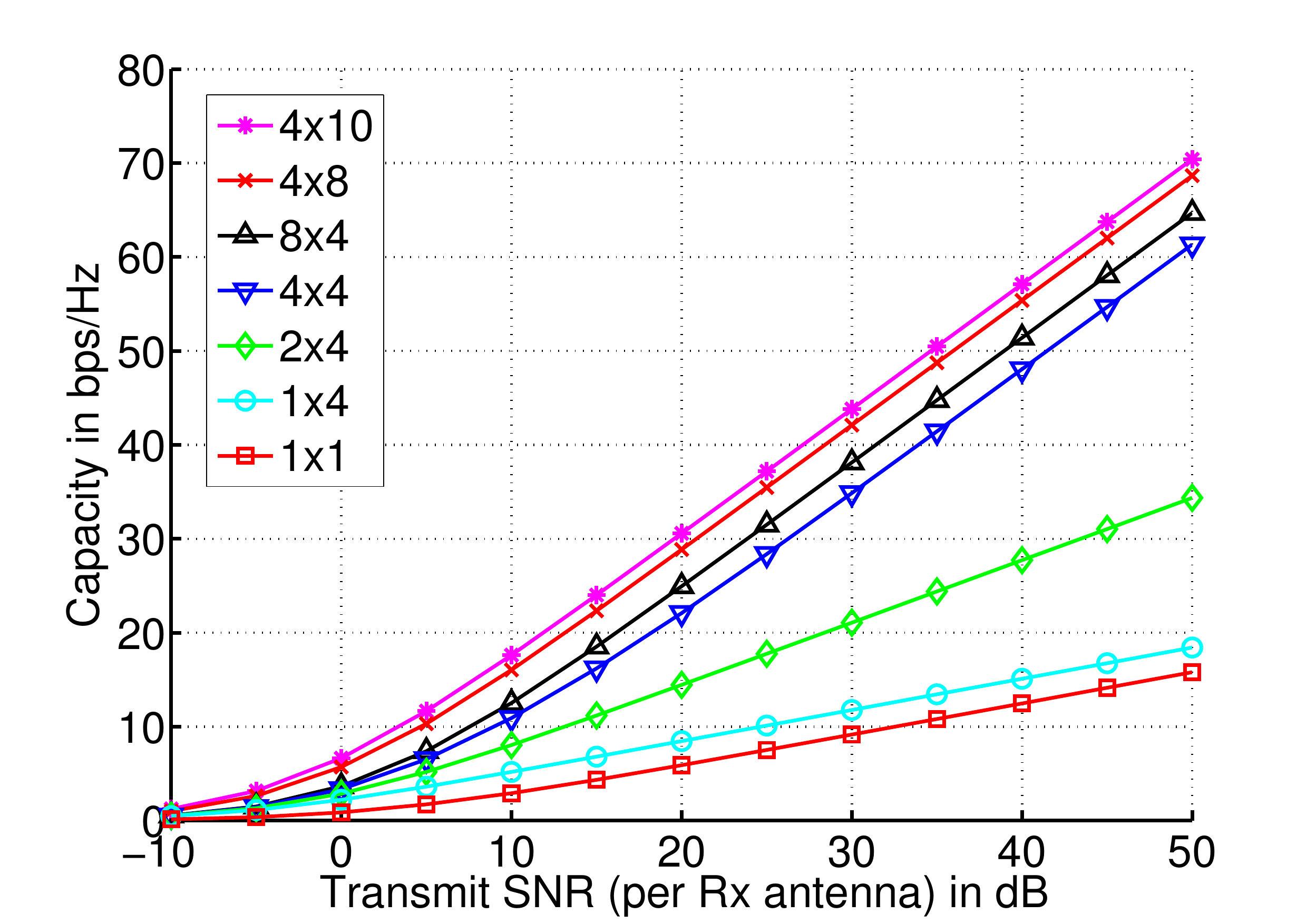}
\caption[Comparison of \ac{MIMO} capacities]{Power-fair \ac{MIMO} capacities at perfect \ac{CSI} at the receiver and statistical \ac{CSI} at the transmitter.}
\label{fig:EPCapacitiesNoTxFullRxCSI}
\end{figure}

Under the assumption of statistical \ac{CSI} at the transmitter, full \ac{CSI} at the receiver and equal power allocated to each antenna, the ergodic \ac{MIMO} link capacity is given by
\begin{subequations}
 \begin{align}
 C &= \mathbb{E}_{\B{H}} \left[  \log_2 \mathrm{det} \left( \B{I}_{M_{\mathrm{R}}} + \frac{P_{\mathrm{Tx}}}{D} \B{H} \B{H}^{\mathrm{H}} \right)   \right],
\label{eq:ergCDITCSIRMIMOcapacity} \\
   &= \displaystyle \sum_{i=1}^{min(D,M_{\textrm{R}})} \log_2 \left( 1 + \frac{P_{\mathrm{Tx}}}{D} \epsilon_i \right)
 \end{align}
\end{subequations}
with the expected value operator $\mathbb{E}_{\B{H}}(\cdot)$, transmit power $P_{\mathrm{Tx}}$, channel state matrix $\B{H}$, the number of transmit and receive antennas $D$ and $M_{\mathrm{R}}$, respectively, identity matrix of size $M_{\mathrm{R}}$, $\B{I}_{M_{\mathrm{R}}}$, and $\epsilon_i$ as the $i$-th singular value of $\B{H} \B{H}^{\mathrm{H}} $~\cite{gjjv0301}. The gains in capacity through the use of multiple antennas can be charted by generating a number of Rayleigh distributed channel coefficients $\B{H}$ and plotting the achieved capacities. Figure~\ref{fig:EPCapacitiesNoTxFullRxCSI} shows an illustration of achievable capacities, revealing two important characteristics of the \ac{MIMO} capacity. First, it can be seen that capacity grows linearly with \ac{SNR} at high \acp{SNR}. Second, the capacity grows linearly with $\mathrm{min}(D, M_{\mathrm{R}})$. Thus, more antennas yield a higher capacity. In this thesis, up to four receive and transmit antennas are considered due to limitations detailed in Section~\ref{step1}.
\nomenclature{$C$}{Ergodic capacity in bps/Hz}
\nomenclature{$\mathbb{E}_{\B{H}}(\cdot)$}{Expected value operator}
\nomenclature{$M_{\mathrm{R}}$}{Number of receiver antennas}
\nomenclature{$\B{H}$}{Channel state matrix}
\nomenclature{$\epsilon_i$}{$i$-th eigenvalue}
\nomenclature{$P_{\mathrm{Tx}}$}{Transmission power of a \ac{BS} at the input of the antenna interface in W}

While raising the number of antennas increases the capacity, it also increases the complexity of transmitters and receivers. As will be shown in Chapter~\ref{ch:device}, this increase in complexity is accompanied by a larger power consumption. It is studied in Chapter~\ref{ch:cell} how to trade off power consumption with capacity for energy efficiency by changing the \ac{MIMO} mode through \ac{AA}.


\subsection{Convex Optimization}
\label{cvx}
An important topic employed in Chapter~\ref{ch:cell} is convex optimization which is briefly introduced here. 

A general optimization problem is often written as
\begin{equation}
\begin{aligned}
 &\text{minimize} 	&f_0(\B{x}) & 			&	\\
 &\text{subject to} 	&f_i(\B{x}) &\leq b_i, 	 	&i=1,\cdots,m\\
 &			&h_i(\B{x}) &=0, 		 	&i=1,\cdots,p.
\end{aligned}
\label{eq:OPboyd}
\end{equation}
The problem consists of the optimization variable $\B{x}=(x_1,\cdots,x_n)$, the objective function $f_0:\mathbb{R}^n \mapsto \mathbb{R}$, the inequality constraint functions $f_i:\mathbb{R}^n \mapsto \mathbb{R}$, $i=1,\cdots,m$ and equality constraint functions $h_i:\mathbb{R}^n \mapsto \mathbb{R},i=1,\cdots,p$, and the constraints $b_1,\cdots,b_m$. $\B{x}^*$ is the solution of the problem~\eqref{eq:OPboyd} if for any $z$, $f_1(\B{z}) \leq b_1,\ldots,f_m(\B{z})\leq b_m$ and $h_1(\B{z})=0,\ldots,h_p(\B{z})=0$, it holds that $f_0(\B{z}) \geq f_0(\B{x}^*)$~\cite{book:bv0401}.

This form of optimization problems is convex, if the functions $f_0,\ldots,f_m$ satisfy
\begin{equation}
 f_j(\alpha \B{x} + \beta \B{y}) \leq \alpha f_j(\B{x}) + \beta f_j(\B{y})
\end{equation}
for all $\B{x},\B{y} \in \mathbb{R}^n$, all $\alpha, \beta \in \mathbb{R}$ with $\alpha + \beta = 1, \alpha \geq 0, \beta \geq 0$, and $j=0,\ldots,m$. In other words, it is convex if the objective function as well as all constraints are convex. Furthermore, any equality constraints $h_i$ have to be affine as any
\begin{equation}
 h(\B{x}) = 0
\end{equation}
can be represented as
\begin{subequations}
\begin{align}
 h(\B{x}) \leq 0,\label{eq:hx}\\
 -h(\B{x}) \geq 0.\label{eq:mhx}
\end{align}
\end{subequations}
If $h(\B{x})$ were convex, $-h(\B{x})$ would be concave. The only way for both \eqref{eq:hx} and \eqref{eq:mhx} to be convex is when $h(\B{x})$ is affine.

\begin{figure}
\centering
\includegraphics[width=0.6\textwidth]{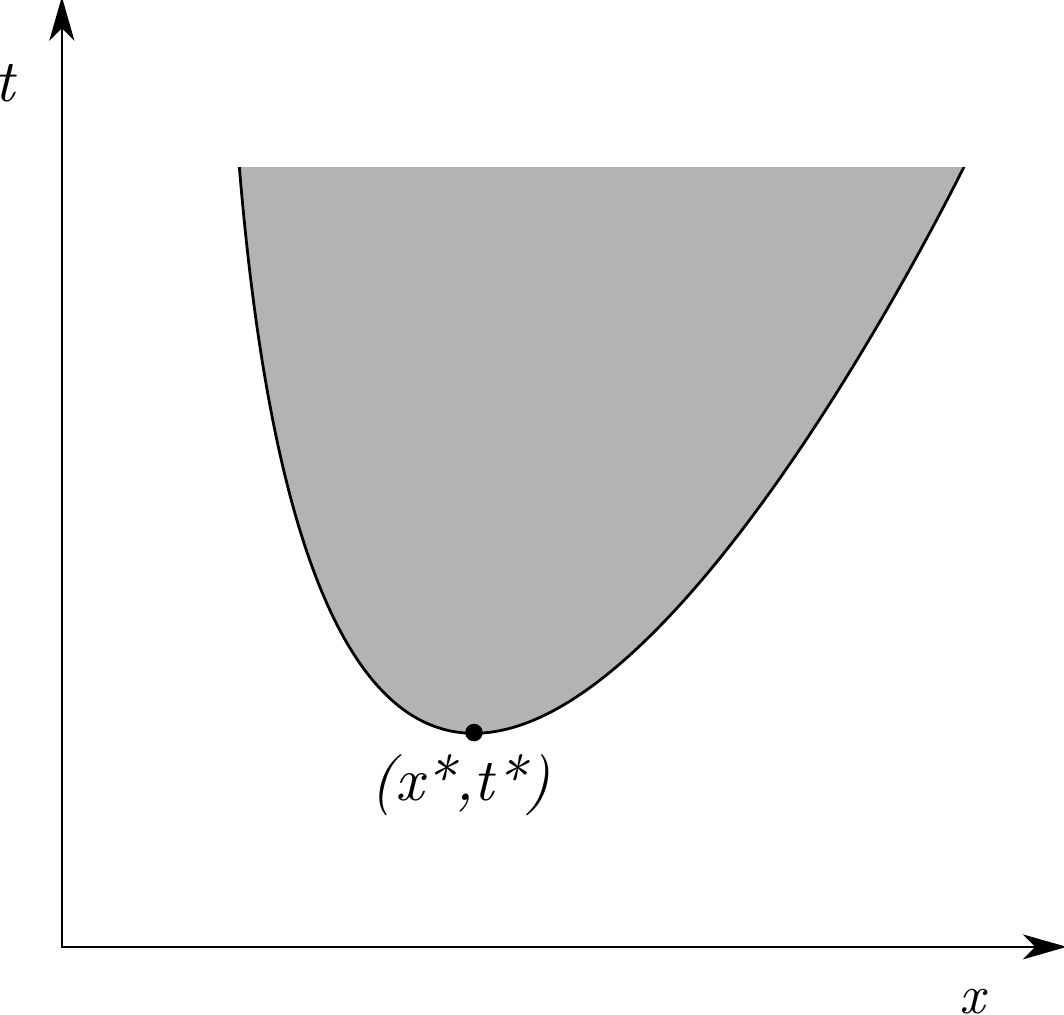}
\caption[Geometric interpretation of a simple convex problem]{Geometric interpretation of a simple convex problem~\cite{book:bv0401}.}
\label{fig:epigraph}
\end{figure}

The important property of convex optimization problems is that they can be solved by interior point methods and that the local optimum is always also the global solution~\cite{book:pe1001}. Furthermore, convex optimization problems can often be solved very efficiently enabling their application even in real-time systems such as resource schedulers. A one-dimensional convex optimization problem without constraints can be graphically illustrated. An example is shown in Fig.~\ref{fig:epigraph} for some $t=f_0(x)$.

A simple example of a two-dimensional convex optimization problem is
\begin{equation}
\begin{aligned}
 &\text{minimize} 	&f_0(\B{z})&=z_1^2 + z_2 + 1		\\
 &\text{subject to} 	&f_1(\B{z})&=-1-z_1 \leq 0\\
 &			&h_1(\B{z})&=z_2 - 1 =0.
\end{aligned}
\label{eq:OPexample}
\end{equation}
The second gradient of $f_0(\B{z})$ is
\begin{equation}
 \nabla^2 f_0(\B{z}) = 
\begin{pmatrix}
 2 & 0\\
 0 & 0
\end{pmatrix}.
\end{equation}
The objective function and its inequality constraints are convex as $\nabla^2 f_0(\B{z})$ is greater or equal to zero on all entries. The equality constraint is affine. The sample problem is thus a convex optimization problem. By inspection of \eqref{eq:OPexample}, it can be found that the solution is
\begin{equation}
 \B{z}^* = (0,1).
\end{equation}
Either proving convexity of a problem or transforming a problem into a convex problem are ways to efficiently solve optimization problems.


\subsection{Network Simulation}

For many problems in communications research, analytical solutions have not yet been found. To still be able to evaluate such problems, researchers resort to repeated independent trial executions of a given problem. The statistical outcomes of these trials provide insight into the problems and potential solutions. These trials are often performed as computer simulations---rather than in hardware---to avoid costly and time consuming prototyping. Such computer simulations allow to model specific problems, vary their nature and generally experiment on them. They can also be used to test solutions and assess whether inventions deliver the anticipated results. As they prove to be such a versatile tool, they are used also in this thesis to assess the proposed solutions, when analytical solutions cannot be found.

\begin{figure}
\centering
\includegraphics[width=0.8\textwidth]{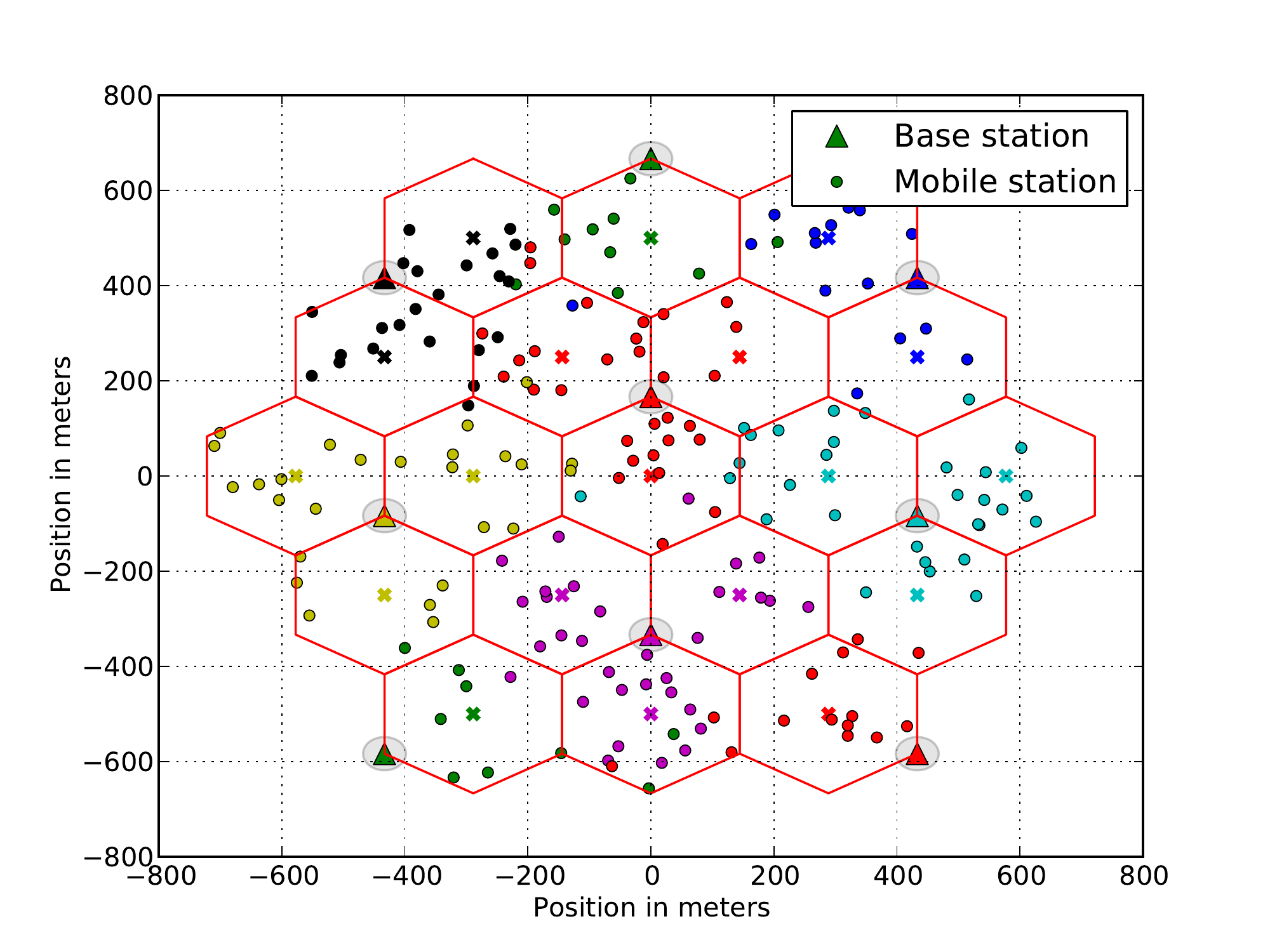}
\caption[A network in the network simulator]{A sample network generated by the network simulator.}
\label{fig:simnetwork}
\end{figure}

More specifically, cellular networks are simulated which are as large as the computational constraints allow. For example, see Fig.~\ref{fig:simnetwork} for the layout of a typical network simulation consisting of \acp{UE} and \acp{BS} distributed over an area. 

\acrodef{IPOPT}[IPOPT]{Interior Point OPTimizer}
The network simulator which produced the results shown in this thesis was developed from the ground up by the author and contains the following elements:
\begin{itemize}
 \item A configuration interface via configuration files,
 \item Network generation consisting of:
 \begin{itemize}
  \item Geometric modules for the distribution of physical entities onto hexagonally arranged three-cell \ac{BS} sites,
  \item Path loss and shadowing modules for the association of \acp{UE} to \acp{BS} as defined by the \ac{3GPP}~\cite{std:3gpp-faeutrapla},
  \item \ac{OFDM} \ac{MIMO} \ac{SINR} calculation,
 \end{itemize}
 \item A frequency-selective fading model calibrated according to the \ac{WINNER} air interface configuration\cite{std:ist-chanmod2},
 \item Margin-adaptive resource allocation,
 \item Several modules for the algorithms described in Chapters~\ref{ch:cell} and~\ref{ch:network},
 \item Interior point convex optimization via the \ac{IPOPT}~\cite{wl01},
 \item Data collection, analysis and plotting tools,
 \item A unit test suite,
 \item Detailed event logging,
 \item Parallelization via the high-throughput parallelization system HTCondor~\cite{t1301}.
\end{itemize}

In addition to the list of simulator elements, it is also important to note what is not modelled. For example, capacity is calculated based on \ac{SINR} by the Shannon limit. Bit errors, retransmission, error correction or protocols are not included; neither are pilots, channel sounding or \ac{LTE} signaling and, thus, imperfect \ac{CSI}. \ac{UE} mobility is modelled as a fast-fading component rather than a change in position on the network map.

\begin{figure}
\centering
\includegraphics[width=0.6\textwidth]{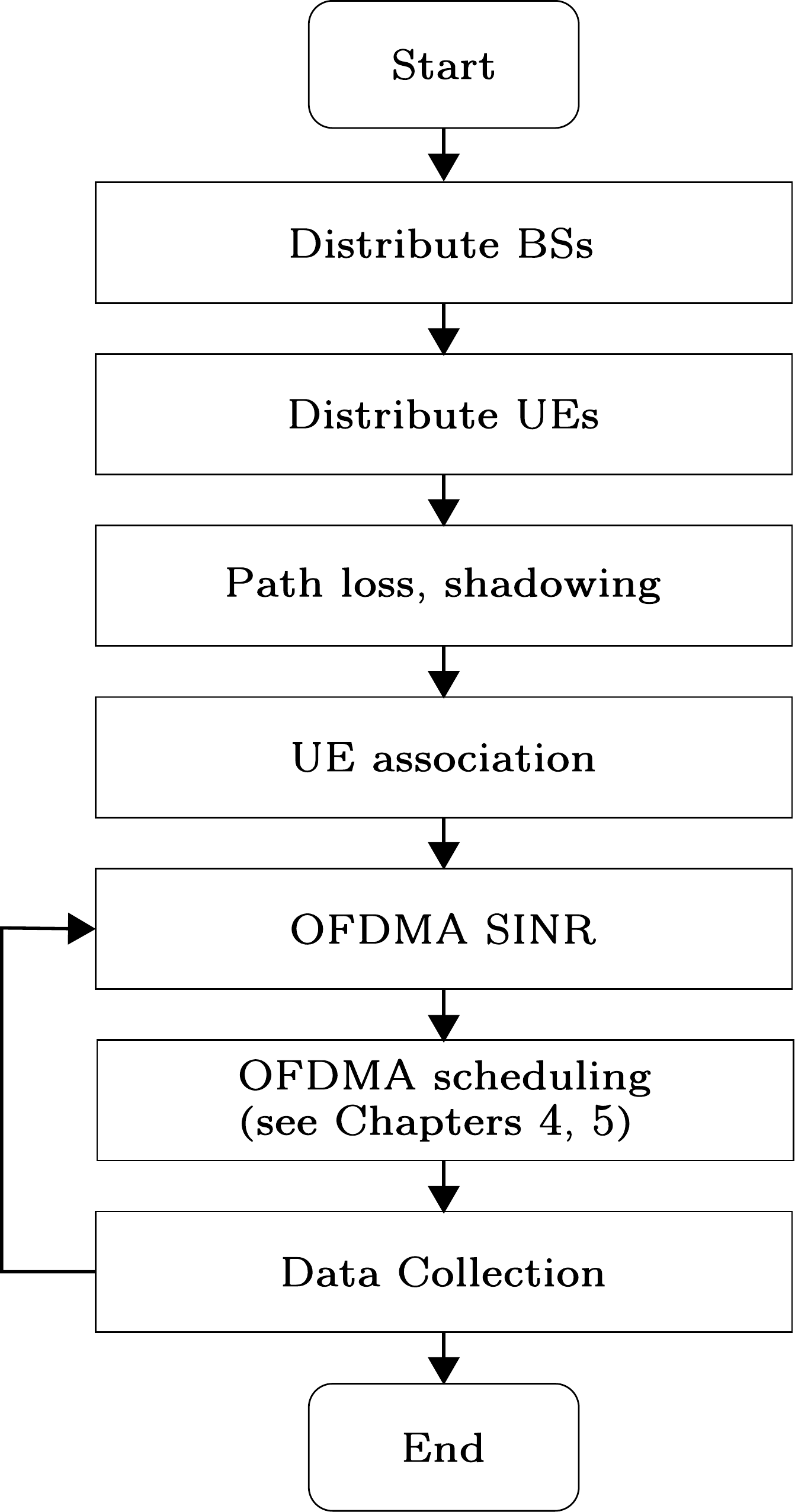}
\caption[Sample simulation flowchart]{Sample simulation flowchart.}
\label{fig:simflowchart}
\end{figure}

A typical simulation consists of a large number of independent trials executed in Monte Carlo fashion, \ie~in parallel, thus yielding statistical outcomes over all trials. A flowchart of the most important steps involved in a simulation run is provided in Fig.~\ref{fig:simflowchart}. To ensure sufficient consideration of interference, data is only collected from cells which are surrounded by at least two layers (tiers) of cells causing interference. 

The simulators which produced the results presented in Chapters~\ref{ch:device} and \ref{ch:cell} were written as MATLAB~\cite{01a} scripts. Required packages are the Optimization and Statistics Toolboxes. An optional package is the Parallel Computing Toolbox. The network simulator employed in Chapter~\ref{ch:network} was written in python~\cite{01} using numpy, scipy, matplotlib, virtualenv, \ac{IPOPT}, ATLAS, LAPACK, MKL and ACML libraries. Depending on the network size, extensive simulation runs can take up to several days on a parallelized system. Unlike the highly optimized convex optimization processes, computation of \acp{SINR} and data handling are very time consuming, causing long simulation run times.

\section{Summary}
This chapter served to provide the motivation for \ac{BS} power saving, establish the literature context, and briefly introduce key technical concepts. To begin with, it was explained that cellular \acp{BS} contribute significantly to the power consumption and \coo~emission of mobile networks and that they do so during operation. Low load situations, such as night time, were identified as opportunities for reducing \ac{BS} power consumption. The necessity of observing \ac{BS} design along with \ac{BS} operation for power saving was emphasized which is reflected in the structure of this thesis. Next, the Green Radio literature was discussed. In the last part of this chapter, key technical concepts were introduced which reappear in the following chapters. \ac{LTE}, multicarrier and multiple antenna technology, and convex optimization are all important topics applied in this thesis. Furthermore, the computer simulation of cellular networks was described in detail as it was used for producing numerical results where analytical solutions could not be found.

\acresetall 

\renewcommand{\dir}{chapter3/chapter/}
\chapter{Power Saving on the Device Level}
\label{ch:device}

\section{Overview}
Power models describe how much power a device consumes in different configurations or scenarios. Depending on their depth, power models may also reveal architectural details and allow comparisons between different scenarios or assist the exploration of configurations which are not available for measurement in laboratories. 

In energy efficiency research, power models are the basis of any power consumption analysis. Solid models are required to produce reliable results, since the model assumptions significantly affect the energy efficiency solutions proposed. For example, if a model were to falsely describe heat dissipation as a major contributor to power consumption, research efforts with regard to the reduction of heat dissipation might be ineffectively spent. Therefore, it is immensely important to have reliable power models such as the ones presented in this chapter.

In this chapter, three levels of detail of a power model for \ac{LTE} \acp{BS} are derived. The reason is that higher levels of detail allow ensuring the reliability of the model, manipulating specific aspects and parameters and recognizing the underlying architecture. In contrast, lower levels of detail abstract the model specifics in favor of applicability. The goal of this chapter is to allow the reader to first follow the construction of the component model and later the simplification towards the affine model.

The remainder of this chapter is structured as follows. Section~\ref{ch3:challenges} summarizes the challenges encountered when modeling \ac{BS} power consumption. Existing power models are discussed in Section~\ref{ch3:existing}. Section~\ref{ch3:componentmodel} describes the \emph{component power model} in detail. This model reveals the sources of power consumption within cellular \acp{BS}. In Section~\ref{parameterizedmodel}, the complex component model is parameterized into a \emph{parameterized power model} such that only those parameters which are often manipulated for power saving remain, greatly reducing the model complexity. In Section~\ref{affinemodel}, an \emph{affine power model} approximation is defined. The affine mapping contained in this model allows integrating it into analytical problems in Chapters~\ref{ch:cell} and \ref{ch:network}. Section~\ref{ch3:summary} summarizes the chapter and compares the three models with regard to their complexity.

The component model was developed in collaboration with the \ac{EARTH} project. Hardware manufacturers have provided the experimental data on individual components. The author of this thesis has contributed to this power model by capturing it in MATLAB source code and aligning its concurrent development between EARTH partners. The model has been published and applied in parts in~\cite{std:earthd23,aggsoisgd1101,aggsoigdb1101,ddgfahwsr1201}. The detailed description of the model in this chapter is an original contribution. The parameterized model has been submitted for publication in~\cite{hag1301}.

\section{Challenges in Power Modeling}
\label{ch3:challenges}
Generic modeling of \ac{BS} power consumption is not a trivial task for multiple reasons. First, in order to protect their industrial designs, equipment vendors and manufacturers are hesitant to reveal the architecture of their products. They go so far as to contractually prohibit the opening of device casings for customers. Consequently, it is difficult to understand the contributing components to power consumption within cellular \acp{BS}. Second, vendors and manufacturers have differing designs, such that a power model may either only be valid for a single brand of devices or that multiple models would have to be developed. Third, architectures and technologies continually evolve, weakening the long-term applicability of power models. As such, a particular model may only be valid for a certain generation of \acp{BS}.

All of these problems have been addressed in the component power model as follows: By the combined effort of experts from multiple manufacturers, general common architectures were agreed upon without revealing competitive technical differences. These common architectures also helped bridge the design differences present between competing manufacturers. Finally, to ensure the lasting applicability of the model, foreseen technical advances were included in the model. As \ac{LTE} is the prominent cellular network technology of the coming decade (see Section~\ref{background}), the component power model describes an \ac{LTE} \ac{BS}.

The data sets provided in this chapter were all obtained through measurements within the EARTH project.

\section{Existing Power Models}
\label{ch3:existing}
In the literature, three distinct power models for wireless transmitters are proposed and applied. 

While in need of a power model for the comparison of the power efficiency of \ac{MIMO} with \ac{SISO} transmission, Cui~\ea~\cite{cgb0401} derived a first technology-independent model for the total power consumption of a generic radio transceiver. It was established that power consumption can be divided into the power consumption of the \ac{PA} and the power consumption of all other circuit blocks. The power consumption of the \ac{PA} was described to vary with transmission settings like the number of antennas, the energy per bit, the bit rate, and the channel gain, while the consumption of circuit blocks was assumed a constant. Note that this power model is not a power model for cellular \acp{BS}, but for a generic wireless transceiver. Yet, it represents the first steps towards modeling the power consumption of wireless devices.

Arnold~\ea~\cite{arfb1001} derive the power consumption of cellular \acp{BS} by considering that a typical \ac{BS} consists of multiple sectors, has multiple \acp{PA}, a cooling unit, and a power supply loss. The remaining components' consumption is considered a constant. This model is accurate in its assumptions. But it only has data available for \ac{GSM} and \ac{UMTS} \acp{BS}. For \ac{LTE}, the model is based on predictions and differs significantly from the values presented in this chapter.

Deruyck~\ea~\cite{djm1201} used extensive experimental measurements to determine the power consumption of several types of \ac{SOTA} \acp{BS}, like the macro cell, micro cell and femto cell. The power consumption of a \ac{BS} is assumed to be a constant, independent of the traffic load. Thus, this model is not suitable for the exploration of \ac{BS} operating power consumption as is required in Chapters~\ref{ch:cell} and \ref{ch:network}.

The model described in the following has partly been published as an element of an \ac{EARTH} project deliverable~\cite{std:earthd23}. It has been published in abbreviated form in~\cite{aggsoisgd1101,aggsoigdb1101,ddgfahwsr1201}.

\section{The Component Power Model}
\label{ch3:componentmodel}
Before the description of the model, it is necessary to make some introductory remarks.

\subsection{Remarks}

First, this model describes the supply power consumption of a \ac{LTE} cellular \ac{BS} and is derived as the sum of the consumption of its subcomponents. The consumption of each subcomponent is derived from experimental measurements. The prediction of the dependence on operating parameters is derived from their architecture.

Second, the model is an instantaneous model. It does not model effects which occur over time, such as power up or power down. Rather it models instantaneous power consumption at a certain configuration. When estimating power consumption over a time duration consisting of different configurations, a weighted averaging of the instantaneous values needs to be performed.

Third, this model encompasses the architecture of an \ac{LTE} \ac{BS} entering the market in the year 2012. For figures in this chapter, the system parameters in Table~\ref{tab:modelparams} are used. It represents the architecture found in most \acp{BS} which is foreseen to remain unchanged for the coming five to ten years. Note that this chapter is not concerned with how the hardware could be improved. Such hardware improvements were published in \cite{std:earthd43}.

\begin{table}
\centering
 \begin{tabular}{c | c}
  Parameter & Value \\
\hline
Feeder loss, $\sigma_{\mathrm{feed}}$ & 0.5 \\
Number of radio chains, $D$ & 2 \\
Transmitted power, $P_{\mathrm{Tx}}$ & 40~W\\
Number of \ac{BS} sectors, $M_{\mathrm{sec}}$ & 3\\
System bandwidth, $W$ & 10~MHz\\
 \end{tabular}
\caption[Model parameters assumed for figures of Chapter~\ref{ch:device}]{Model parameters used in figures of this chapter.}
\label{tab:modelparams}
\end{table}

\subsection{The Components of a \ac{BS}}
\begin{figure}
\centering
\includegraphics[width=0.8\textwidth]{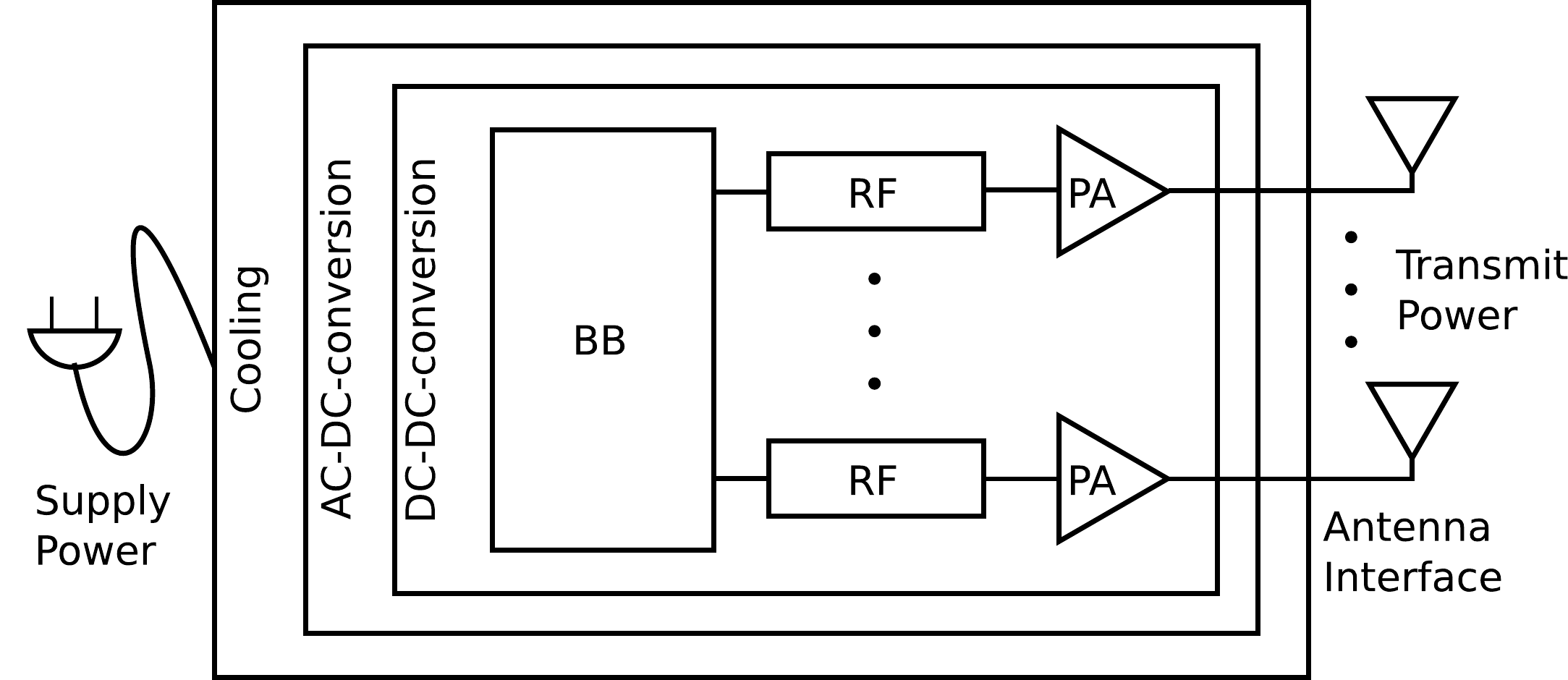}
\caption[The components of the modelled \ac{BS}]{The components of the modelled \ac{BS}.}
\label{fig:BSparts}
\end{figure}

The modelled \ac{LTE} macro \ac{BS} consists of a transceiver with multiple antennas. The transceiver comprises one or more lossy antenna interfaces, \acp{PA} and \ac{RF} transceiver sections as well as one \ac{BB} interface including both a receiver section for uplink and transmitter section for downlink, one \ac{DC}-\ac{DC} power supply regulation, one active cooling system and one mains \ac{AC}-\ac{DC} power supply for connection to the electrical power grid. Fig.~\ref{fig:BSparts} shows an illustration.

In practice, a \ac{BS} may be serving multiple cells by sectorization. In that case, there may be overlap between the components and the sectors. For example, a single cooling cabinet may be used to house the electronics for all sectors. However, for purposes of modeling, it is assumed that a \ac{BS} only serves a single cell. Thus, the power consumption of a \ac{BS} for multiple cells can be found as the sum of the individuals.

In currently prevailing architecture as portrayed in Fig.~\ref{fig:BSparts}, the \ac{BS} contains one or multiple so-called \emph{radio chains} each consisting of one \ac{RF} transceiver, one \ac{PA} and one antenna. Thus, the number of radio chains, antennas, \acp{PA}, and \ac{RF} transceivers is identical and terminology can be used interchangeably.


In the following, the power consumption of each individual component is inspected to generate the component power model for the entire \ac{BS}. Components are treated from right to left as depicted in Fig.~\ref{fig:BSparts}.

\subsubsection{Antenna Interface}
This passive component is listed here for completeness, although it has no power consumption characteristics. For macro \ac{BS} with long feeder cables between the \ac{PA} and the antenna, a feeder loss has to be compensated for, which typically reduces the signal power by around 3~dB. For the model, this feeder loss is included in the modeling of the \ac{PA}.

\subsubsection{\ac{PA}}
The \ac{PA} amplifies the signal before it is transmitted by the antenna. Due to low achievable efficiencies, \acp{PA} contribute significantly to the overall power consumption if high powers are targeted at the antenna, as described in Section~\ref{sect:eebs}.



Note that when comparing transmission powers in this study, it is referred to as the sum measured at the inputs of all antenna interfaces. Thus, feeder cable losses are included and \acp{BS} with differing numbers of antennas can be compared fairly with regard to their transmission power. 

\begin{figure}
\centering
\includegraphics[width=0.6\textwidth]{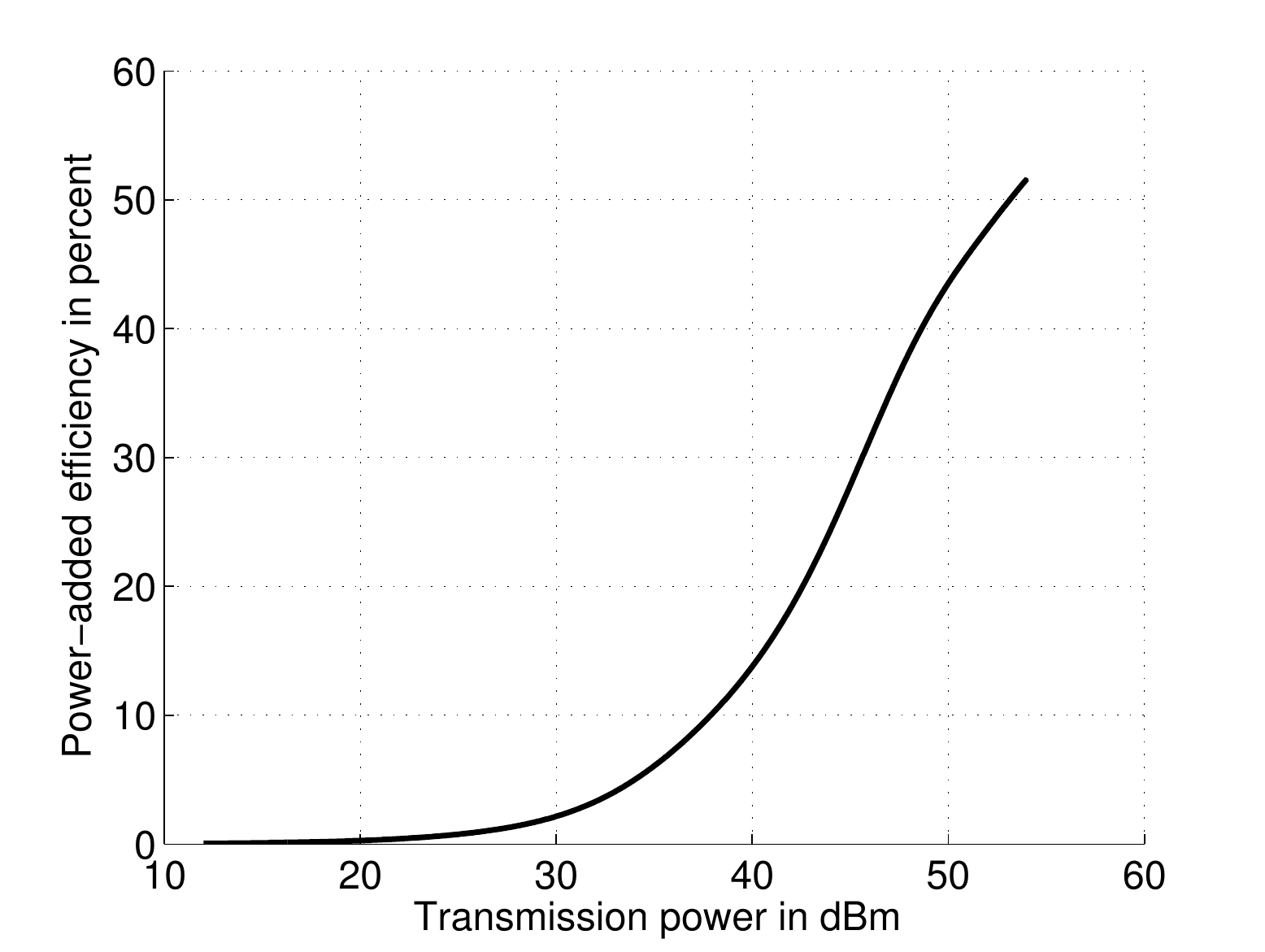}
\caption[The power-added efficiency over the maximum output power]{The power-added efficiency, $l_{\mathrm{PA}}(\cdot)$, over the maximum output power as measured within the \ac{EARTH} project.}
\label{fig:PAeff}
\end{figure}

Due to its non-linearity, the \ac{PA} modeling involves measurement results of the power-added efficiency. The power-added efficiency of the \ac{PA} is looked up during modeling via the function $l_{\mathrm{PA}}(\cdot)$. The efficiency function is shown in Fig.~\ref{fig:PAeff}. The efficiency of the \ac{PA} can reach up to 54\% at very high transmission powers. However, these high efficiencies can rarely be achieved in operation due to the strong transmission power fluctuation of \ac{OFDMA} signals and power control. At low transmission powers, \acp{PA} have very low efficiencies. This effect cancels the power saving of power control to some degree.

At constant power spectral density, the power output of each \ac{PA}, $P_{\mathrm{out}}$ in W, is a function of the transmission power, $P_{\mathrm{Tx}}$ in W, the number of antennas, $D$, the share of the maximum bandwidth used for transmission, $f\in[0,1]$, and an adjustment for feeder losses, $\sigma_{\mathrm{feed}}\in[0,1]$ is given as
\begin{equation}
 P_{\mathrm{out}} =  \frac{ P_{\mathrm{Tx}} f (1-\sigma_{\mathrm{feed}})}{D}.
\end{equation}
\nomenclature{$P_{\mathrm{out}}$}{The power output of a single \ac{PA} in W}%
\nomenclature{$f$}{Share of available bandwidth used for transmission}%
\nomenclature{$\sigma_{\mathrm{feed}}$}{Feeder loss}%
\nomenclature{$D$}{Number of \ac{BS} antennas/radio chains}%

The power consumption of the \acp{PA} in W is then determined by the efficiency and the number of \acp{PA}, such that
\begin{equation}
 P_{\mathrm{PA}} = D \frac{P_{\mathrm{out}}}{l_{\mathrm{PA}}(P_{\mathrm{out}})}.
\end{equation}
\nomenclature{$P_{\mathrm{PA}} $}{Power consumption of all \acp{PA} in a \ac{BS} in W}%
\nomenclature{$l_{\mathrm{PA}}$}{\ac{PA} efficiency}%

The power consumption in sleep mode, $P_{\mathrm{PA,S}}$ in W, scales with the number of \acp{PA}. The power consumption of one \ac{PA} in sleep mode, $P_{\mathrm{PA,S}}'$, was found by experimental measurement, such that
\begin{equation}
  P_{\mathrm{PA,S}} = D P_{\mathrm{PA,S}}',
\end{equation}
with $P_{\mathrm{PA,S}}' = 27.75$\,W.
\nomenclature{$P_{\mathrm{PA,S}}$}{Power consumption of all \acp{PA} in a \ac{BS} in sleep mode in W}%
\nomenclature{$P_{\mathrm{PA,S}}'$}{Power consumption of a \ac{PA} in sleep mode in W}%

\begin{figure}
\centering
\includegraphics[width=0.6\textwidth]{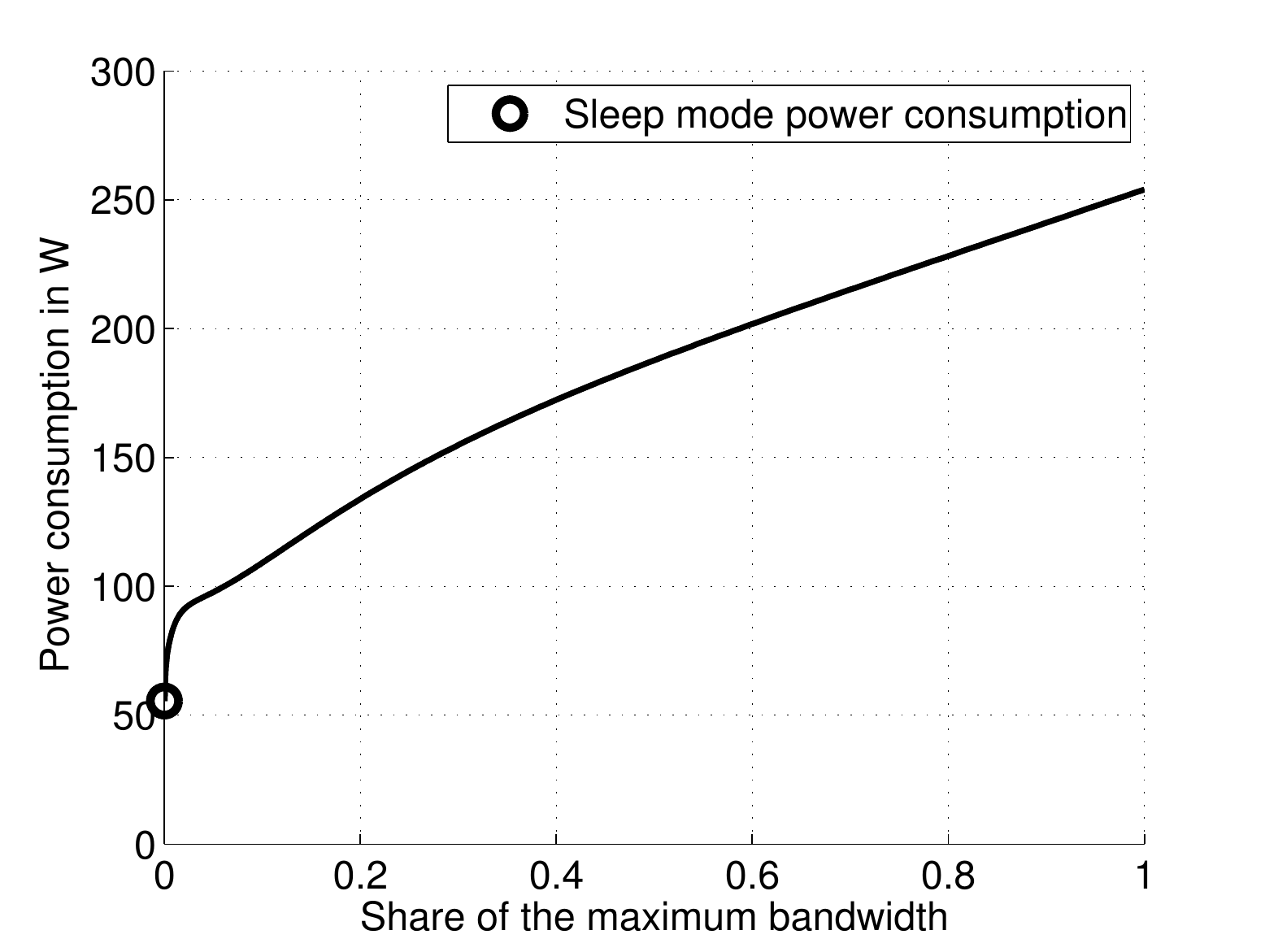}
\caption[$P_{\mathrm{PA}}$ as a function of bandwidth used]{$P_{\mathrm{PA}}$ as a function of bandwidth used.}
\label{fig:P_PA}
\end{figure}

The power consumption of the \ac{PA} over $f$ is shown in Fig.~\ref{fig:P_PA}. The effect of the reduced \ac{PA} efficiency at lower transmission powers results in an almost linear relationship of bandwidth/transmission power to power consumption. At a maximum of 250~W, the \acp{PA} can consume a significant amount of power. However, this can be mitigated by reducing the transmission bandwidth. Also, as an analog component, a \ac{PA} can be switched to a sleep mode with low power consumption quickly which has just above 50~W power consumption.

\subsubsection{\ac{RF} Transceiver}
The \ac{RF} transceiver provides the signal conversion between the digital and analog domain as well as frequency conversion between the \ac{BB} and \ac{RF} frequencies. It consists of a large number of elements such as \ac{IQ}-Modulators, voltage-controlled oscillators, mixers, \acp{DAC} and \acp{ADC}, low-noise and variable gain amplifiers and clocks. Power consumption of each of these elements is independent of external factors in measurement. The values and the resulting sum consumption are shown in Table~\ref{tab:P_RF}. Thus, the power consumption of the \ac{RF} transceiver is modelled as a constant. As one \ac{RF} transceiver is installed in each antenna chain, \ac{RF} power consumption scales with the number of antenna chains.

The power consumption of all \ac{RF} transceivers, $P_{\mathrm{RF}}$ in W, is modelled as a linear function of an individual transceiver, $P_{\mathrm{RF}}^\prime$, with
\begin{equation}
 P_{\mathrm{RF}} = D P_{\mathrm{RF}}^\prime,
\end{equation}
where $P_{\mathrm{RF}}^\prime$ in W is comprised of reference values for commercially available components as shown in Table~\ref{tab:P_RF}. A technology scaling factor, $a_{\mathrm{TECH}}$, is included to account for different generations of \ac{RF} transceivers.
\nomenclature{$P_{\mathrm{RF}}$}{Power consumption of all \ac{RF} transceivers in W}%
\nomenclature{$P_{\mathrm{RF}}^\prime$}{Power consumption of an \ac{RF} transceiver in W}%

\begin{table}
\centering
\begin{tabular}{l |l c}
Block element &  & Power consumption in mW \\ 
\hline
\ac{IQ} Modulator, DL& & 1000\\
Variable attenuator, DL & & 10 \\
Buffer, DL & &300 \\
\ac{VCO}, DL &  &340 \\
Feedback mixer, DL && 1000 \\
Clock Generation, DL && 990 \\
\ac{DAC}, DL && 1370 \\
\ac{ADC}, DL & &730 \\
\ac{LNA} 1, UL && 300 \\
Variable Attenuator, UL & &10\\
\ac{LNA} 2, UL & &1000 \\
Dual mixer, UL & &1000 \\
Variable Gain Amplifier, UL && 650 \\
Clock generation, UL & &990 \\
\ac{ADC} 1, UL & &1190 \\
\hline
Sum $P_{\mathrm{RF}}''$ && 10880\\
Technology scaling $a_{\mathrm{TECH}}$&& 1.19\\
\hline \hline
$P_{\mathrm{RF}}^\prime= a_{\mathrm{TECH}}  P_{\mathrm{RF}}''$& & 12940\\
\end{tabular}
\nomenclature{$a_{\mathrm{TECH}}$}{Technology scaling for \ac{RF} transceiver}%
\nomenclature{$P_{\mathrm{RF}}''$}{Sum of \ac{RF} blocks power consumption in W}%

\caption[Reference power consumption values of \acs{RF} transceiver blocks]{Reference power consumption values of \ac{RF} transceiver blocks, where UL and DL refer to blocks which are part of the uplink or downlink operation, respectively.}
\label{tab:P_RF}
\end{table}

The \ac{RF} transceiver as present in currently manufactured \acp{BS} is incapable of sleep mode operation. Therefore, its power consumption in sleep mode is equal to its power consumption when active.,
\begin{equation}
 P_{\mathrm{RF,S}}= P_{\mathrm{RF}}.
\end{equation}
\nomenclature{$P_{\mathrm{RF,S}}$}{Power consumption of all \ac{RF} transceivers in sleep mode in W}%

\begin{figure}
\centering
\includegraphics[width=0.6\textwidth]{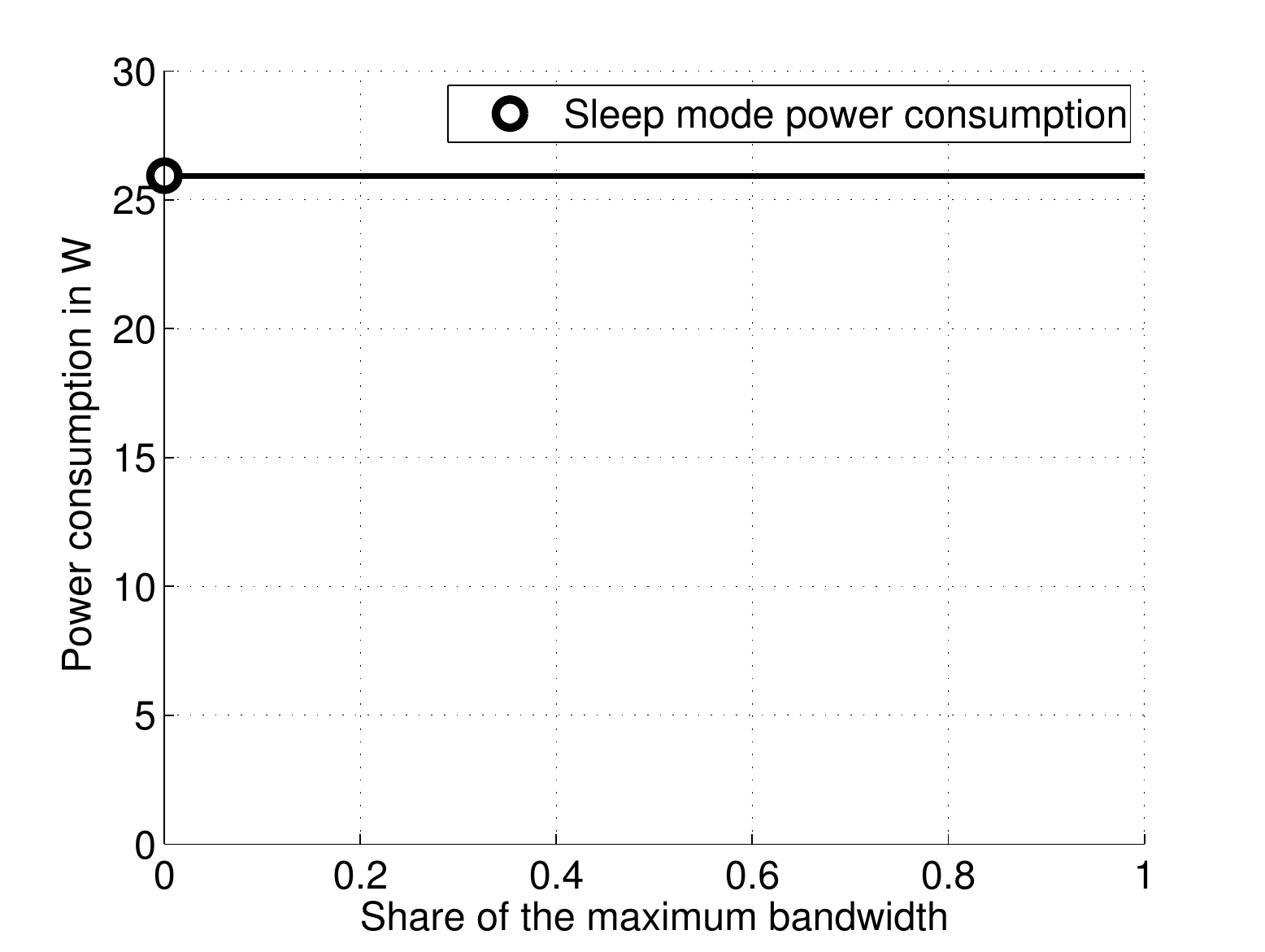}
\caption[$P_{\mathrm{RF}}$ as a function of bandwidth used]{$P_{\mathrm{RF}}$ as a function of bandwidth used.}
\label{fig:P_RF}
\end{figure}

The power consumption of the \ac{RF} transceiver is shown as a function of $f$ in Fig.~\ref{fig:P_RF}. Since it is constant, it can only be manipulated through the number of \ac{RF} transceivers in operation. Sleep mode operation is not yet foreseen for \ac{RF} transceivers.

\subsubsection{\ac{BB} Unit}
The \ac{BB} unit generates and processes the digital signal before it is passed to the \ac{RF} transceiver. The \ac{BB} unit of macro \acp{BS} is typically a stand-alone system-on-chip based on flexible \ac{FPGA} architecture. Since this component is a digital processor, it is best modelled by its operations, namely by \ac{GOPS}. As the power cost of \ac{GOPS} can be estimated at 40~\ac{GOPS}/W and a set of different functions of the \ac{BB} unit, $I_{\mathrm{BB}}$, can be associated with the number of \ac{GOPS} required, the power consumption can be modelled. \nomenclature{$I_{\mathrm{BB}}$}{Set of subcomponents of the \ac{BB} unit} Unlike in other \ac{BS} components, the \ac{BB} unit scales non-linearly with $D$ and $f$. For example, the frequency domain processing complexity grows exponentially with the number of antennas. The exact scaling exponents along with the reference complexities for the modeling of \ac{BB} operations are shown in Table~\ref{tab:P_BB}. 

\begin{table}
\centering
 \begin{tabular}{l | c c c c c}
Type of processing, $i$ 	& \ac{GOPS} 	& \begin{tabular}{@{}c@{}}$P_{i,\mathrm{BB}}^{\mathrm{ref}}$ \\ in W\end{tabular}& $y_i^D$ 	& $y_i^f$ 	& $z_{\mathrm{s}}$\\
\hline
Time Domain Processing		& 360 		&9.0					&1		&0		&0\\
Frequency Domain Processing	& 60 		&1.5					&2		&1		&0\\
Forward Error Correction	& 60 		&1.5					&1		&0		&1\\
Central Processing Unit		& 400 		&10.0					&1		&0		&1\\
Common Public Radio Interface	& 300 		&7.5					&1		&0		&1\\
Leakage				& 118 		&3.0					&1		&0		&1\\
 \end{tabular}
\caption[Complexity of \acs{BB} operations]{Complexity of \ac{BB} operations and their scaling with the number of transmit antennas and the transmission bandwidth.}
\label{tab:P_BB}
\end{table}

These reference values and scaling exponents are used to find the \ac{BB} unit power consumption in W, with
\begin{equation}
 P_{\mathrm{BB}} = \displaystyle \sum_{i \in I_{\mathrm{BB}}} P_{i,\mathrm{BB}}^{\mathrm{ref}} D^{y_i^D} f^{y_i^f},
\end{equation}
where $P_{i,\mathrm{BB}}^{\mathrm{ref}}$ in W is the power consumption per sub-component, $y_i^D$ is the scaling exponent for the number of radio chains, and $y_i^f$ is the scaling exponent for the share of bandwidth used, each as provided in Table~\ref{tab:P_BB}.
\nomenclature{$P_{\mathrm{BB}}$}{Power consumption of the \ac{BB} unit in W}%
\nomenclature{$P_{i,\mathrm{BB}}^{\mathrm{ref}}$}{Reference power consumption of a \ac{BB} subcomponent $i$ in W}%
\nomenclature{$y_i^D$}{Scaling exponent for \ac{BB} power consumption which depends on the number of antennas}%
\nomenclature{$y_i^f$}{Scaling exponent for \ac{BB} power consumption which depends on the share of available bandwidth used}%

The sleep mode power consumption of the \ac{BB} units, $P_{\mathrm{BB,S}}$ in W, is lower than when in active mode, as time domain and frequency domain processing can be disabled as indicated by the sleep mode switch, $z_{\mathrm{s}}$. Like the active power consumption, sleep mode power consumption is found by the combination of reference values and scaling exponents,
\begin{equation}
 P_{\mathrm{BB,S}} = \displaystyle \sum_{i \in I_{\mathrm{BB}}} P_{i,\mathrm{BB}}^{\mathrm{ref}} D^{y_i^D} f^{y_i^f} z_{\mathrm{s}}.
\end{equation} 
\nomenclature{$P_{\mathrm{BB,S}}$}{Power consumption of the \ac{BB} unit during sleep mode in W}%
\nomenclature{$z_{\mathrm{s}}$}{Sleep mode switch for the calculation of \ac{BB} power consumption}%

\begin{figure}
\centering
\includegraphics[width=0.6\textwidth]{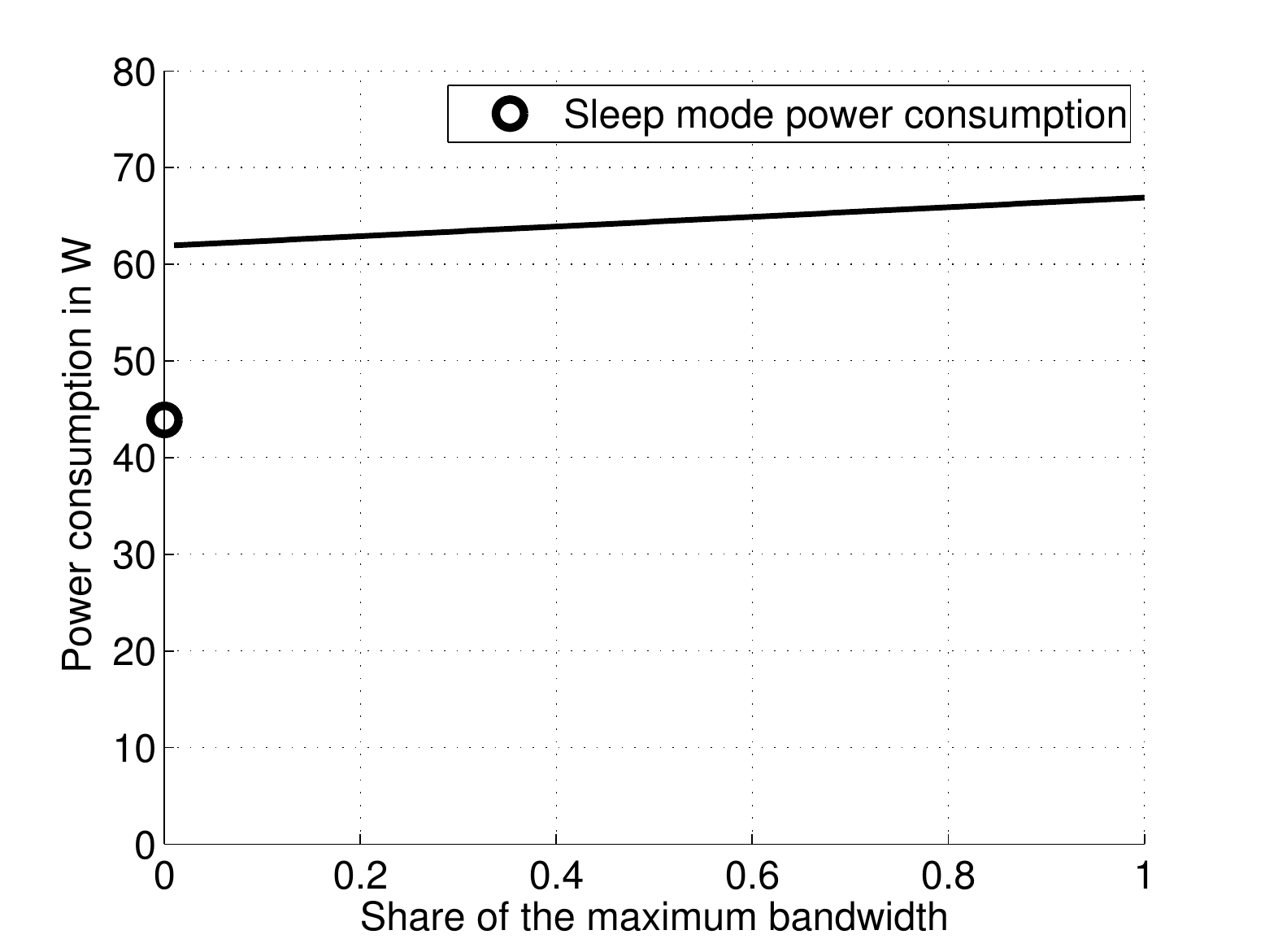}
\caption[$P_{\mathrm{BB}}$ as a function of bandwidth used]{$P_{\mathrm{BB}}$ as a function of bandwidth used.}
\label{fig:P_BB}
\end{figure}

When drawn over the transmission bandwidth, see Fig.~\ref{fig:P_BB}, the power consumption of the \ac{BB} unit is found to be only lightly affected by load, \ie~the bandwidth used. However, sleep modes are available for the \ac{BB} which can reduce its power consumption by more than 30\%. \ac{MIMO} operation has a significant effect on the \ac{BB} power consumption, as it is modelled with a quadratic scaling factor due to the added computational complexity of \ac{MIMO} processing.

\subsubsection{\ac{DC}-\ac{DC}-Conversion}
\begin{figure}
\centering
\includegraphics[width=0.6\textwidth]{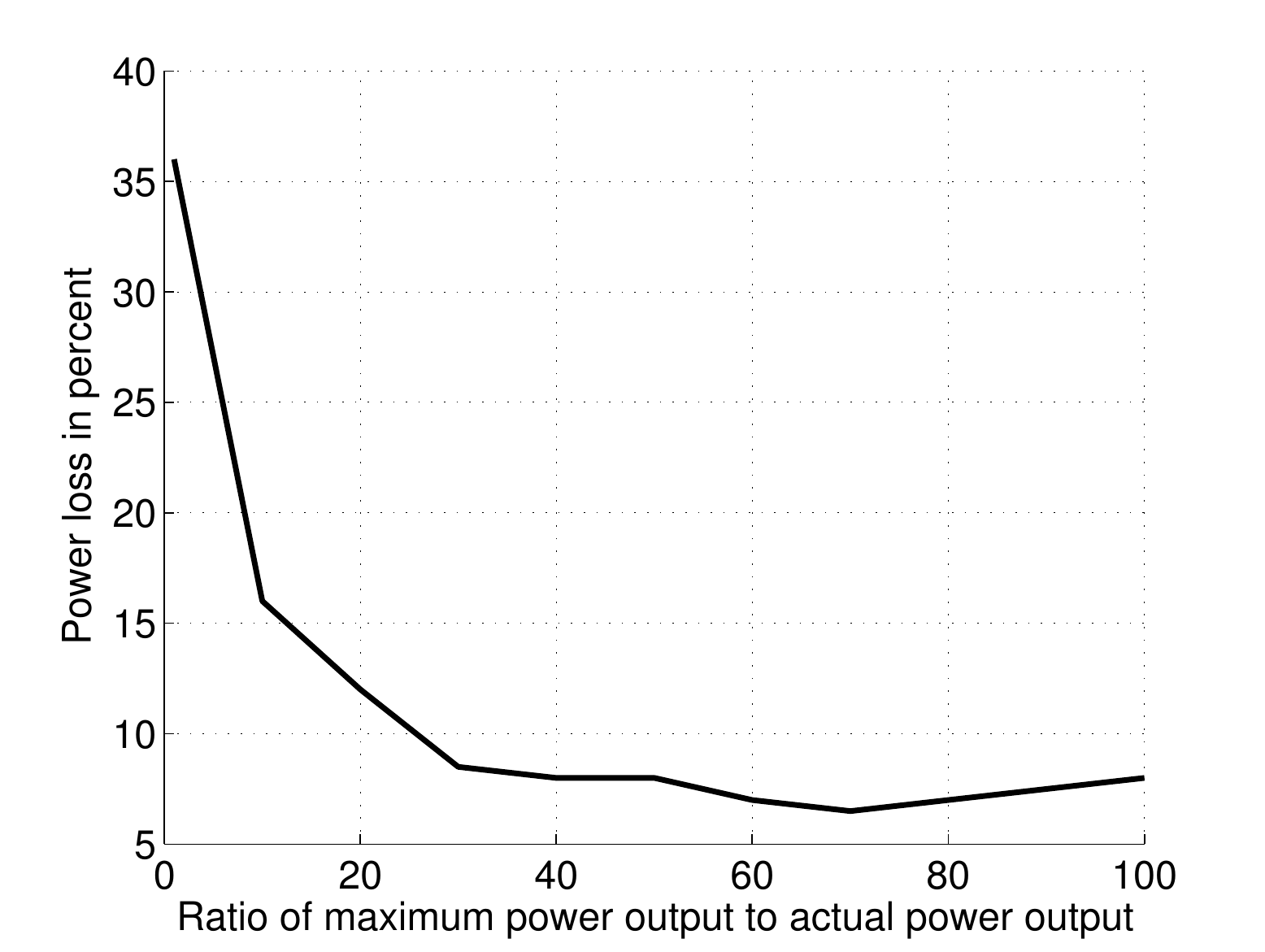}
\caption[\acs{DC} conversion loss function]{\ac{DC} conversion loss function $l_{\mathrm{DC}}(\zeta_{\mathrm{DC}})$ as measured within the \ac{EARTH} project.}
\label{fig:Loss_DC}
\end{figure}

To supply the required \ac{DC} voltages for the various components mentioned above, a \ac{BS} is equipped with several \ac{DC}-\ac{DC}-converters. As the architectures of these converters may vary across vendors, the conversion is modelled as a single component. Laboratory measurements have been taken from one representative \ac{DC}-\ac{DC}-converter during the \ac{EARTH} project as a function of the required load. Conversion has an efficiency of less than one, which can also be represented as a loss. With the \ac{DC}-\ac{DC}-converter built for a maximum power output of $P_{\mathrm{DC, out, max}}$ in W, the loss is larger the further the actual required power by the \ac{PA}, \ac{RF} transceiver and \ac{BB} unit is from $P_{\mathrm{DC, out, max}}$. See Fig.~\ref{fig:Loss_DC} for the measured losses function, $l_{\mathrm{DC}}(\zeta_{\mathrm{DC}})$, of the ratio of maximum power output to actual power output, $\zeta_{\mathrm{DC}}$, with
\begin{equation}
 \zeta_{\mathrm{DC}} = \frac{P_{\mathrm{DC, out, max}}} {P_{\mathrm{PA}} + P_{\mathrm{RF}} + P_{\mathrm{BB}}}.
\end{equation}
\nomenclature{$\zeta_{\mathrm{DC}}$}{Ratio of used to maximum output power by the \ac{DC}-\ac{DC}-converter}%
\nomenclature{$P_{\mathrm{DC, out, max}}$}{Maximum output power of the \ac{DC}-\ac{DC}-converter in W}%

The power consumption caused by \ac{DC}-\ac{DC}-conversion in W is then found as
\begin{equation}
 P_{\mathrm{DC}} = l_{\mathrm{DC}}(\zeta_{\mathrm{DC}}) \left( P_{\mathrm{PA}} + P_{\mathrm{RF}} + P_{\mathrm{BB}} \right).
\end{equation}
\nomenclature{$P_{\mathrm{DC}}$}{Power consumption of \ac{DC}-\ac{DC}-conversion in W}%
\nomenclature{$l_{\mathrm{DC}}$}{Efficiency of \ac{DC}-\ac{DC}-conversion; a function of $\zeta_{\mathrm{DC}}$}%

\ac{SOTA} \ac{DC}-\ac{DC}-converters are assumed to possess no capability for sleep modes and, thus, remain activated during \ac{BS} sleep mode, with power consumption during sleep mode, $P_{\mathrm{DC,S}}$ in W,
\begin{equation}
 P_{\mathrm{DC,S}} = P_{\mathrm{DC}}.
\end{equation}
\nomenclature{$P_{\mathrm{DC,S}}$}{Power consumption of \ac{DC}-\ac{DC}-conversion in sleep mode in W}%

\begin{figure}
\centering
\includegraphics[width=0.6\textwidth]{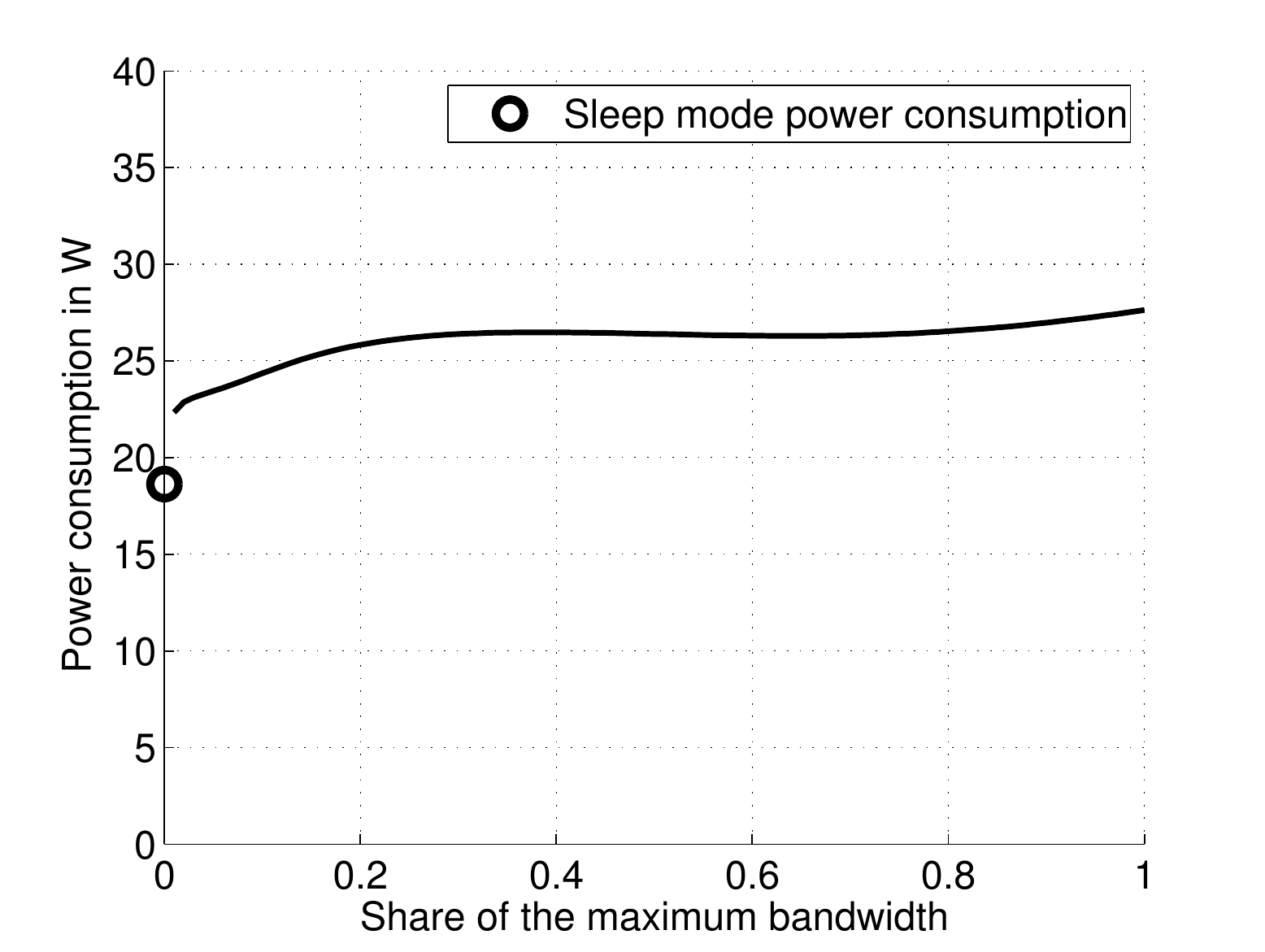}
\caption[$P_{\mathrm{DC}}$ as a function of bandwidth used]{\ac{DC}-\ac{DC}-converter power consumption as a function of bandwidth used.}
\label{fig:P_DC}
\end{figure}

The power consumption of the \ac{DC}-\ac{DC}-converter is shown as a function of $f$ in Fig.~\ref{fig:P_DC}. It is directly dependent on the components the \ac{DC}-\ac{DC}-converter powers and varies by less than 20\% over all bandwidths. The reduced power consumption in sleep mode is only a passive effect and not an active switching process.

\subsubsection{Mains Supply/\ac{AC}-\ac{DC}-Conversion}
\begin{figure}
\centering
\includegraphics[width=0.6\textwidth]{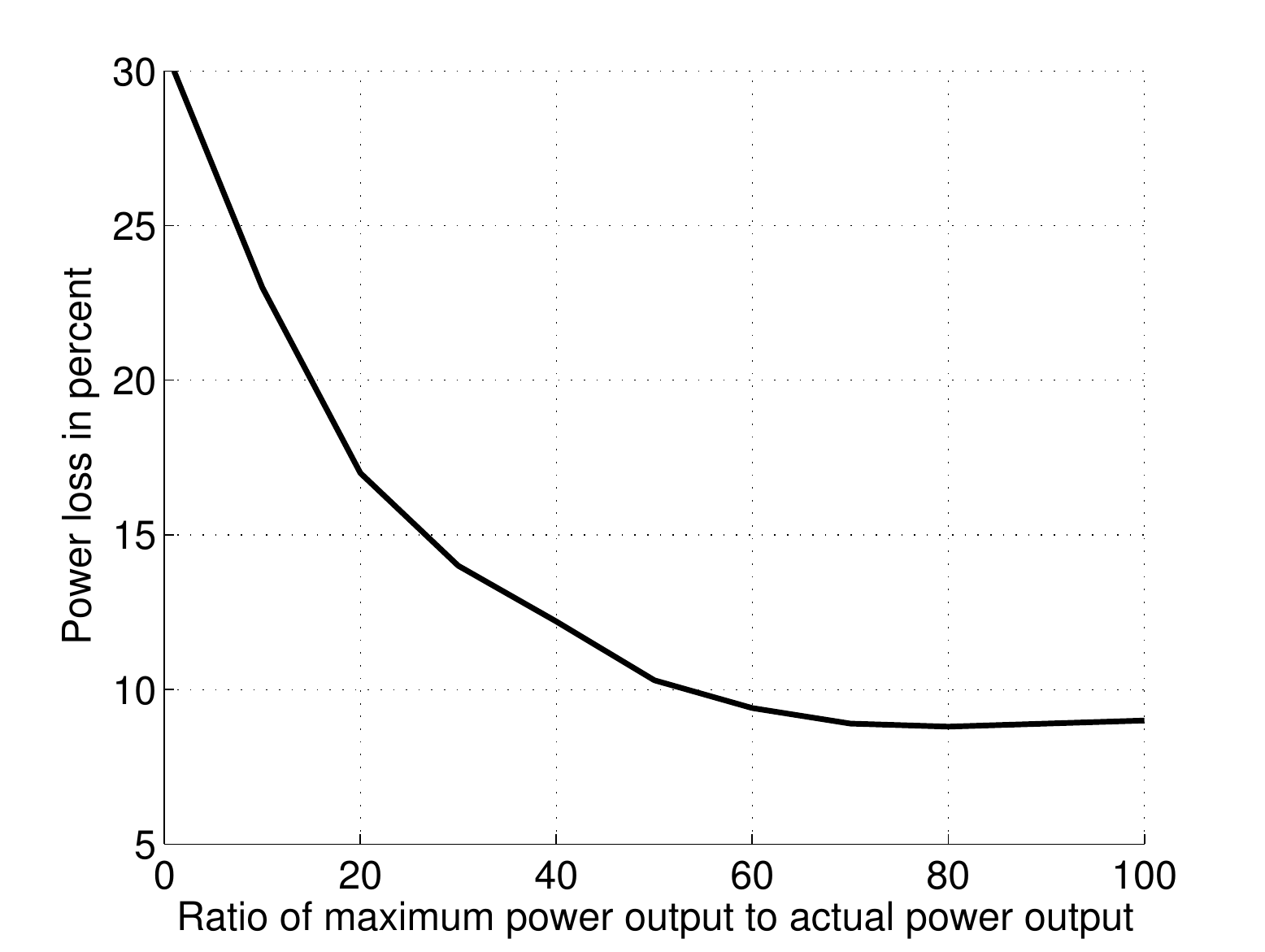}
\caption[\acs{AC} conversion loss function]{\ac{AC} conversion loss function $l_{\mathrm{AC}}(\zeta_{\mathrm{AC}})$ as measured within the \ac{EARTH} project.}
\label{fig:Loss_AC}
\end{figure}

Power from the \ac{AC} mains grid is converted to \ac{DC} by the mains power supply unit. This unit is measured and modelled in the same fashion as \ac{DC}-\ac{DC}-conversion with conversion loss. See Fig.~\ref{fig:Loss_AC} for the measured losses as a function of the ratio of maximum power output to actual power output, $\zeta_{\mathrm{AC}}$, with
\begin{equation}
 \zeta_{\mathrm{AC}} = \frac{P_{\mathrm{AC, out, max}}} {P_{\mathrm{PA}} + P_{\mathrm{RF}} + P_{\mathrm{BB}} + P_{\mathrm{DC}}}.
\end{equation}
\nomenclature{$\zeta_{\mathrm{AC}}$}{Ratio of used to maximum output power by the \ac{AC}-\ac{DC}-converter}%
\nomenclature{$P_{\mathrm{AC, out, max}}$}{Maximum output power of the \ac{AC}-\ac{DC}-converter in W}%

The power consumption caused by \ac{AC}-\ac{DC}-conversion is then found as
\begin{equation}
 P_{\mathrm{AC}} = l_{\mathrm{AC}}(\zeta_{\mathrm{AC}}) \left( P_{\mathrm{PA}} + P_{\mathrm{RF}} + P_{\mathrm{BB}} + P_{\mathrm{DC}} \right).
\end{equation}
\nomenclature{$P_{\mathrm{AC}}$}{Power consumption of \ac{AC}-\ac{DC}-conversion in W}%
\nomenclature{$l_{\mathrm{AC}}$}{Efficiency of \ac{AC}-\ac{DC}-conversion; a function of $\zeta_{\mathrm{AC}}$}%

\ac{SOTA} \ac{AC}-\ac{DC}-converters are assumed to possess no special adaptation to sleep modes and, thus, remain activated during \ac{BS} sleep mode, with power consumption during sleep mode, $P_{\mathrm{AC,S}}$ in W,
\begin{equation}
 P_{\mathrm{AC,S}} = P_{\mathrm{AC}}.
\end{equation}
\nomenclature{$P_{\mathrm{AC,S}}$}{Power consumption of \ac{AC}-\ac{DC}-conversion in sleep mode in W}%

\begin{figure}
\centering
\includegraphics[width=0.6\textwidth]{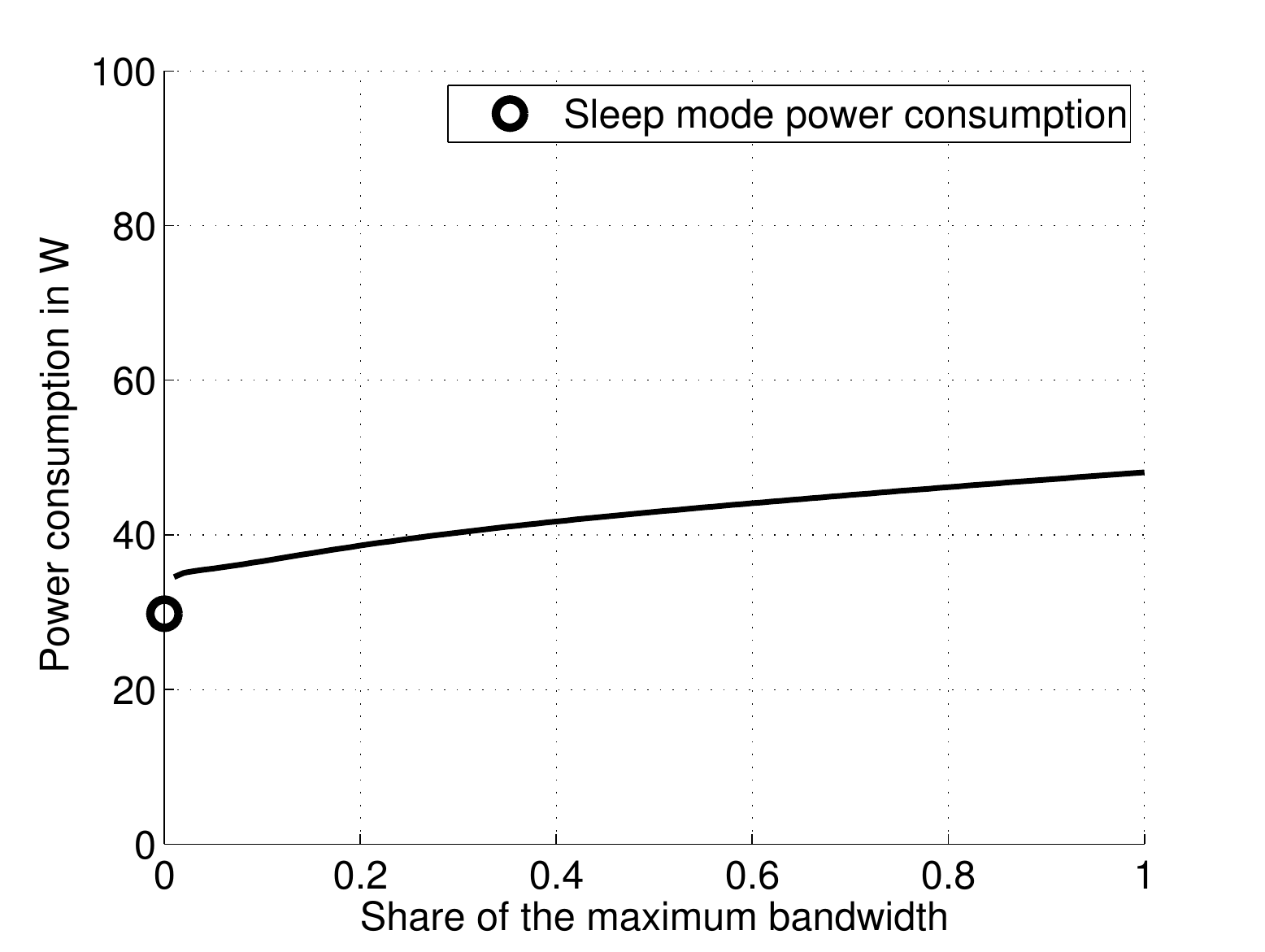}
\caption[$P_{\mathrm{AC}}$ as a function of bandwidth used]{\ac{AC}-\ac{DC}-converter power consumption as a function of bandwidth used.}
\label{fig:P_AC}
\end{figure}

The power consumption of the \ac{AC}-\ac{DC}-converter is shown as a function of $f$ in Fig.~\ref{fig:P_AC}. Although the converter has non-linear efficiencies, these seem to compensate for the non-linearities of the  components it powers. As a result, the power consumption curve is close to a straight line. 

\subsubsection{Cooling}
Macro \acp{BS} are typically housed in a cooled cabinet. This cooling is a very difficult effect to model. Cooling requirements vary greatly between different geographic locations, positioning in or on buildings and the size of the storage cabinet. Different vendors recommend different operating temperatures for their devices and these recommendations may not be followed by the network operators. Furthermore, cooling is generally a very slow process compared to the orders of timing in a \ac{BS}. Thus, considering cooling at its instantaneous power consumption may be misleading. As there is no simple solution to this problem, cooling is modelled as a fixed power loss.

The heat dissipation of all other components needs to compensated for by active cooling. This heat dissipation can be assumed to be proportional to the power consumed. Thus, cooling power consumption is modelled to be proportional to the consumption of all other components. With cooling loss, $\xi_{\mathrm{COOL}}=0.12$, the power consumption of the cooling unit in W is
\begin{equation}
 P_{\mathrm{COOL}} = \xi_{\mathrm{COOL}} \left( P_{\mathrm{PA}} + P_{\mathrm{RF}} + P_{\mathrm{BB}} + P_{\mathrm{DC}} + P_{\mathrm{AC}} \right).
\label{eq:P_COOL}
\end{equation}
\nomenclature{$P_{\mathrm{COOL}}$}{Power consumption of active \ac{BS} cooling in W}%
\nomenclature{$\xi_{\mathrm{COOL}}$}{Cooling loss factor}%

In this model, there are no further assumptions with regard to cooling during sleep mode. The power consumption of the cooling unit during sleep mode, $P_{\mathrm{COOL,S}}$ in W, is therefore identical, with
\begin{equation}
 P_{\mathrm{COOL,S}} = P_{\mathrm{COOL}}.
\end{equation}
\nomenclature{$P_{\mathrm{COOL,S}}$}{Power consumption of active \ac{BS} cooling in sleep mode in W}%

\begin{figure}
\centering
\includegraphics[width=0.6\textwidth]{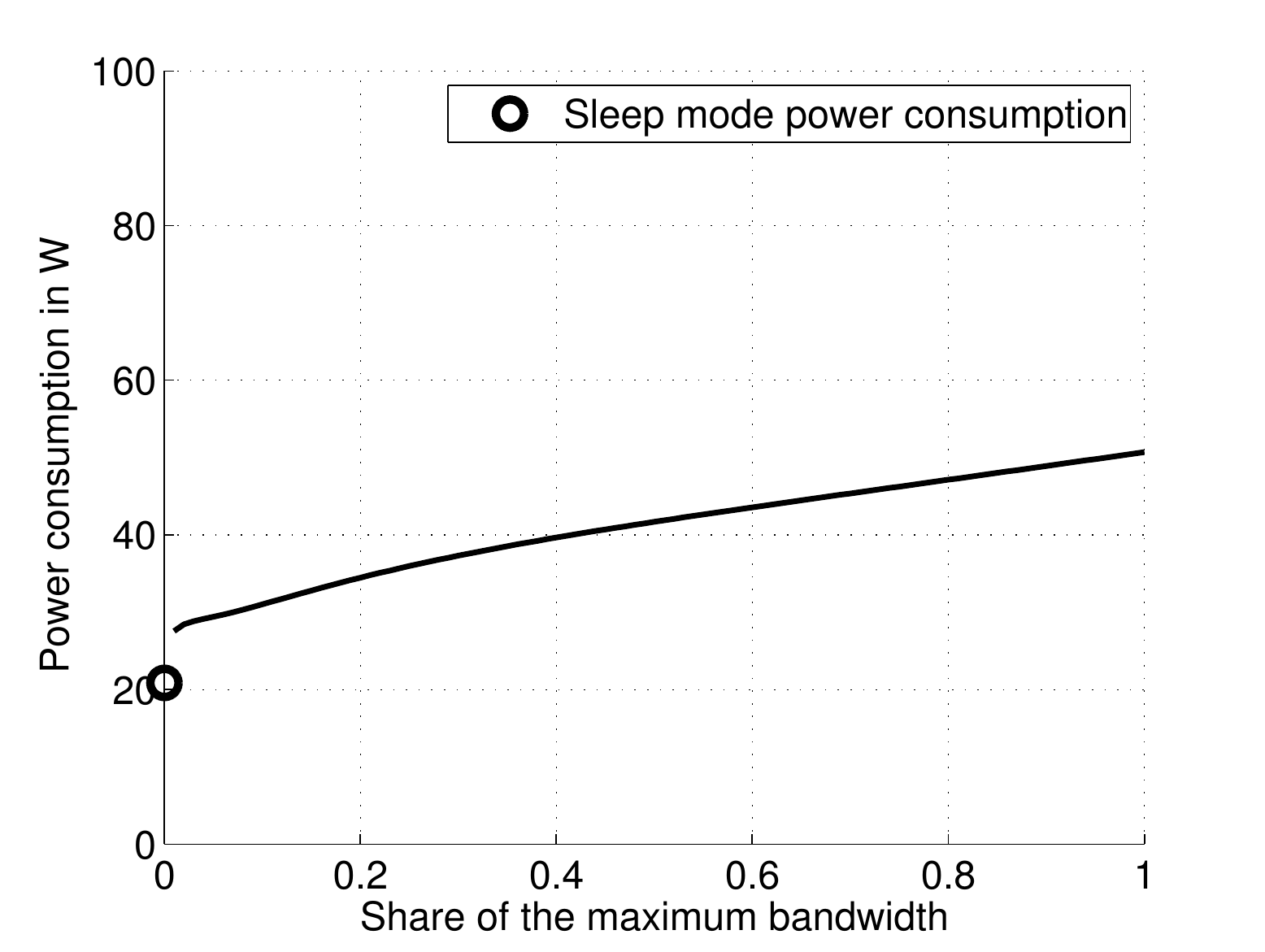}
\caption[$P_{\mathrm{COOL}}$ as a function of bandwidth used]{Cooling power consumption as a function of bandwidth used.}
\label{fig:P_COOL}
\end{figure}

The power consumption of the cooling unit as a function of $f$ is shown in Fig.~\ref{fig:P_COOL}. As it depends on all other components, it follows their power consumption behavior and is almost affine.

\subsection{\ac{BS} Power Consumption}
The power consumption, $P_{\mathrm{supply}}$ in W, of an entire \ac{LTE} \ac{BS} is found as the sum of its components. With $M_{\mathrm{sec}}$ the number of sectors,
\begin{equation}
 P_{\mathrm{supply}} = M_{\mathrm{sec}} \left( P_{\mathrm{PA}} + P_{\mathrm{RF}} + P_{\mathrm{BB}} + P_{\mathrm{DC}} + P_{\mathrm{AC}} + P_{\mathrm{COOL}}\right).
\end{equation}
\nomenclature{$P_{\mathrm{supply}}$}{\ac{BS} supply power consumption in W}%
\nomenclature{$M_{\mathrm{sec}}$}{Number of sectors on a macro \ac{BS}}%

\begin{figure}
\centering
\begin{subfigure}[t]{.8\textwidth}
\includegraphics[width=\textwidth]{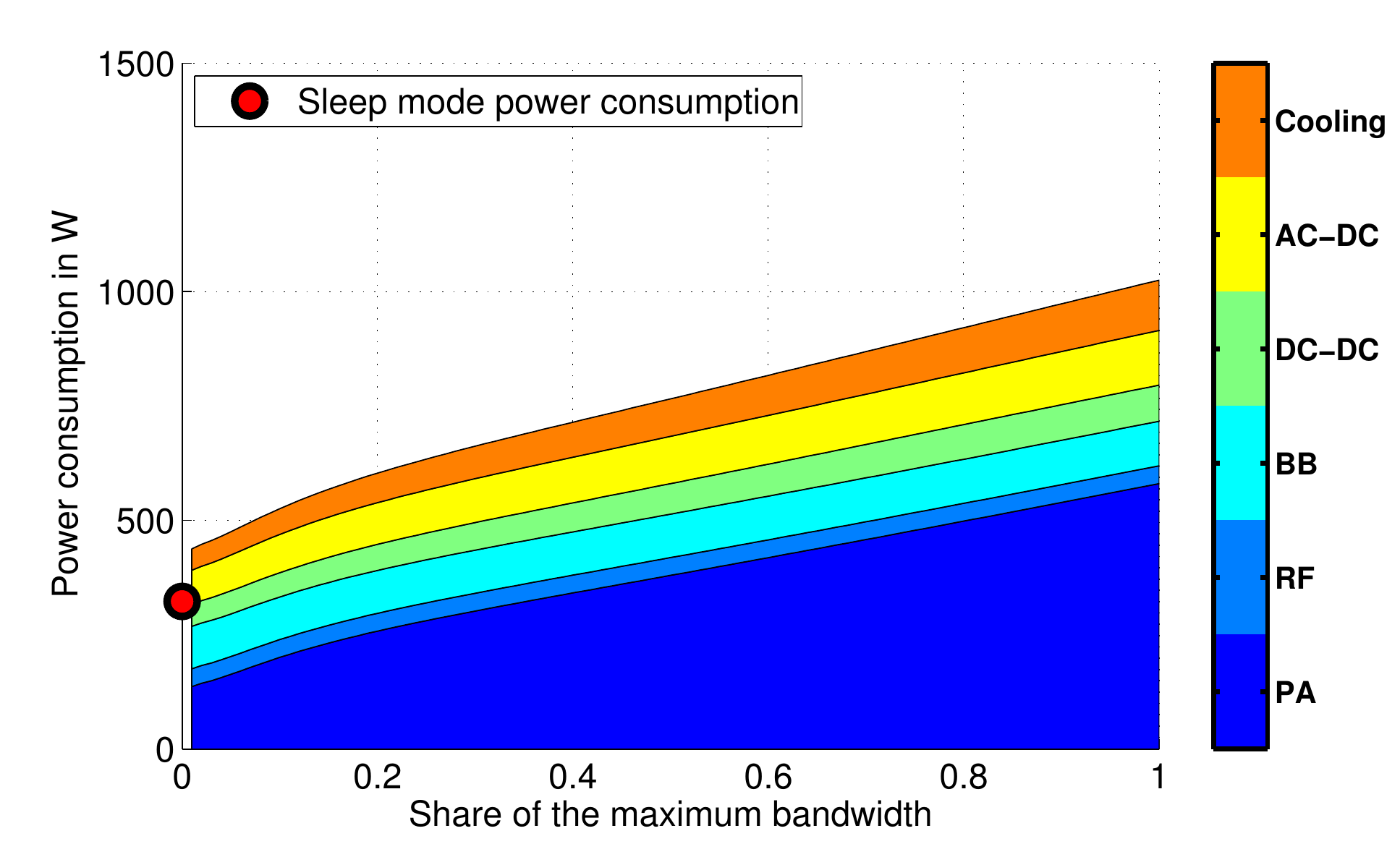}
\caption[\acs{BS} power consumption per component, $D=1$]{One antenna.}
\label{fig:P_all_1x}
\end{subfigure}

\begin{subfigure}[t]{.8\textwidth}
\includegraphics[width=\textwidth]{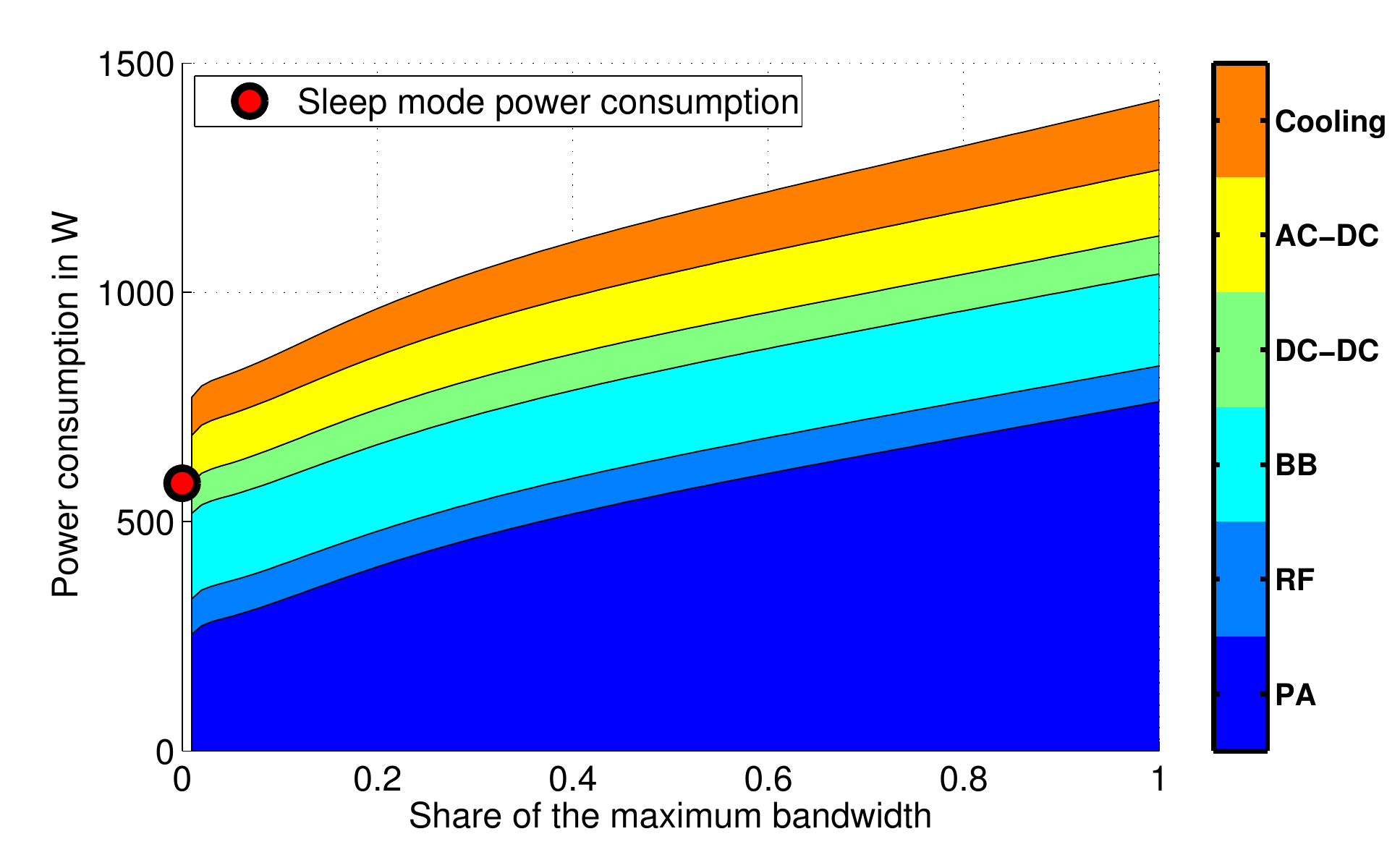}
\caption[\acs{BS} power consumption per component, $D=2$]{Two antennas.}
\label{fig:P_all_2x}
\end{subfigure}
\caption[$P_{\mathrm{supply}}$ as a function of bandwidth used]{\acs{BS} power consumption per component as a function of bandwidth used.}
\label{fig:P_all}
\end{figure}

\begin{figure}
\centering
\begin{subfigure}[t]{0.65\textwidth}
\includegraphics[width=\textwidth]{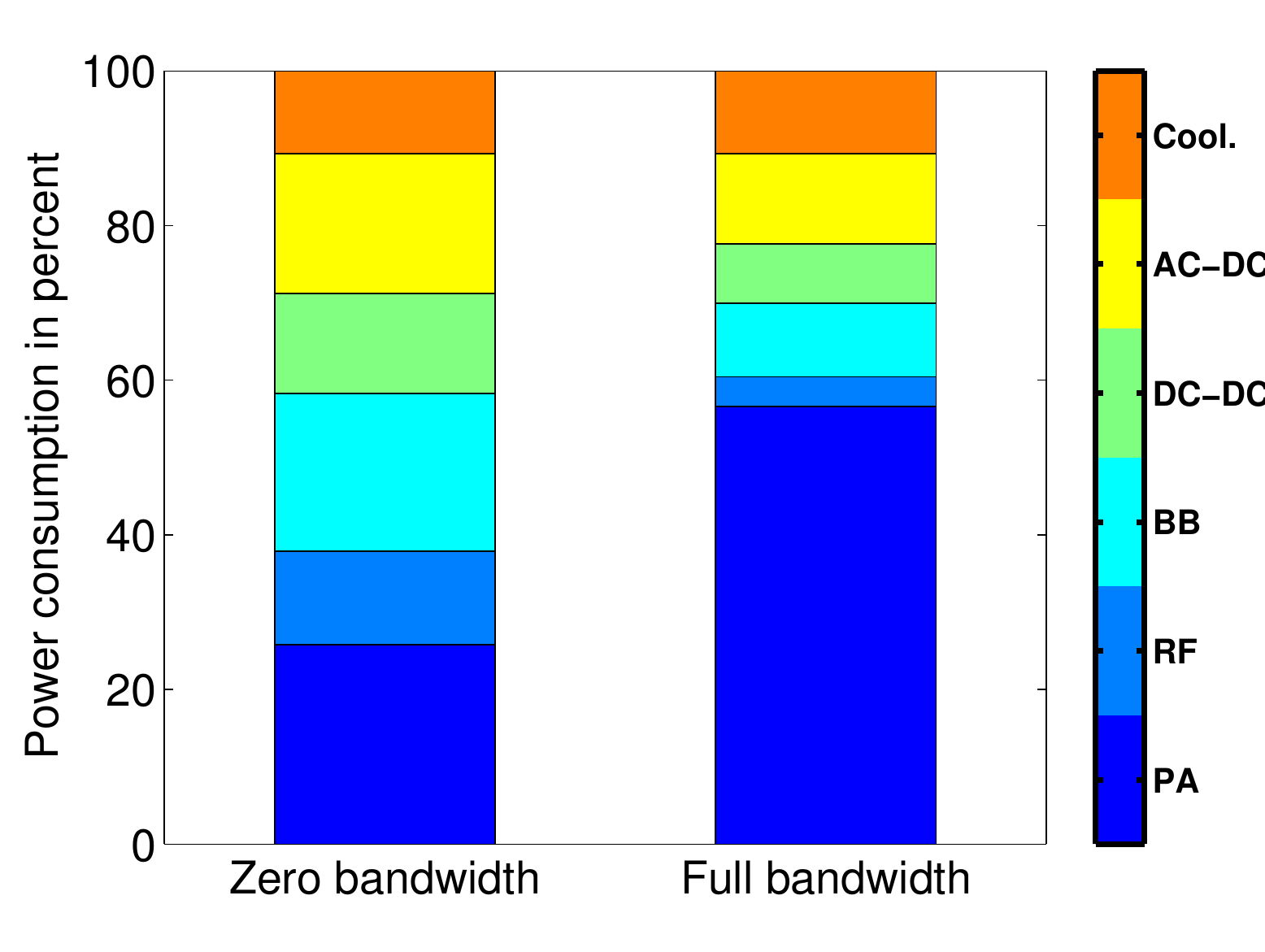}
\caption[]{One antenna}\label{fig:P_all_percent_1x}
\end{subfigure}

\begin{subfigure}[t]{0.65\textwidth}
\includegraphics[width=\textwidth]{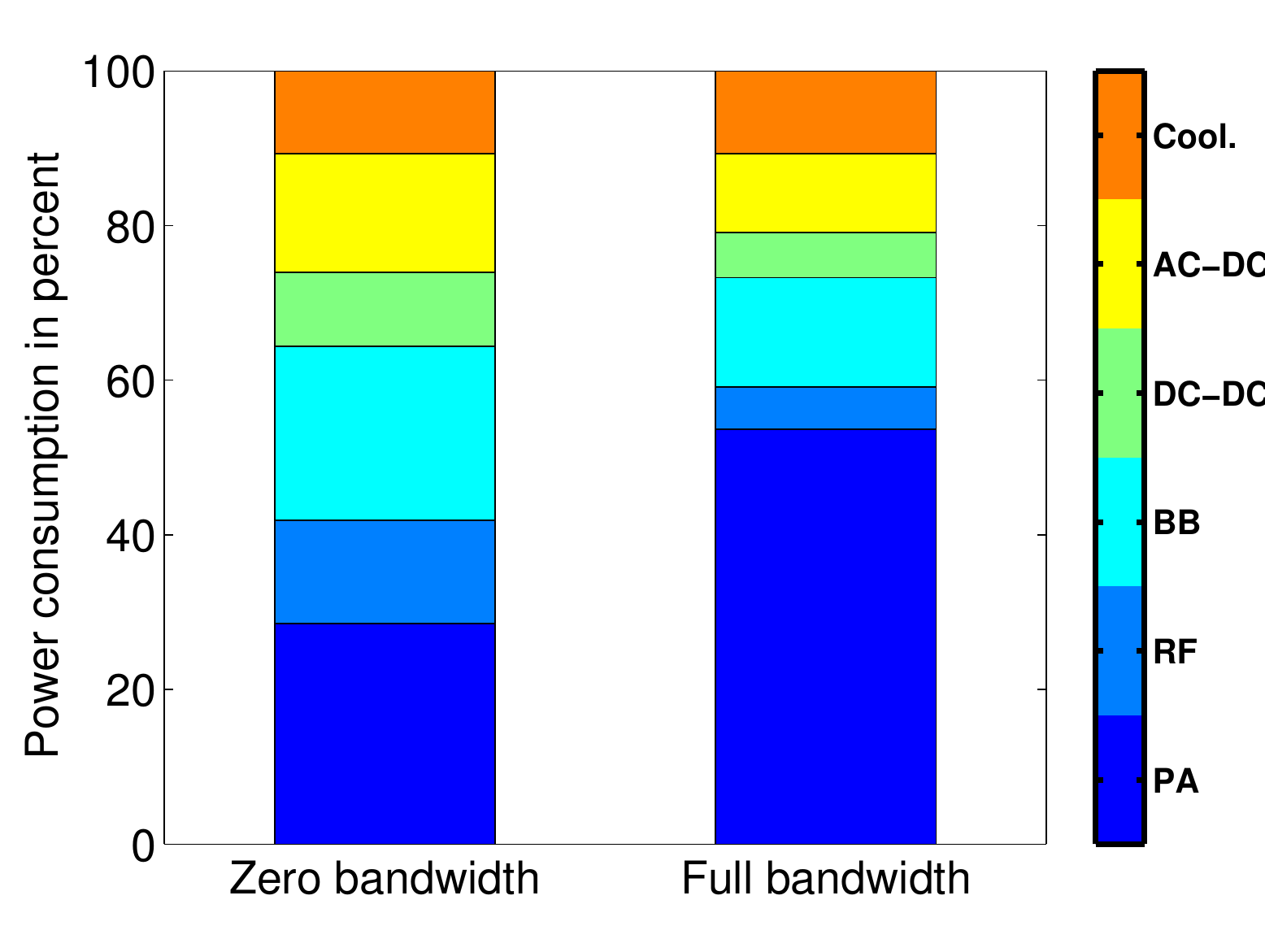}
\caption[]{Two antennas}\label{fig:P_all_percent_2x}
\end{subfigure}

\caption[$P_{\mathrm{supply}}$ per component in percent]{\ac{BS} power consumption per component at zero and full bandwidth.}
\label{fig:P_all_percent}
\end{figure}

It is illustrated in Fig.~\ref{fig:P_all} for $D=\{1,2\}$. Fig.~\ref{fig:P_all_percent} provides a normalized comparison of the shares of power consumption at the two edge points of Fig.~\ref{fig:P_all}, namely at zero and at full bandwidth. 

In both cases, for $D=1$ and $D=2$, respectively, the largest contributors to power consumption, especially at full bandwidth transmission, are found to be the \acp{PA} with 56.6\% and 53.7\%. At zero bandwidth, with 25.8\% and 28.5\%, the \acp{PA} still consume a larger share of the power than other components, but significantly less than for higher transmission bandwidths. Therefore, reducing the \ac{PA} power consumption significantly affects the power consumption of the entire device. 

With regard to sleep modes, for $D=1$ and $D=2$, respectively, they are estimated to lower consumption to 322~W and 583~W compared to an overall consumption of 437~W and 771~W at zero bandwidth, and 1025~W and 1419~W at full bandwidth usage. That corresponds to a reduction of power consumption by 26\% and 24\% at zero bandwidth and by 69\% and 59\% at full bandwidth usage. Thus, sleep modes can reduce the supply power consumption significantly, especially in high load operation.

Overall, it is important to recognize that the power consumption of a \ac{BS} is far from constant and can be varied through operation parameters by as much as 69\%. To reduce its power consumption, it should be operated with low power, low bandwidth or in sleep mode. In addition, the slope of the variation is nearly constant over $f$ for both one and dual antenna operation. This allows for a greatly simplified power consumption model. 

The power model described in this section can be generated for other than macro \ac{BS} types such as micro, pico, femto or \ac{RRH} in similar fashion. However, as this thesis is only concerned with the power consumption of macro type \acp{BS}, these types are not discussed here.

\section{The Parameterized Power Model}
\label{parameterizedmodel}
The power model described above is very useful for purposes of derivation, understanding, calibration and exploration of changes in architecture. The effect of changes to individual components on the power consumption of the entire \ac{BS} can be studied. However, it is often unnecessary to derive the power consumption of a \ac{BS} from the component level up. When exploring power saving techniques, only few parameters are varied regularly while many remain constant for an entire study. To address this need, this section proposes a simpler, parameterized power model. It abstracts architectural details for simplicity and only keeps those parameters which are regularly altered. It is based on the identification of power saving techniques.

In the literature, specific power saving techniques are suggested: namely, the effects of the number of antennas on power consumption~\cite{cgb0401}, the reduction of transmission power~\cite{hah1101,lcvc1101,mhl0801}, the deactivation of unneeded antennas~\cite{hafm1201,xylzcx1101}, the adaptation of the transmission bandwidth~\cite{abh1101,a1201} and the use of low power consumption sleep modes of varied durations~\cite{fmmjg1101,acc1201,vh1101,d1201}.

The parameterized model encompasses all of these approaches to allow a direct comparison while abstracting parameters which can either be assumed to be constant or have little effect in the studied scenarios, such as \ac{GOPS}, lookup tables, equipment manufacturing details or leakage powers. To this extent, only the following is covered in the proposed model:
\begin{itemize}
 \item The different \ac{BS} types of a heterogeneous network are modelled by applying different parameter sets to the same model equations. 
 \item The number of transmission antennas and radio chains affects consumption during design and operation.
 \item The same holds true for transmission power, which affects the design indirectly by choice of a suitable power amplifier as well as the operation directly. 
 \item Also, transmission bandwidth and sleep modes are modelled to allow the investigation of their impact on \ac{BS} power consumption.
\end{itemize}

For parameterization, the maximum supply power consumption, $P_1$ in W, is first found establishing how power consumption scales with bandwidth, $W$, the number of \ac{BS} radio chains/antennas, $D$, and the maximum transmission power, $P_{\mathrm{max}}$. This requires the consideration the main units of a \ac{BS} as presented earlier: \ac{PA}, \ac{RF} small-signal transceiver, \ac{BB} unit, DC-DC converter, mains supply and active cooling. Summarizing Section~\ref{ch3:componentmodel}, the dependence of the \ac{BS} units on $W$, $D$ and $P_{\mathrm{max}}$ can be approximated as follows:
\begin{itemize}
 \item Both \ac{BB} and \ac{RF} power consumption, $P_\mathrm{BB}^{\mathrm{ppm}}$ in W and $P_\mathrm{RF}^{\mathrm{ppm}}$ in W, respectively, are assumed to scale nearly linearly with bandwidth, $W$, and the number of \ac{BS} antennas, $D$.
 \item The \ac{PA} power consumption, $P_{\mathrm{PA}}^{\mathrm{ppm}}$, depends on the maximum transmission power per antenna, $P_{\mathrm{max}}/D$, and the \ac{PA} efficiency, $\eta_\mathrm{PA}$. Also, possible feeder cable losses, $\sigma_\mathrm{feed}$, have to be accounted for, such that
\begin{equation}
\label{ppa}
 P_{\mathrm{PA}}^{\mathrm{ppm}} = \frac{P_{\mathrm{max}}}{D \eta_\mathrm{PA}(1{-}\sigma_\mathrm{feed})}.
\end{equation}
 \item Losses incurred by \ac{DC}-\ac{DC} conversion, \ac{AC}-\ac{DC} conversion and active cooling scale linearly with the power consumption of other components and may be approximated by the constant loss factors $\sigma_\mathrm{DC}$, $\sigma_\mathrm{AC}$, and $\sigma_\mathrm{COOL}$, respectively.
\end{itemize}
\nomenclature{$\eta_\mathrm{PA}$}{\ac{PA} efficiency in the parameterized model}%
\nomenclature{$P_{\mathrm{max}}$}{\ac{BS} maximum sum transmission power in W}%

These assumptions allow calculating the maximum power consumption of a \ac{BS} sector,
\begin{equation}
\label{p1}
\begin{aligned}
 P_1 = \frac{\left[ D \frac{W}{10~\mathrm{MHz}} \left( P_\mathrm{BB}^{\mathrm{ppm}} + P_\mathrm{RF}^{\mathrm{ppm}}   \right) + P_{\mathrm{PA}}^{\mathrm{ppm}} \right]}
      {\displaystyle (1{-}\sigma_{\mathrm{DC}})(1{-}\sigma_{\mathrm{AC}})(1{-}\sigma_{\mathrm{COOL}}) }.
\end{aligned}
\end{equation}%
\nomenclature{$\sigma_{\mathrm{AC}}$}{Loss factor for \ac{AC}-\ac{DC}-conversion in the parameterized model}%
\nomenclature{$\sigma_{\mathrm{DC}}$}{Loss factor for \ac{DC}-\ac{DC}-conversion in the parameterized model}%
\nomenclature{$\sigma_{\mathrm{COOL}}$}{Loss factor for active cooling in the parameterized model}%
\nomenclature{$P_{\mathrm{PA}}^{\mathrm{ppm}}$}{Power consumption of the \acp{PA} in the parameterized model}%
\nomenclature{$P_\mathrm{RF}^{\mathrm{ppm}}$}{Power consumption of the \acp{RF} transceivers in the parameterized model}%
\nomenclature{$P_\mathrm{BB}^{\mathrm{ppm}}$}{Power consumption of the \ac{BB} units in the parameterized model}%

An important characteristic of a \ac{PA} is that operation at lower transmit powers reduces the efficiency of the \ac{PA} and that, consequently, power consumption is not a linear function of the \ac{PA} output power. This is resolved by taking into account the ratio of maximum transmission power of a \ac{PA} from its data sheet, $P_{\mathrm{PA,limit}}$ in W, to the maximum transmission power of the \ac{PA} during operation, $\frac{P_{\mathrm{max}}}{D}$. In typical \acp{PA}, the current transmission power can be adjusted by adapting the DC supply voltage, which impacts the offset power of the \ac{PA}. The efficiency is assumed to decrease by a factor of $\theta$ for each halving of the transmission power. The efficiency is thus maximal when $P_{\mathrm{max}}=P_{\mathrm{PA,limit}}$ in single antenna transmission and is heuristically well-described by
\begin{equation}
 \eta_\mathrm{PA}= \eta_\mathrm{PA,max}\left[ 1-\theta \log_2 \left( \frac{P_{\mathrm{PA,limit}}}{P_{\mathrm{max}} / D} \right)  \right],
\end{equation}
where $\eta_\mathrm{PA,max}$ is the maximum \ac{PA} efficiency.
\nomenclature{$\eta_\mathrm{PA,max}$}{Maximum \ac{PA} efficiency in the parameterized model}%
\nomenclature{$\theta$}{\ac{PA} efficiency decrease}%
\nomenclature{$P_{\mathrm{PA,limit}}$}{Maximum output power of the \ac{PA} in W}%

The reduction of power consumption in sleep mode is achieved by powering off \acp{PA} and reduced computations necessary in the \ac{BB} engine. For simplicity, only the dependence on $D$ is modelled as each \ac{PA} is powered off.  Thus, $P_{\mathrm{S}}$, is approximated as
\begin{equation}
 P_{\mathrm{S}} = D P_{\mathrm{S},0},
\label{eq:p_sleep}
\end{equation}
where $P_{\mathrm{S},0}$ is a reference value for the single antenna \ac{BS} chosen such that $P_{\mathrm{S}}$ matches the complex model power consumption for two antennas.
\nomenclature{$P_{\mathrm{S},0}$}{Reference value for the single antenna \ac{BS} sleep mode power consumption in W}%

\begin{figure}
\centering
\includegraphics[width=0.6\textwidth]{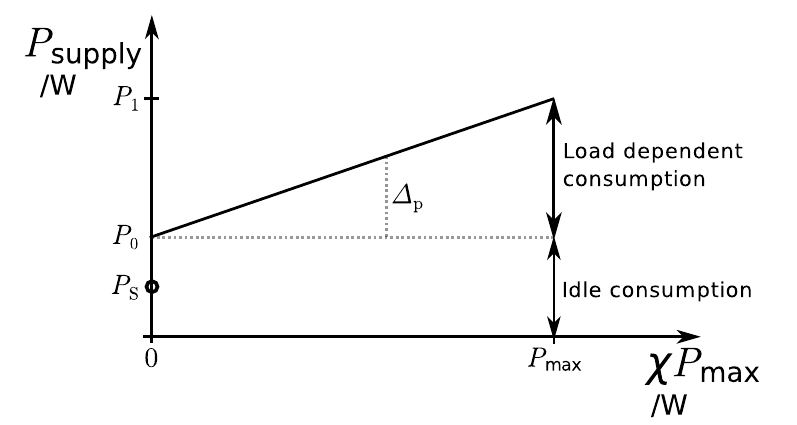}
 \caption[Load-dependent power model for an \acs{LTE} \acs{BS}]{Load-dependent power model for an \ac{LTE} \ac{BS}.}
\label{fig:loaddeppm}
\end{figure}

It can be approximated from Fig.~\ref{fig:P_all} to treat the supply power consumption as an affine function of the bandwidth and, thus,---given constant power spectral density---of the transmission power. In other words, the consumption can be represented by a static (load-independent) share, $P_0$ in W, with an added load-dependent share that increases linearly by a power gradient, $\Delta_{\mathrm{p}}$, see Figure~\ref{fig:loaddeppm}. The maximum supply power consumption, $P_1$ in W, is reached when transmitting at maximum total transmission power, $P_{\mathrm{max}}$. Furthermore, a \ac{BS} may enter a sleep mode with lowered consumption, $P_{\mathrm{S}}$, when it is not transmitting. Total power consumption considering the number of sectors, $M_{\mathrm{sec}}$, is then formulated as
\begin{equation}
\label{eq:psupply}
 P_{\mathrm{supply}}(\chi) = 
  \begin{cases}
   M_{\mathrm{sec}}  \left( P_1 + \Delta_{\mathrm{p}} P_{\mathrm{max}} (\chi - 1) \right)			& \text{if } 0 < \chi \leq 1  \\
   M_{\mathrm{sec}} P_{\mathrm{S}}      	& \text{if } \chi = 0,
  \end{cases}
\end{equation}
where $P_1 =  P_0 + \Delta_{\mathrm{p}} P_{\mathrm{max}}$. The scaling parameter, $\chi$, is the load share, where $\chi=1$ indicates a fully loaded system, \eg~transmitting at full power and full bandwidth, and $\chi=0$ indicates an idle system.
\nomenclature{$P_1$}{\ac{BS} power consumption at maximum load in W}%
\nomenclature{$\Delta_{\mathrm{p}}$}{Power consumption load factor}%
\nomenclature{$\chi$}{Transmission load share}%
\nomenclature{$P_{\mathrm{S}}$}{\ac{BS} power consumption in sleep mode in W}%
\nomenclature{$P_0$}{Idle \ac{BS} power consumption (at zero transmission power) in W}%

\begin{figure}
\centering

\begin{subfigure}[t]{0.8\textwidth}
\includegraphics[width=\textwidth]{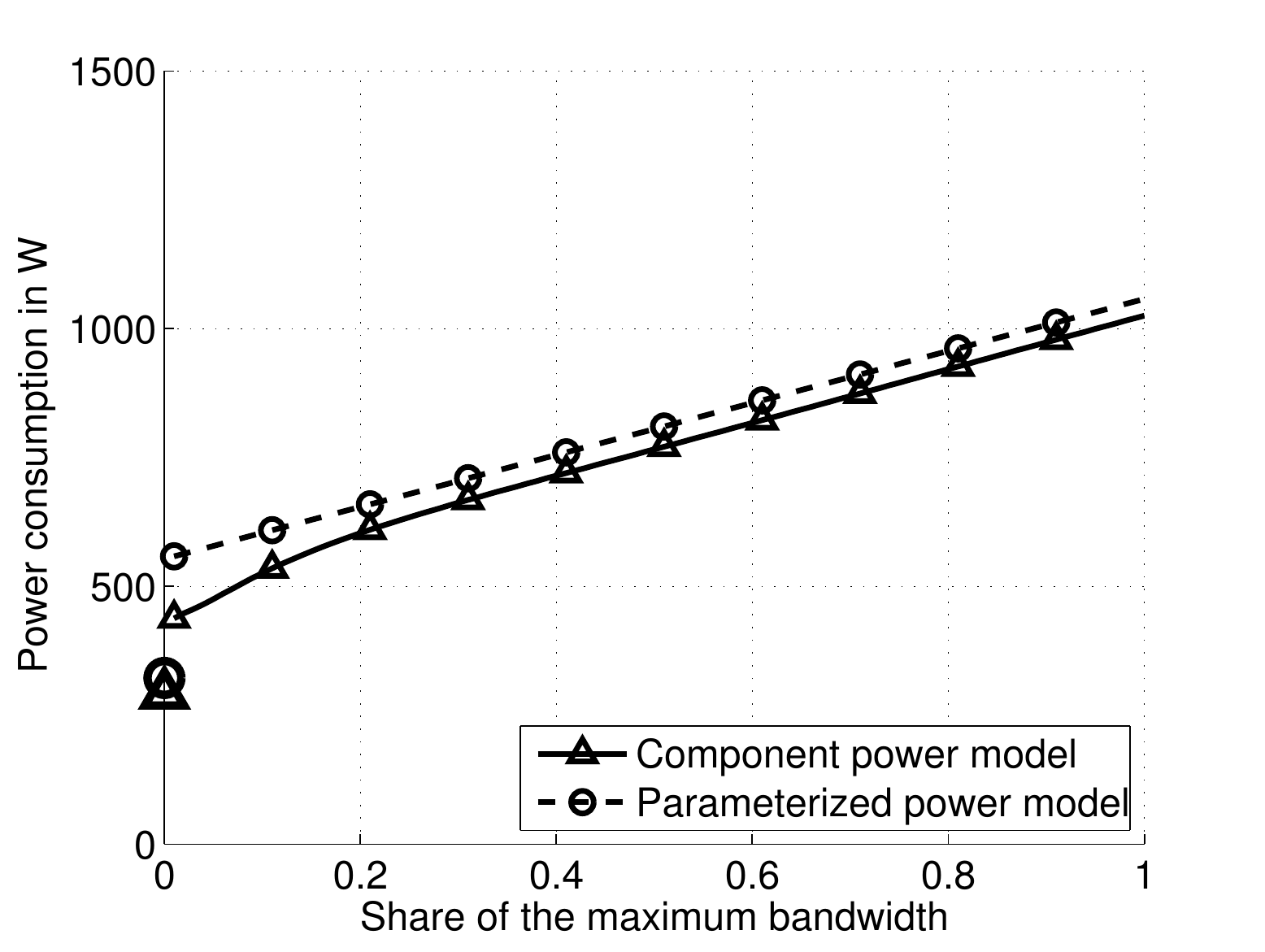}
\caption[]{One antenna}\label{fig:comparison_cmplx_param_1x}
\end{subfigure}

\begin{subfigure}[t]{0.8\textwidth}
\includegraphics[width=\textwidth]{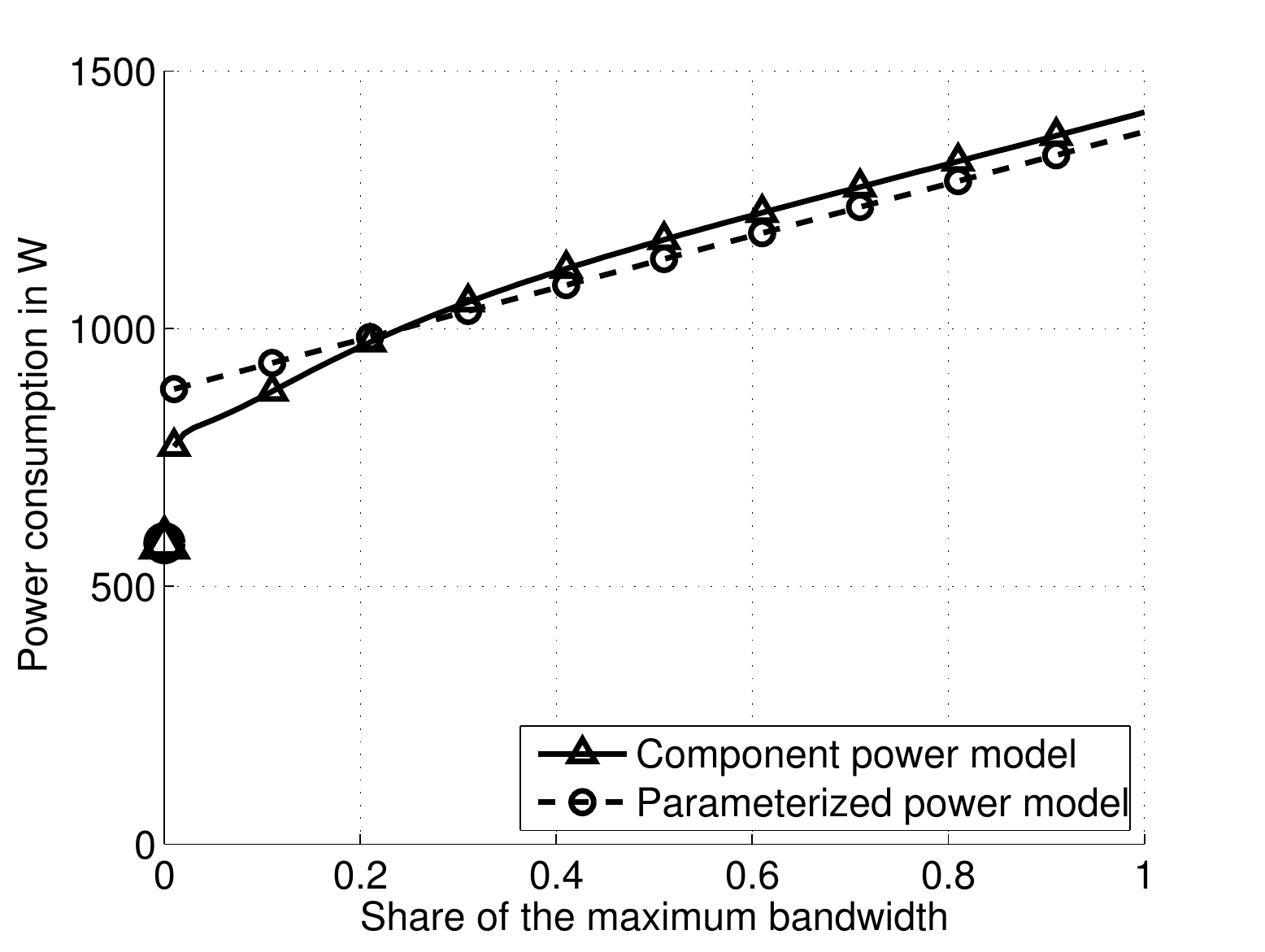}
\caption[]{Two antennas}\label{fig:comparison_cmplx_param_2x}
\end{subfigure}

 \caption[Comparison of the parameterized with the complex model]{Comparison of the parameterized with the complex model~\cite{ddgfahwsr1201} power models for the macro \ac{BS} type with 40~W transmission power.}
\label{fig:comparison_cmplx_param}
\end{figure}

The parameterized power model is applied to approximate the consumption of the \ac{BS} which was described in the component model. Parameters are chosen where possible according to~\cite{aggsoigdb1101}, such as losses, efficiencies and power limits. The remaining parameters are adapted such that a closer match to the component model could be achieved. The resulting parameter breakdown is provided in Table~\ref{tab:powerbreakdownmaximumload}. The two power models are compared over $f$ for one and two antennas in Figure~\ref{fig:comparison_cmplx_param}. Although two parameters, the bandwidth and the number of \ac{BS} antennas, are varied, the parameterized model can be seen to closely approximate the complex model for all \ac{BS} types. The largest deviation of the parameterized model from the complex model occurs when modeling four transmit antennas. This is caused by the fact that the parameterized model considers a constant slope, $\Delta_{\mathrm{p}}$, which is independent of $D$. In contrast, the \ac{PA} efficiency in the complex model decreases with rising $D$, leading to an increasing slope which can not be matched by a constant slope. This deviation is a trade-off between simplicity and model accuracy. 

In addition to providing a solid reference, the model and the parameters can provide a basis for exploration. Individual parameters can be changed to observe the resulting variation in power consumption. With regard to the number of antennas, the parameterized model can only be verified up to four antennas, which is the extent of the complex model. Extending the system bandwidth, for example to 20~MHz, is expected to increase the \ac{BB} and \ac{RF} power consumption. The other parameters such as the transmission power and losses are expected to remain unaffected by different system bandwidths. Adapting the design maximum transmission power, $P_{\mathrm{max}}$, affects the \ac{PA} efficiencies, which decrease with $P_{\mathrm{max}}$.


\begin{table*}[t]
\centering
\begin{tabular}{|c|c|c|c|c|c|c|c|c|c|}
\hline
 $P_{\mathrm{PA,limit}}$	& $\eta_\mathrm{PA,max}$	& $\theta$	& $P_\mathrm{BB}$	& $P_\mathrm{RF}$	& $\sigma_\mathrm{feed}$	& $\sigma_\mathrm{DC}$	& $\sigma_\mathrm{COOL}$	& $\sigma_\mathrm{AC}$ 	& $M_{\mathrm{Sec}}$ 	\\
 /W			&				&		&	/W		&	/W		&				&			&	&&\\
\hline
 	80.00		& 	0.36			& 	0.15	& 	29.4		& 	12.9		& 	0.5			& 	0.075		& 	0.1
& 	0.09	& 	3			\\
\hline
\end{tabular}
\\
\begin{tabular}{|c|c|c|c|}
\hline
 $P_{\mathrm{max}}$  	& $P_{1}$  		& $\Delta_{\mathrm{p}}$ & $P_{\mathrm{S},0}$\\
		/W		&	/W		& *10\,MHz		& /W\\
\hline
	40.00		& 	460.4 		& 	4.2 		&	324.0 \\
\hline
\end{tabular}

\caption[Parameter breakdown]{Parameter breakdown.}
\label{tab:powerbreakdownmaximumload}
\end{table*}

\section{The Affine Power Model}
\label{affinemodel}
In preparation for the following chapters of this thesis, one more simplification to power modeling is made. When the input parameters for \eqref{p1} and \eqref{eq:p_sleep} are not changed, the power model can be reduced to \eqref{eq:psupply} as a function of the load $\chi$. When considering \ac{AA}, a separation has to be made depending on $D$. This simplification carries the benefit of turning the power model into an affine function. As will be shown in the next chapter, the treatment of power consumption as an affine function maintains convexity. 

Let the general affine model be
\begin{equation}
\label{eq:affinemodel}
 P_{\mathrm{supply}}(\chi) = 
  \begin{cases}
   3  \left( P_{1,D} + \Delta_{\mathrm{p}} P_{\mathrm{max}} (\chi - 1) \right)			& \text{if } 0 < \chi \leq 1,  \\
   3 \cdot P_{\mathrm{S},D}      	& \text{if } \chi = 0,
  \end{cases}
\end{equation}
with $P_{1,D} =  P_{0,D} + \Delta_{\mathrm{p}} P_{\mathrm{max}}$.
\nomenclature{$P_{1,D}$}{\ac{BS} power consumption at maximum load in the affine model in W}%
\nomenclature{$P_{0,D}$}{Idle \ac{BS} power consumption (at zero transmission power) in the affine model in W}%
\nomenclature{$P_{\mathrm{S},D}$}{\ac{BS} power consumption in sleep mode in the affine model in W}%

For the remainder of this thesis, power consumption will therefore be modelled according to \eqref{eq:affinemodel}. This is a combination of the parameterized affine function and the component model sleep mode value in Fig.~\ref{fig:comparison_cmplx_param}. 

For later reference, the affine power model for a single antenna \ac{BS} is explicitly written as
\begin{equation}
\label{eq:affinemodel1x}
 P_{\mathrm{supply}}(\chi) = 
  \begin{cases}
   3 \left( 354 + 4.2 P_{\mathrm{max}} (\chi - 1) \right)			& \text{if } 0 < \chi \leq 1,  \\
   321      	& \text{if } \chi = 0.
  \end{cases}
\end{equation}

The affine power model for a two-antenna \ac{BS} is
\begin{equation}
\label{eq:affinemodel2x}
 P_{\mathrm{supply}}(\chi) = 
  \begin{cases}
   3 \left( 460 + 4.2 P_{\mathrm{max}} (\chi - 1) \right)			& \text{if } 0 < \chi \leq 1,  \\
   648      	& \text{if } \chi = 0.
  \end{cases}
\end{equation}

A tabular summary is provided in Table~\ref{tab:affinepmparameters}.

\begin{table}
 \centering
\begin{tabular}{c | c c c c}
$D$ 	& $P_{1,D}$ 	&$P_{0,D}$ 	&$P_{\mathrm{S},D}$	&$\Delta_{\mathrm{p}}$  \\
	&	/W	&	/W	&	/W		& 			\\
 \hline
1	& 354 		& 186 		& 107 			&4.2			\\
2 	& 460		& 292		& 216			&4.2			\\
\end{tabular}
\caption[Summary of affine power model parameters]{Summary of affine power model parameters.}
\label{tab:affinepmparameters}
\end{table}

\section{Summary}
\label{ch3:summary}
In this chapter, three different power models have been derived, namely the component model, the parameterized model, and the affine model. The component model was derived by first defining the architecture of a typical \ac{LTE} \ac{BS} and then inspecting each component with regard to its power consumption. The \ac{PA} was found to have a power consumption which strongly depends on the transmission power with some capability for sleep mode. The \ac{RF} transceiver in its current implementation has limited adaptability and, therefore, constant power consumption. The \ac{BB} unit's power consumption is determined by the internal computing operations, some of which can be reduced for power saving. The power consumption components such as power conversions and cooling, strongly depends on the power consumption of other components. Altogether, the power consumption of the individual components comprises the component \ac{BS} power model. While very detailed and necessary for derivation, the component model is too elaborate for many uses. To address this complexity, the parameterized power model was derived from it, which only considers typical operating parameters while abstracting architectural and technical details. It was shown to closely approximate the component model for a range of parameters. Instead of using measured data ranges containing many values in the model, all input parameters are generalized to scalars (single values). When employed in an analytical context, a further fixation of input parameters is advisable, yielding the affine model. Overall, the number of input parameters including experimental data was reduced from 15~parameters (out of which 8~are data ranges) in the component model to 14~scalar parameters in the parameterized model and to four scalar parameters in the affine model. See Table~\ref{tab:pmcomparison} on page~\pageref{tab:pmcomparison} for the detailed comparison.

\begin{table}
\centering
\begin{tabular}{l c c c}
Parameter 			& \begin{tabular}{@{}c@{}}Component \\ model\end{tabular}  	& \begin{tabular}{@{}c@{}}Parameterized \\ model\end{tabular}  	& \begin{tabular}{@{}c@{}}Affine \\ model\end{tabular} \\
\hline
Feeder loss 			& S		& S		& \\
\ac{PA} efficiency 		& V		& S		& \\
\ac{PA} sleep consumption	& S		& 		& \\
\ac{RF} consumption 		& V		& S		& \\
\ac{BB} consumption 	 	& V		& S		& \\
Scaling exponent for $D$	& V		& 		& \\
Scaling exponent for $f$ 	& V		& 		& \\
Sleep switching \ac{BB} 	& V		& 		& \\
\ac{DC}-\ac{DC} loss		& V		& S		& \\
Maximum \ac{DC}-\ac{DC} output 	& S		& 		& \\
\ac{AC}-\ac{DC} loss 		& V		& S		& \\
Maximum \ac{AC}-\ac{DC} output	& S		& 		& \\
Cooling efficiency 		& S		& S		& \\
Number of sectors 		& S		& S		& \\
Number of radio chains 		& S		& S		& S\\
Maximum transmission power 	& 		& S		& \\
\ac{PA} efficiency decrease	& 		& S		& \\
\ac{PA} limit	 		& 		& S		& \\
Sleep consumption reference	&		& S		& S\\
Power consumption load factor 	& 		& S		& S\\
Maximum power consumption	&		&		& S\\
\hline\hline
Number of data range vectors (V)& 8		&0		&0 \\
Number of scalars (S)		& 7		&14		&4\\
\end{tabular}
\caption[Comparison of required input parameters for different power models]{Comparison of required input parameters for different power models.}
\label{tab:pmcomparison}
\end{table}

\acresetall 

\renewcommand{\dir}{chapter4/chapter/}
\chapter{Power Saving on the Cell Level (Single-cell)}
\label{ch:cell}

\section{Overview}
As introduced in Chapter~\ref{ch:motivation}, current cellular systems are capacity optimized with little consideration of the power consumed. In contrast, for power saving, networks need to be efficiency optimized. The difference between a power efficient and a capacity maximizing system is that a power efficient system only consumes the resources that are \emph{needed} while a capacity maximizing system consumes those that are \emph{available}. As described in Section~\ref{sect:eebs}, cellular systems are rarely fully loaded and at times completely unloaded~\cite{s1001}. This implies that not all of the bandwidth, time or transmission power are needed. The power-saving \ac{RRM} mechanisms proposed in this chapter address this issue and exploit spare capacity to increase the system efficiency. 

In this chapter the previously established power models are applied to study power-saving \ac{BS} operation. Section~\ref{ch4:literature} discusses the \ac{SOTA} of power-saving \ac{RRM} in the literature. It is then identified in Section~\ref{ch4:pctdma} how a transmission consumes power and which parameters should be optimized to minimize consumption on the basis of general communication theory. Trade-offs are identified between \ac{TDMA} and \ac{PC}. In Section~\ref{ch4:PRAIS}, the solution to multi-user \ac{BS} \ac{PC} is proposed and studied. Once the system is optimized with regard to \ac{PC}, \ac{DTX} is added to further drive down power consumption via joint optimization. The general real-valued \ac{TDMA} solution is derived. In Section~\ref{ch4:RAPS}, the convex optimization problem is then extended into a single-cell multi-user \ac{MIMO} \ac{OFDMA} power-saving algorithm. The chapter is summarized in Section~\ref{ch4:summary}.

The material presented in this chapter has previously been published or submitted for publication in~\cite{ha1101,hah1101a,hah1101,habh1301,wrz1201}.

\section{Power-saving \ac{RRM} in Literature}
\label{ch4:literature}
In the literature, \ac{PC} is the most prominent power-saving \ac{RRM} mechanism with numerous proposals for a range of different problems, for example the related works in~\cite{ssha0701,wclm9901,mhlb0801,kll0301,aw1001}. This prominence is due to the fact that, in addition to reducing power consumption, \ac{PC} is also beneficial to link adaptation and interference reduction. Sinanovi\'{c} \ea~\cite{ssha0701} explore the optimal power points in a network with two interfering links and provide a communication theoretic basis for this work. Wong \ea~\cite{wclm9901} minimize transmit power for multi-user \ac{OFDM} under rate constraints. Although their work is pioneering, it has the drawbacks that the \ac{PC} algorithm is computationally complex and does not consider a transmit power constraint or a power model. Miao \emph{et al.}~\cite{mhlb0801} presented an early work on efficiency in which they derive the uplink data rate which maximizes the transmitted information per unit energy (bit per joule). Kivanc \ea~\cite{kll0301} allocate subcarriers such that transmission powers are reduced. Their algorithm is modified and applied in this chapter. Al-Shatri \ea~\cite{aw1001} approach sum rate maximization by using margin-adaptive power allocation, a method which is also used towards the end of this chapter.

More recently, acknowledging the significance of transmission-power-independent power consumption for energy efficiency, \ac{DTX} and \ac{AA} have been identified as promising energy saving \ac{RRM} techniques. The term \emph{\ac{DTX}} refers to an interruption of transmission which can be used to enter a short sleep mode. This interruption can be short enough to be unnoticeable to a receiver and fit into an \ac{OFDMA} frame. This allows to schedule \ac{DTX} flexibly without affecting the \ac{UE}. Thus, \ac{DTX} carries the benefit of not increasing delay or inhibiting discovery compared to longer sleep modes. \ac{DTX} is first proposed in~\cite{fmmjg1101}. To the best of the author's knowledge, there are no prior works which consider \ac{DTX} in combination with other power saving \ac{RRM} mechanism. With regard to \ac{AA}, Cui \emph{et al.} analyze the energy efficiency of \ac{MIMO} transmissions, being the first to consider the supply power consumption in energy efficient operation~\cite{cgb0401}. They show that in terms of energy efficiency, the optimal number of transmit antennas used for \ac{MIMO} transmission is not the highest by default, but rather depends on the \ac{SNR}. Kim \ea~\cite{kcvh0901} establish \ac{AA} as a \ac{MIMO} resource allocation problem and adapt the number of transmit antennas on a single link. However, solutions provided for \ac{AA} on the link level cannot be directly applied to the \ac{BS}, since for the latter multiple transmissions are active simultaneously on the same set of antennas. Xu \ea~\cite{xylzcx1101} propose a convex optimization scheme to reduce the number of transmission antennas during transmission. A related approach is applied in this chapter to the optimization of sleep modes. Hedayati \emph{et al.}~\cite{hafm1201} propose to perform \ac{AA} without informing the receiver, thus mimicking deep fades on some antennas. This is predicted to reduce power consumption by up to 50\%. 

Note that---as mentioned in Chapter~\ref{ch:motivation}---there is also a number of works concerned with \emph{long sleep} modes, such as~\cite{ok1001,abh1101,vdcjpmd1201,xpvgc1101}. In comparison with \ac{DTX}, such long sleep modes are applied for minutes, hours or days. This increases the achievable power saving, as devices are completely turned off. However, long sleep modes cause problems with coverage and discovery, as two communication devices may be completely unaware of one another. Also, such sleep modes clearly affect the \ac{QoS}, unless redundant network elements exist. Thus, they are not considered here. The work presented in this chapter can be combined with long sleep modes as part of future work as the two are not mutually exclusive.

\section{\ac{PC} and \acs{TDMA}}
\label{ch4:pctdma}

\begin{figure}
\begin{center}
\includegraphics[width=0.6\textwidth]{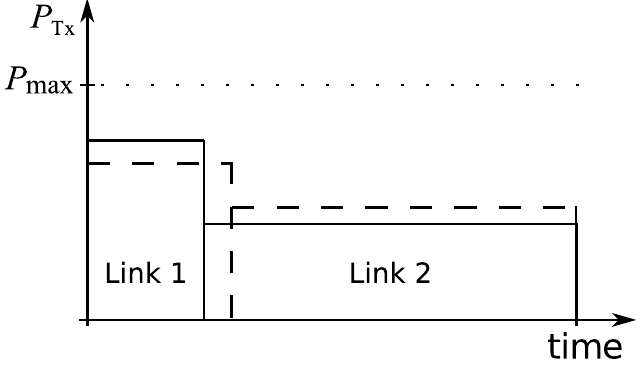}
\end{center}
\caption[Illustration of two possible power/time trade-offs]{Illustration of two possible power/time trade-offs that provide equal rates on both links.}
  \label{fig:PC}
\end{figure}

When the number of bits transmitted on a link with fixed bandwidth and optimal modulation and coding is to be increased, there are two options: either to increase the link rate by raising the transmission power or to transmit for a longer time. From a power efficiency perspective, providing a fixed rate is thus a trade-off between transmit power and transmission duration. A higher power for smaller duration can provide the same rate as a lower power with higher transmission duration. Therefore, for power minimization on a single link, transmission should always be as long as possible to allow for the lowest transmit power. However, in a shared multi-user channel with orthogonal access, such as \ac{TDM}, \ac{FDM}, or \ac{OFDM}, all links have to be considered which each have individual rate requirements that have to be fulfilled in a given time. This problem is illustrated in Fig.~\ref{fig:PC} for two links. Which combination of transmission duration and transmission power achieves the lowest overall power consumption is the problem discussed in this section.

\subsubsection{Power allocation on the link level}
It follows a derivation of the power optimal allocation strategy of transmission power and time for a single link. The Shannon bound~\cite{s4801}---as one of the most fundamental laws in communications---is employed to map transmission power to achievable rate. It is assumed that in a cellular system a set of mobiles with known channel gains has a set of rate requirements that needs to be fulfilled. 

A cell consists of one \ac{BS} and $K$ links (mobiles). The capacity in bps per link $k$ is upper bounded by the Shannon bound as
\begin{equation}
R_k = W \log_2 \left( 1 + \gamma_k \right),
\label{eq:Shannon}
\end{equation}
\nomenclature{$R_k$}{Capacity on link $k$ in bps}%
\nomenclature{$\gamma_k$}{\ac{SNR} on link $k$}%
\nomenclature{$G_{k}$}{Channel gain on link $k$}%
\nomenclature{$P_k$}{Transmission power on link $k$ in W}%
\nomenclature{$P_{\mathrm{N}}$}{Thermal noise power in W}%
\nomenclature{$q_{\mathrm{B}}$}{Boltzmann constant in $\mathrm{m}^2 \mathrm{kg} \mathrm{s}^{-2} \mathrm{K}^{-1}$}%
\nomenclature{$\vartheta$}{Operating temperature in K}%
\nomenclature{$k$}{Link or user index}%
where $W$ is the channel bandwidth in Hz and $\gamma_k = \frac{G_{k} P_k}{P_{\mathrm{N}}}$ is the \ac{SNR} with $G_{k}$ the link channel gain, $P_k$ the transmit power on link $k$ in W and $P_{\mathrm{N}}$ thermal noise power in W. The noise power is defined as $P_{\mathrm{N}} = W q_{\mathrm{B}} \vartheta$ with Boltzmann constant~$q_{\mathrm{B}}$ and operating temperature~$\vartheta$ in Kelvin. While capacity and bandwidth are linearly related, capacity and transmit power have a logarithmic relationship. As a consequence it is much more expensive in terms of power to increase the channel capacity than in terms of bandwidth. In other words, if there is a choice between 
\begin{enumerate}[label=\emph{\alph*})]
 \item leaving idle spectrum and transmitting at higher power, and
 \item using all available spectrum and transmitting at the lowest required power,
\end{enumerate}
then the latter will always consume less overall transmit power. 

\subsubsection{Power allocation on the cell level}
Next, an optimization problem is proposed which describes this notion for $K$ users on a shared channel. Since all links in the system share the available resources, the \ac{RRM} of one link affects all others. Thus, power-optimization can not be performed on the link level only, but has to be approached from the cell level.
\nomenclature{$K$}{Number of users/links}%

Let the normalized transmission time per link be given by
\begin{equation}
\label{eq:muk}
  \mu_k = \frac{\tau_k}{\tau_{\mathrm{frame}}},
\end{equation}
\nomenclature{$\mu_k$}{Normalized transmission time on link $k$}%
\nomenclature{$\tau_k$}{Transmission time on link $k$ in s}%
\nomenclature{$\tau_{\mathrm{frame}}$}{Frame duration in s}%
where $\tau_k$ is the transmission time on link $k$ in seconds, $\tau_{\mathrm{frame}}$ is the frame duration in seconds. Optimization of normalized time rather than absolute time is chosen here as it results in average power minimization and is more illustrative than energy minimization (which is only meaningful for a known $\tau_{\mathrm{frame}}$). 

The average capacity, $\overline{R}_k$, on link~$k$ depends on $\mu_k$ and $R_k$, with
\begin{equation}
 \overline{R}_k = \mu_k R_k,
\label{eq:overlineR}
\end{equation}
where $\overline{R}_k$ indicates the average of $R_k$.

Solving \eqref{eq:Shannon} for $P_k$ and combining with \eqref{eq:overlineR}, the transmission power on link $k$ in W as a function of the required average rate is found to be
\begin{equation}
 P_k(\overline{R}_k) = \frac{P_{\mathrm{N}}}{G_k} \left( 2^\frac{\overline{R}_k}{W \mu_k} - 1 \right),
\end{equation}
where $0 < P_k(\overline{R}_k) < P_{\mathrm{max}}$ for some $P_{\mathrm{max}}$.

To account for the fact that all links are served by the \ac{BS} orthogonally within some time $\tau_{\mathrm{frame}}$, the system average transmission power at the \ac{BS} for all links, $\overline{P}_{\mathrm{Tx}}$, is the sum of individual transmit powers weighted with the transmit duration, with
\begin{subequations}
\begin{align}
 \overline{P}_{\mathrm{Tx}}  			&= \sum_{k=1}^{K} \mu_k  P_{\mathrm{Tx},k}(\overline{R}_k)  \\
						&= \sum_{k=1}^{K} \mu_k  \frac{P_{\mathrm{N}}}{G_k} \left( 2^\frac{\overline{R}_k}{W \mu_k} - 1  \right). \label{PSYS}
\end{align}
\end{subequations}
The combined duration of all transmissions must be less or equal to $\tau_{\mathrm{frame}}$. But since it is most efficient to use the entire available $\tau_{\mathrm{frame}}$ as derived from \eqref{eq:Shannon}, it holds that
\begin{equation}
\label{muconstraint}
\displaystyle\sum_{k=1}^{K} \mu_k = 1.
\end{equation}

Furthermore, time has positive values and a power constraints must be obeyed with
\begin{subequations}
\begin{align} 
\mu_k &> 0 \quad \forall k,\\
0 &< P_k(\overline{R}_k) < P_{\mathrm{max}}\quad \forall k.
\end{align}
\end{subequations}

The allocation vector $\bm{\mu}^*=(\mu_1,\cdots,\mu_K)$ of transmission durations which minimizes \eqref{PSYS} is power optimal. 
\nomenclature{$\bm{\mu}^*$}{Solution vector to the time share problem}%

\subsubsection{Evaluation}
Next, the affine power model is taken into account, as it is important to consider transmission power in relationship with the consumption of the entire \ac{BS}. Instead of the transmission power, the supply power consumption is minimized. With the power model from \eqref{eq:affinemodel}, the \ac{PC} cost function \eqref{PSYS} and $\chi P_{\mathrm{max}}=P_k(\overline{R}_k)$, the supply power consumption in W is found to be
\begin{equation}
\label{eq:PSYSPM}
\begin{aligned}
 P_{\mathrm{supply}}(\overline{R}_k)  &= \sum_{k=1}^{K} \mu_k \left( P_0 + \Delta_{\mathrm{p}} P_{\mathrm{Tx},k}(\overline{R}_k)  \right)\\
					     &= \sum_{k=1}^{K} \mu_k \left( P_0 + \Delta_{\mathrm{p}} \frac{P_{\mathrm{N}}}{G_k} \left( 2^\frac{\overline{R}_k}{W \mu_k} - 1 \right) \right).
\end{aligned}
\end{equation}

Note that, against intuition, the power model has no effect on the solution of \eqref{eq:PSYSPM}, \ie~$\bm{\mu}^*$ is equal for \eqref{PSYS} and \eqref{eq:PSYSPM}. As the power model affects all links equally, the \ac{PC} solution is the same for any power model and, thus, any \ac{BS}. However, although the \emph{solution} is independent of the hardware, the \emph{benefits} of \ac{PC} do strongly depend on hardware. As a metric for characterizing hardware by the share of load-dependent consumption of the overall consumption, the load dependence of a \ac{BS} is defined as
\begin{equation}
 \eta_{\mathrm{ld}} = \frac{\Delta_{\mathrm{p}} P_{\mathrm{max}}}{P_0+\Delta_{\mathrm{p}} P_{\mathrm{max}}}.
\end{equation} \nomenclature{$\eta_{\mathrm{ld}}$}{Load dependence of a \ac{BS}}%

\begin{figure}
\centering
 \includegraphics[width=0.8\textwidth]{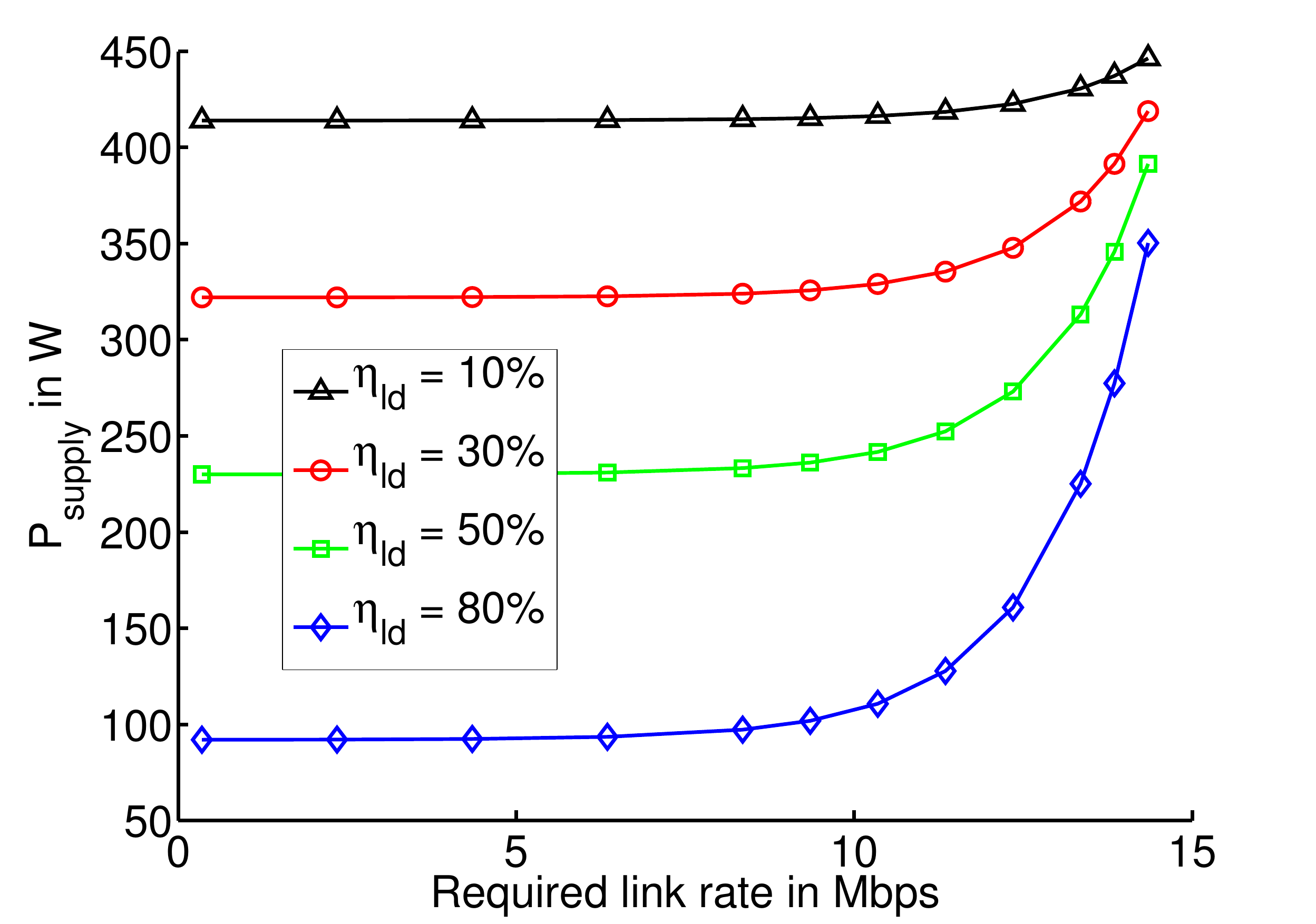}
\caption[Comparison of the effect of load dependence]{Comparison of the effect of load dependence on achievable power savings by \ac{PC}.}
\label{fig:PMcomparisonOverRate}
\end{figure}

Using this load dependence, the benefits of \ac{PC} can be evaluated. Figure~\ref{fig:PMcomparisonOverRate} shows the supply power in a cell with ten users for a range of load dependence values after the application of the \ac{PC} strategy in~\eqref{PSYS}. The slope of the power model $\Delta_{\mathrm{p}}$ was chosen for a representative set of load dependences with $\eta_{\mathrm{ld}}=\{0.1,0.3,0.5,0.8\}$ for $P_1=460$\,W and $P_{\mathrm{max}}=40$\,W, from~\eqref{eq:affinemodel}. Channel gains were calculated based on distance by the \ac{3GPP} path loss model~\cite{std:3gpp-faeutrapla} after dropping users uniformly on a disc with 250\,m radius around the \ac{BS} in the center. The problem \eqref{eq:PSYSPM} was solved iteratively using the interior-point algorithm contained in the MATLAB Optimization Toolbox. The remainder of the simulation parameters was set as described in Table~\ref{tab:uniformdistrscenarioparameters}.

As shown in Fig.~\ref{fig:PMcomparisonOverRate}, the effectiveness of \ac{PC} strongly depends on the underlying hardware. A higher load dependence factor yields a lower supply power consumption for \ac{PC}. The effectiveness of \ac{PC} in future \acp{BS} therefore strongly depends on hardware developments.

\begin{table}      
\centering
      \begin{tabular}{c|c}
	Parameter          			& Value\\
	\hline
	Carrier frequency  			& 2\,GHz\\
	Cell radius  				& 250\,m\\
	Path loss model  			& \ac{3GPP} UMa, NLOS, shadowing~\cite{std:3gpp-faeutrapla}\\
	Shadowing standard deviation		& 8\,dB\\
	Iterations  				& 1,000 \\
	Bandwidth $W$ 				& 10\,MHz \\
	Maximum transmission power $P_{\mathrm{max}}$	& 46\,dBm \\
	Operating temperature $T$ 		& 290\,K \\
	System/subcarrier bandwidth 		& 10\,MHz/200\,kHz\\
	Thermal noise power per subcarrier	& $4w\times 10^{-21}$\,W\\
	\end{tabular}
\caption[Simulation parameters]{Simulation parameters.}
\label{tab:uniformdistrscenarioparameters}
\end{table}

Although \ac{PC} is thus shown to provide significant saving potential in \acp{BS} with high $\eta_{\mathrm{ld}}$, like macro \acp{BS}, the power consumption of a device employing \ac{PC} is always lower bound by its idle power consumption, $P_0$. Even when transmission is stopped, a device employing only \ac{PC} still consumes a supply power consumption of $P_0$. Thus, to further reduce power consumption, a method is required which allows power consumption to reach lower values than $P_0$. This can be achieved by employing sleep modes or---in the case of \ac{OFDMA} scheduling---\ac{DTX}.

\section{\ac{PRAIS}}
\label{ch4:PRAIS}

\begin{figure}
\centering
\includegraphics[width=0.6\textwidth]{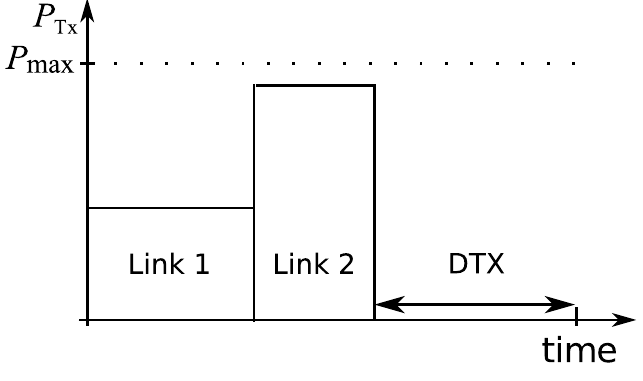}
\caption[Illustration of \acs{PRAIS} for two links]{Illustration of \ac{PRAIS} for two links: transmit power, resource share and \ac{DTX} are allocated optimally in order to serve the requested rate at minimal supply power consumption.}
\label{fig:PRAIS}
\end{figure}

This section introduces the \ac{PRAIS} scheme, which combines three techniques: \ac{PC}, \ac{TDMA} and \ac{DTX}. See Figure~\ref{fig:PRAIS} for an illustration of \ac{PRAIS} on two links. 

The parameterized power model in Section~\ref{parameterizedmodel} derived three general mechanisms which affect the power consumption of \acp{BS}: 
\begin{enumerate}
 \item The overall power emitted at the antenna(s), $P_{\mathrm{Tx}}$, affinely relates to supply power consumption, see~\eqref{eq:affinemodel}. 
 \item If the same \ac{BS} is operated with fewer \ac{RF} chains, it consumes less power, see Fig.~\ref{fig:comparison_cmplx_param}. 
 \item If the \ac{BS} is put into sleep mode, it consumes less power than in the active (idle) state with zero transmission power, see Fig.~\ref{fig:comparison_cmplx_param}.
\end{enumerate}

These observations lead to the following intuitive saving strategies: 
\begin{description}
 \item[\ac{PC}]: Reduce transmission power.
 \item[\ac{AA}]: Reduce the number of \ac{RF} chains.
 \item[\ac{DTX}]: Increase the time the \ac{BS} spends in \ac{DTX}.
\end{description}

The first and the last strategy are clearly opposing each other as lower transmission powers lead to lower link rates and thus longer transmission duration (for transmitting a certain number of bits), whereas it would be beneficial for long \ac{DTX} to have short transmissions. The second and third strategy are related, as \ac{AA} can be considered a weak form of \ac{DTX} which still allows transmission on a subset of antenna elements. In this section, \ac{PC} and \ac{DTX} are jointly addressed. \ac{AA} is treated in the following section, Section~\ref{ch4:RAPS}.

\subsubsection{Joint \ac{PC} and \ac{DTX}}
When employing \ac{DTX} individually to minimize power consumption, the optimal strategy is to serve all links at the maximum available transmission power until target rates are fulfilled, and then set the \ac{BS} to \ac{DTX}. Alternatively, when employing \ac{PC} with \ac{TDMA} individually for maximum efficiency, the transmission duration is stretched over the available time frame with the lowest power necessary to serve the required rates. When \ac{PC} and \ac{DTX} are combined, there is a trade-off. Between the two extremes of maximum transmit power with longest \ac{DTX} and lowest transmit power with no \ac{DTX}, there are configurations with medium transmit power and medium \ac{DTX} duration that consume less supply power. 

\begin{figure}
\centering
 \includegraphics[width=0.8\textwidth]{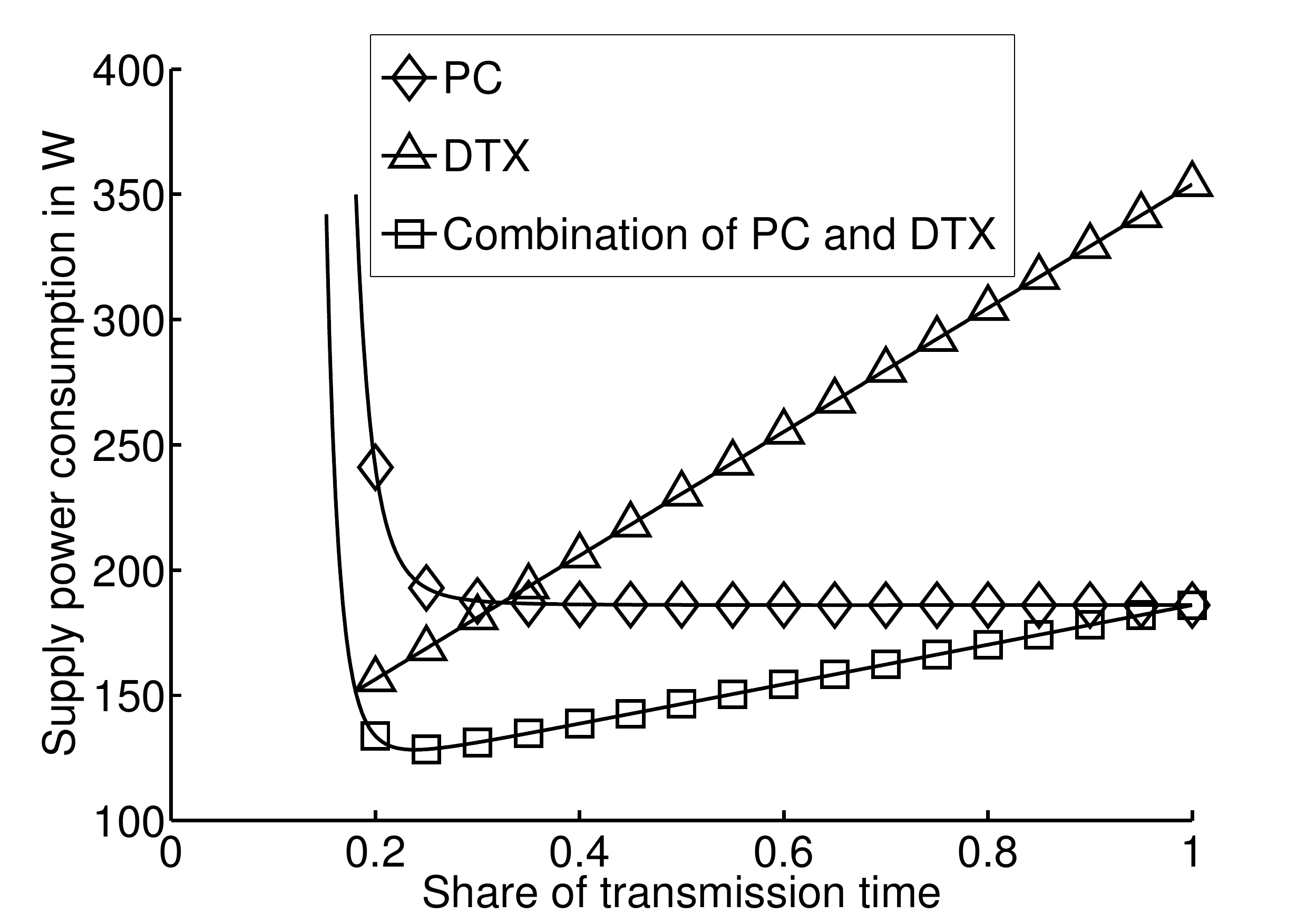}
\caption[Supply power consumption for a target spectral efficiency]{Supply power consumption for transmission of a target rate as a function of time spent transmitting, $\mathit{\Phi}$. The combination of \ac{DTX} and \ac{PC} achieves lower power consumption than exclusive operation of either DTX or \ac{PC}.}
\label{fig:PCsleeptradeoff}
\end{figure}

This can be illustrated graphically for a single link example, as shown in Fig.~\ref{fig:PCsleeptradeoff}. Plotted is the supply power consumption caused by transmitting a fixed number of bits. Let $\mathit{\Phi}\in[0,1]$ denote the share of time spent transmitting. Three operation modes are defined:
\nomenclature{$\mathit{\Phi}$}{Share of time spent transmitting instead of idling or \ac{DTX}}%
\begin{itemize}
 \item \ac{PC}: Only \ac{PC} but no \ac{DTX} is available. To this extent, the transmission power is adjusted depending on the transmission time normalized to the time slot duration, $\mathit{\Phi}$, such that the target rate is achieved. The \ac{BS} consumes idle power $P_0$ when there is no data transmission. Clearly, the point of lowest power consumption in this case is at $\mathit{\Phi}=1$ where the supply power consumption approaches $P_0$. Lower values of $\mathit{\Phi}$ increase the required transmission power, until at $\mathit{\Phi}=0.18$, $P_{\mathrm{Tx}} = P_{\mathrm{max}}$.
\item \ac{DTX}: The \ac{BS} transmits with full power $P_{\mathrm{Tx}} = P_{\mathrm{max}}$. Thus, the supply power is $P_0 + \Delta_{\mathrm{p}} P_{\mathrm{max}}$ when transmitting or $P_{\mathrm{S}}$ when in \ac{DTX} mode, yielding an affine function of $\mathit{\Phi}$. At $\mathit{\Phi}=1$, a higher rate than the target rate is achieved. Reducing $\mathit{\Phi}$ from $\mathit{\Phi}=1$ decreases the supply power consumption linearly up to the point at which the target rate is met with equality. In this mode of operation, the best strategy is to minimize the time transmitting (small $\mathit{\Phi}$), in order to maximize the time in \ac{DTX}, ($1-\mathit{\Phi}$). In Fig.~\ref{fig:PCsleeptradeoff}, the point of lowest power consumption for this mode is at $\mathit{\Phi}=0.19$ with 154~W.
\item Joint application of \ac{PC} and \ac{DTX}: In this mode of operation, the \ac{DTX} time ($1-\mathit{\Phi}$) is gradually reduced. The transmission power is adjusted to meet the target rate within~$\mathit{\Phi}$. Here, the point of lowest power consumption is at $\mathit{\Phi}=0.25$ with 135~W.
\end{itemize}
Inspection of Fig.~\ref{fig:PCsleeptradeoff} reveals that the joint operation of \ac{PC} and \ac{DTX} always consumes less power than either individual mode of operation with an optimal point at $\mathit{\Phi}=0.25$. This is the power-saving benefit provided by the joint optimization of \ac{PC} and \ac{DTX}.

\subsubsection{Problem formulation}
This section proceeds to derive the power consumption optimization problem for a multi-user single-cell allocation of transmit powers, \ac{DTX} times and transmit durations contained within the \ac{PRAIS} scheme. The normalized time spent transmitting, $\mu_k$, is defined as in \eqref{eq:muk} and the normalized duration spent in \ac{DTX} is
\begin{equation}
 \label{eq:nu}
 \nu = \frac{\tau_{\mathrm{S}}}{\tau_{\mathrm{frame}}},
\end{equation}
\nomenclature{$\nu$}{Normalized time spent in \ac{DTX} mode}%
where $\tau_{\mathrm{S}}$ is the time spent in \ac{DTX} and 
\nomenclature{$\tau_{\mathrm{S}}$}{Time spent in \ac{DTX} in s}%
\begin{equation}
 \displaystyle\sum_{k=1}^{K} \mu_k + \nu = 1.
\end{equation}

Combining this with the consumption during \ac{DTX} from \eqref{eq:affinemodel} and \eqref{PSYS} yields the following \ac{PRAIS} optimization problem.
\begin{subequations}
\begin{align}
& \underset{\mu, \nu}{\text{minimize}} 		& 	P_{\mathrm{supply}}(\overline{R}_k) 	&=  \left[ \sum_{k=1}^{K} \mu_k \left( P_0 + \Delta_{\mathrm{p}} \frac{P_{\mathrm{N}}}{G_k} \left( 2^\frac{\overline{R}_k}{W \mu_k} - 1 \right) \right) \right] + \nu P_{\mathrm{S}} \\
& \text{subject to} 				& 	 \sum_{k=1}^{K} \mu_k + \nu &= 1 \,,\label{eq:OPnumu} \\
& 						&	\nu 						&\ge 0,\label{eq:OPnu}\\
&						&	\mu_k 						&\ge 0 \quad \forall\,k, \label{eq:OPmu}\\
&						&	 0 \leq P_k 					&= \frac{P_{\mathrm{N}}}{G_k} \left( 2^{\frac{\overline{R}_k}{W \mu_k}} - 1 \right) \leq P_{\mathrm{max}} \quad \forall\,k.\label{eq:OPpmax}%
\end{align}%
\label{eq:OP}%
\end{subequations}%

The constraints \eqref{eq:OPnumu}, \eqref{eq:OPnu}, \eqref{eq:OPmu} reflect the facts that normalized time has to be positive and its sum unity. Transmission powers are positive and bounded by a maximum transmit power in~\eqref{eq:OPpmax}. The cost function is a non-negative sum of functions that are convex within the constraint domain and is therefore still convex. See Appendix~\ref{appendix:proof1} for a proof. It can be solved efficiently with appropriate software like the MATLAB optimization toolbox. The solution of \eqref{eq:OP} determines the vector $(\mu_1,\dots,\mu_K,\nu)$, which minimizes the overall power consumption under target rates. Note that selecting $P_{\mathrm{S}}\geq P_0$ is equivalent to disabling \ac{DTX}, \ie\ employing \ac{PC} individually similar to \eqref{eq:PSYSPM}, whereas fixating $P_{\mathrm{Tx}} = P_{\mathrm{max}}$ is equivalent to disabling \ac{PC}. 

The resource allocation problem~\eqref{eq:OP} can be solved for any number of users. Without loss of generality, a ten user scenario for numerical evaluation is selected.

\subsubsection{Evaluation}
For the numerical analysis, the \ac{PRAIS} scheme is evaluated in a Monte Carlo simulation using the parameters shown in Table~\ref{tab:uniformdistrscenarioparameters}. Users are dropped uniformly onto a disk and the associated channel gains~$G_k$ are generated by applying the \ac{3GPP} urban macro path-loss model including shadowing with a standard deviation of~$8$\,dB~\cite{std:3gpp-faeutrapla}. In addition to the individual \ac{DTX} and \ac{PC} allocation schemes and the \ac{PRAIS} scheme, two references that serve as upper limits are presented. First, the maximum transmission power as defined by the \ac{LTE} standard provides the theoretical reference. Second, the reference against which gains are measured is the power behavior of a \ac{BS} as defined in the power models. This is a \ac{BA} scheme where each user receives a share of the frequency band as well as the entire considered time slot. This can also be interpreted as `\ac{DTX} in the frequency domain' where $P_{\mathrm{S}} = P_0$. 

For power modeling, representative of current and future developments, the affine power model from Chapter~\ref{ch:device}, a model from the literature and a purely theoretical model are selected; the parameters are normalized for a single sector, single sector and listed in Table~\ref{tab:powermodelparameters}. 

For an assessment of the relevance of \ac{DTX}, an assumption is that \ac{DTX} will greatly improve, while standby component consumption cannot be reduced substantially in the coming years~\cite{fmmjg1101}. This model is labelled the \emph{Frenger model}. This model particularly emphasizes \ac{DTX} effects.

As a best-case example, the theoretical power model assumes idealized components which scale perfectly with load. Power consumption is set to scale almost linearly with load with near-zero stand-by consumption. This model is not a prediction of technology advances, but provides theoretical limits. Parameters are selected such that the power consumption at full load matches the affine model.

\begin{table}
\centering
 \begin{tabular}{c | c c c}
  Model & $P_0$/W & $\Delta_{\mathrm{P}}$ & $P_{\mathrm{S}}$/W \\
\hline
  Affine model in \eqref{eq:affinemodel1x} & 186 & 4.2 & 107\\
  Frenger model~\cite{fmmjg1101} & 170 & 3.4 & 10 \\
  Linear model & 1 & 8.8 & 1 \\
 \end{tabular}
\caption[Power model parameters used in Section~\ref{ch4:PRAIS}]{Single-sector power model parameters used in Section~\ref{ch4:PRAIS}.}
\label{tab:powermodelparameters}
\end{table}

\begin{figure*}
\centering
\begin{subfigure}[b]{.6\textwidth}
\includegraphics[width=\textwidth]{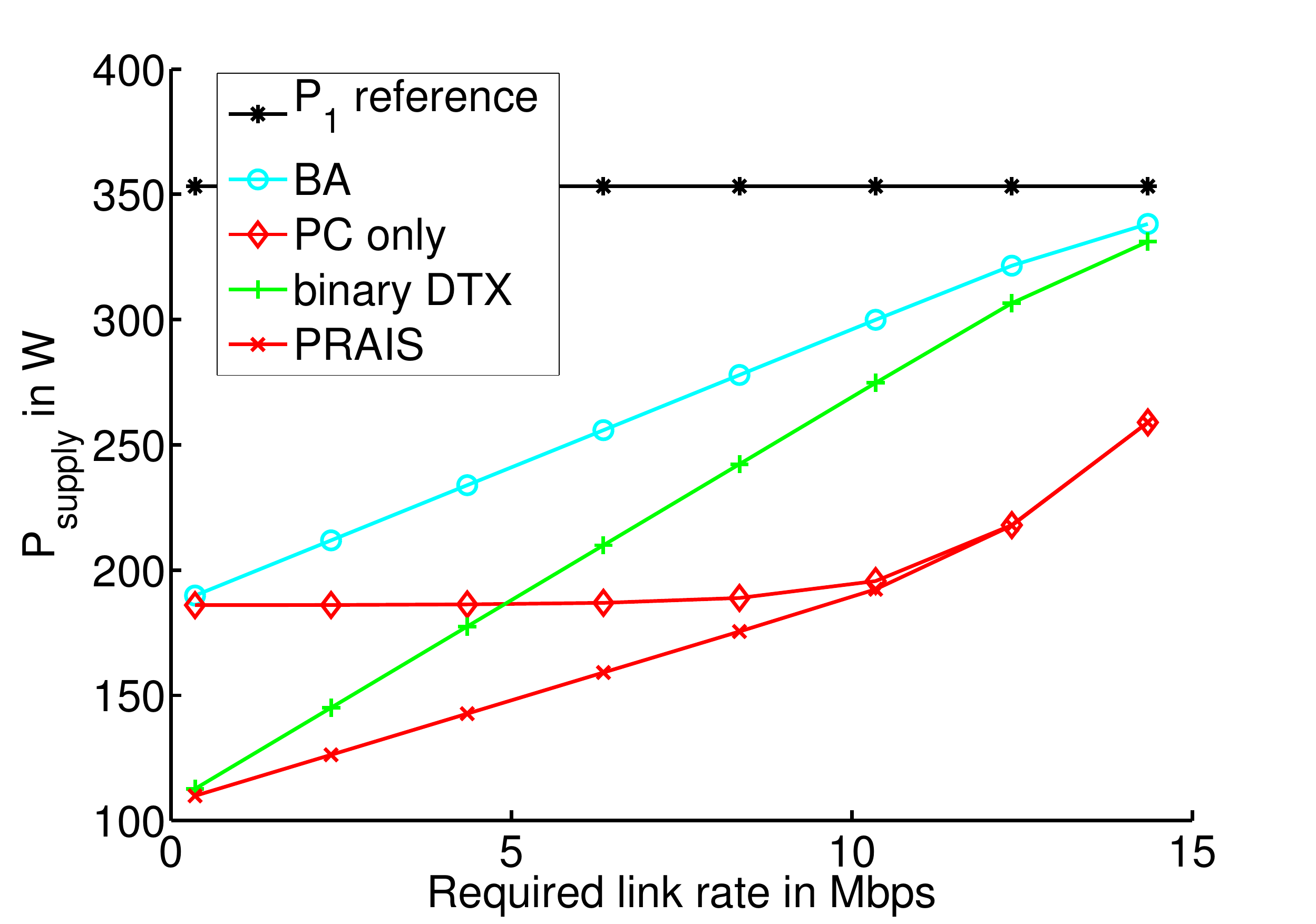}
\caption{Supply power consumption under the affine model.}\label{fig:resultsPM1}
\end{subfigure}
\begin{subfigure}[b]{.6\textwidth}
\includegraphics[width=\textwidth]{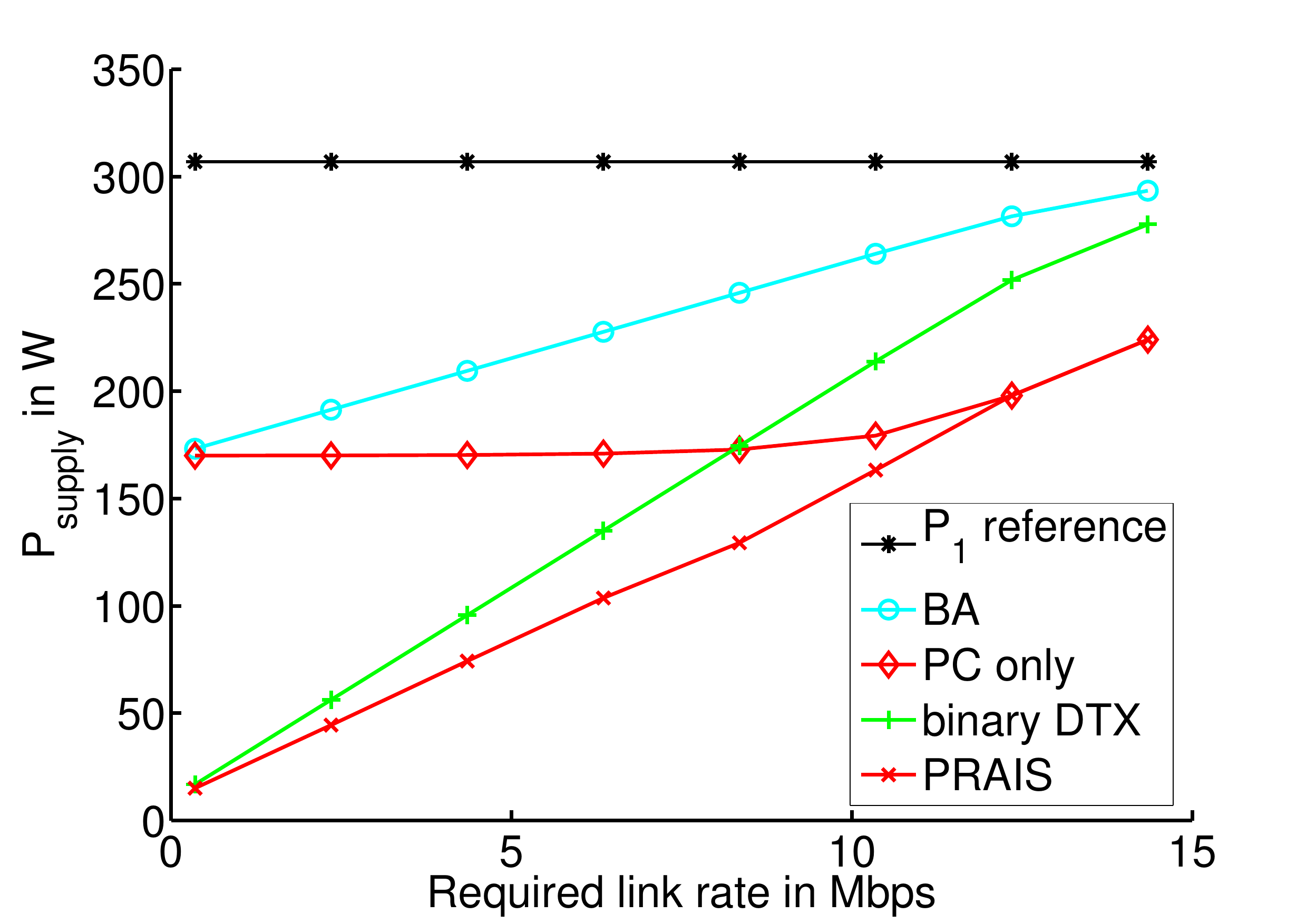}
\caption{Supply power consumption under the Frenger model.}\label{fig:resultsPM3}
\end{subfigure}
\begin{subfigure}[b]{.6\textwidth}
\includegraphics[width=\textwidth]{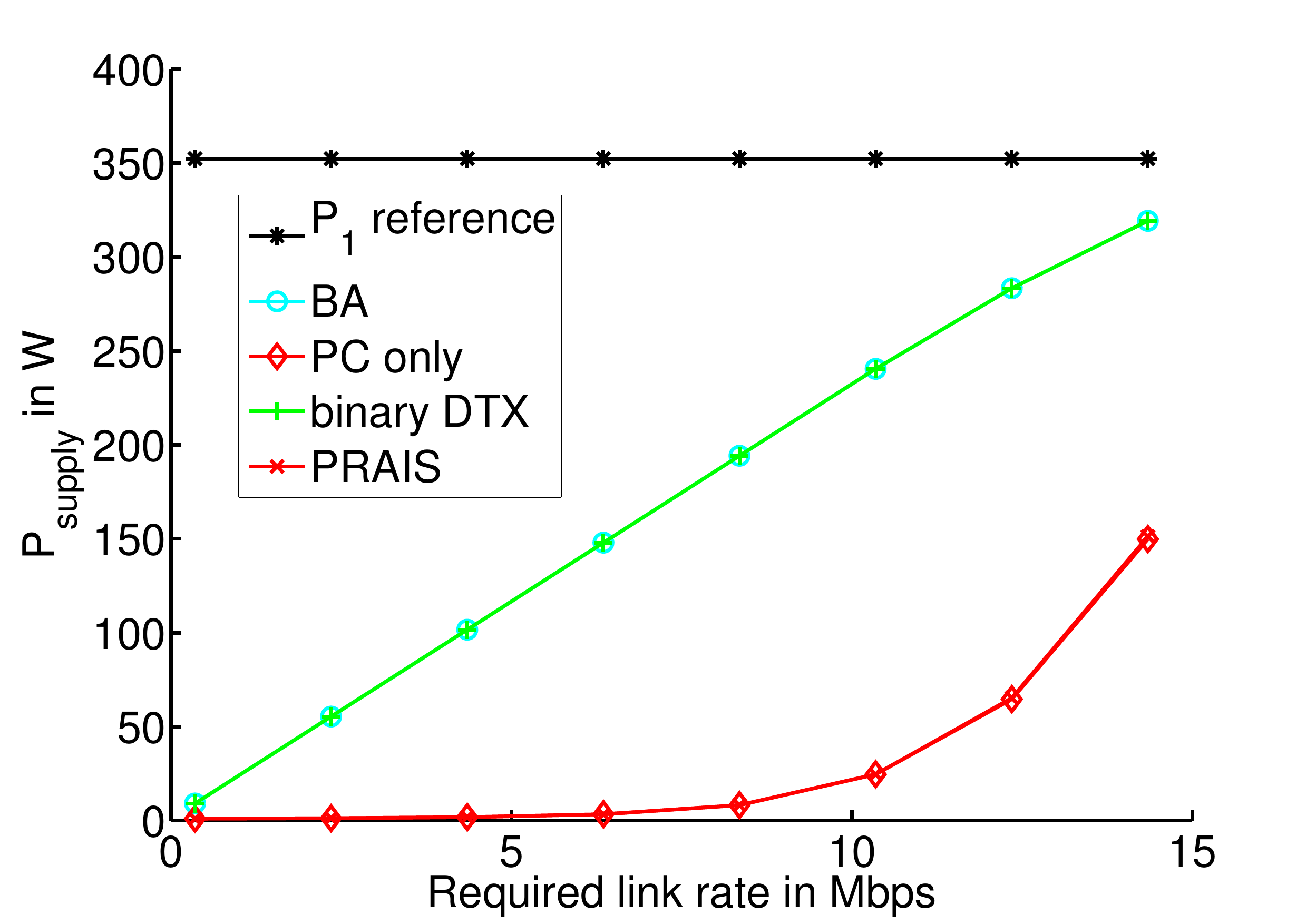}
\caption{Supply power consumption under the linear model.}\label{fig:resultsPM4}
\end{subfigure}
\caption[Fundamental limits for power consumption in \acsp{BS}]{Fundamental limits for power consumption in \acp{BS}.\label{fig:results}}
\end{figure*}

The simulation results are shown in Figure~\ref{fig:results} where the per-user link rate is plotted against the average supply power consumption under the different schemes. Figures~\ref{fig:resultsPM1},  \ref{fig:resultsPM3}, \ref{fig:resultsPM4} reflect the three chosen power models. 

It is found in Figure~\ref{fig:resultsPM1} that a \ac{SISO} \ac{LTE} \ac{BS} consumes 186\,W up to 340\,W employing \ac{BA}, which is considered the operation of the \ac{SOTA}. An important albeit foreseen result is that the consumption curves of \ac{BA} and \ac{PC} as well as \ac{DTX} and \ac{PRAIS} originate in the same values of $P_{\mathrm{supply}}$ at low load. The higher one is $P_0$ at 186~W and the lower one $P_{\mathrm{S}}$ at 107~W. This is true for all power models and confirms that in an empty cell, power consumption is determined by hardware and resource allocation has no effect. 

As shown in Fig.~\ref{fig:resultsPM1}, the application of \ac{PC} only allows keeping the overall consumption constant for a large set of low target rates. In this rate region, transmit powers are very low compared to standby consumption. Only when rates above 10~Mbps are targeted is the transmission power high enough to make a noticeable difference compared to the standby consumption. This is reflected in the rising \ac{PC} only curve at target rates above 10~Mbps. In contrast, the binary \ac{DTX} scheme (which transmits either at full power or not at all) has much lower power consumption (up to $-45$\%) than \ac{PC} only or \ac{BA} at low rates. However, it rises much quicker than the \ac{PC} only curve. There is a crossover point at 5.6~Mbps between \ac{DTX} and \ac{PC}. The \ac{PRAIS} scheme, which joins \ac{PC} and \ac{DTX}, benefits from both individual schemes and has minimal consumption over all rates. At higher target rates, the \ac{PRAIS} consumption curve joins the \ac{PC} only curve. This reflects that it is not feasible to put the \ac{BS} to sleep at high rates. Employing the affine power model, the \ac{PRAIS} scheme achieves savings between 42\% and 23\% over \ac{BA}.

A surprising result is found in Figure~\ref{fig:resultsPM3}. Although application of this model has a strong bias towards \ac{DTX} effects, there remains a cross-over point between \ac{DTX} and \ac{PC} after which the use of \ac{PC} can still save 19\% of the supply power consumption in addition to \ac{DTX}. In the Frenger model, the \ac{PRAIS} scheme offers between 91\% savings at near-zero rates and 23\% savings at 15~Mbps.

In the future linear model shown in Figure~\ref{fig:resultsPM4}, the behavior of \ac{DTX} and \ac{BA}, as well as \ac{PC} and \ac{PRAIS}, are identical, due to the fact that there is no gain of sleep modes over standby consumption. \ac{BA} consumption is significantly lower than for all other power models. All benefits are owed to \ac{PC} which is strongly amplified by the linear model behavior, delivering significant savings over all target rates.

Another finding evident from all figures is that there is no point of `full load' in a cell which causes maximum power consumption. In theory, when a cell is fully loaded---regardless of the resource allocation scheme---the \ac{BS} is expected to consume the maximum supply power, $P_1$. However, the chance that the set of target rates matches the given channel conditions to generate a situation of `full load' is extremely low. It is much more probable that either the channels conditions are good enough to have `almost full load' or that the channel conditions are too bad to fulfill the target rates, generating an overload or outage situation. As target rates increase, the probability of the latter increases. To cover only a representative set of target rates, only those data points are plotted which contain less than 10\% overload/outage. Due to this outage, the simulated power consumption never reaches the theoretical maximum. Thus, the selection of the resource allocation scheme does in fact matter, even at near-full loads.

\section{\acf{RAPS}}
\label{ch4:RAPS}
To extend the previous analytical work into a comprehensive, practical mechanism for current cellular systems like \ac{3GPP} \ac{LTE}, the \ac{RAPS} algorithm is proposed in this section, which reduces the \ac{BS} supply power consumption of multi-user \ac{MIMO}-\ac{OFDM}. Given the channel states and target rates per user, \ac{RAPS} finds the number of transmit antennas, the number of \ac{DTX} time slots and the resource and power allocation per user. 
The \ac{RAPS} solution is found in two steps: First, \ac{PC}, \ac{DTX} and resource allocation are joined into a convex optimization problem which can be efficiently solved. Second, subcarrier and power allocation for a frequency-selective time-variant channel pose a combinatorial problem, which is solved by means of a heuristic solution.

\begin{figure}
\centering
\includegraphics[width=0.6\textwidth]{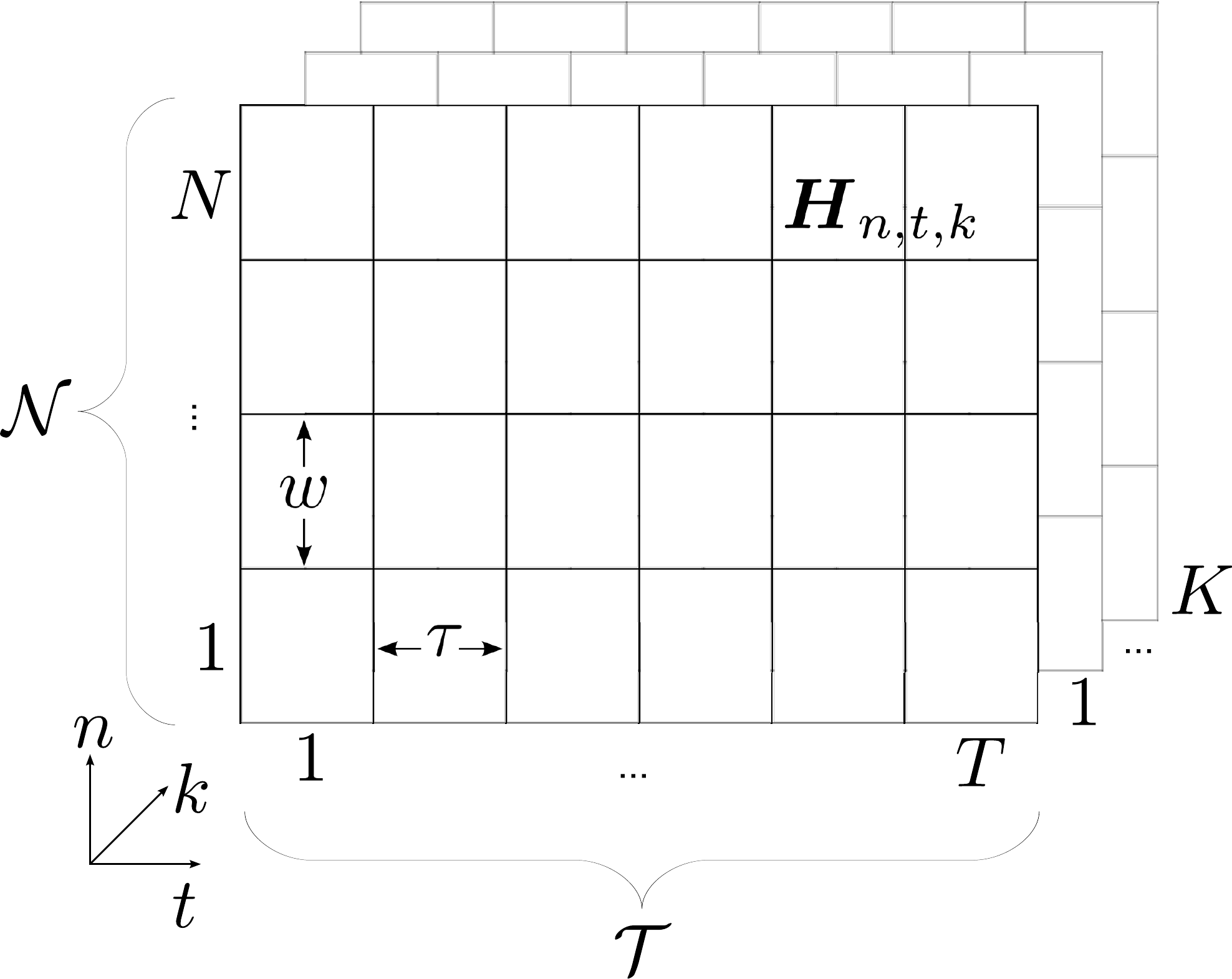}
\caption[\acs{OFDM} frame structure]{\ac{OFDM} frame structure.}
\label{fig:framestructure}
\end{figure}

One transmission frame is considered in the downlink of a point-to-multipoint wireless communication system, comprising one serving \ac{BS} and multiple mobile receivers. The \ac{BS} transmitter is equipped with $D$ antennas and all antennas share the transmit power budget. Mobile receivers have $M_{\mathrm{R}}$ antennas and system resources are shared via \ac{OFDMA} between $K$ users on $N$ subcarriers and $T$ time slots. In total, there are $N T$ resource blocks as introduced in Section~\ref{sect:multicarrier}. A frequency-selective time-variant channel is assumed, with each resource unit characterized by the channel state matrix, $\B{H}_{{n,t,k}} \in \mathbb{C}^{M_{\mathrm{R}} \times D}$, with subcarrier index $n = 1,\dots,N$, time slot index $t = 1,\dots,T$, user index $k = 1,\dots,K$. 
The vector of spatial channel eigenvalues per resource unit~$a$ and user~$k$ is $\mathcal{E}_{a,k}$ and its cardinality is $\min\{D,M_{\mathrm{R}}\}$. The system operates orthogonally such that individual resource units cannot be shared among users. \ac{MIMO} transmission with a variable number of spatial streams is assumed over the set of resources assigned to each user, $\mathcal{U}_{k}$. This frame structure is illustrated in Fig.~\ref{fig:framestructure}. 
\nomenclature{$n$}{Subcarrier index}%
\nomenclature{$t$}{Time slot index}%
\nomenclature{$\B{H}_{{n,t,k}}$}{Matrix of channel states on subcarrier $n$, time slot $t$, link $k$}%
\nomenclature{$\mathcal{E}_{a,k}$}{Vector of spatial channel eigenvalues on resource block $a$ scheduled to user $k$}%
\nomenclature{$\mathcal{U}_{k}$}{Set of resources scheduled to user/link $k$}%

The \ac{PC} and \ac{DTX} trade-off is extended to \ac{MIMO}-\ac{OFDM} serving multiple users over frequency-selective channels. The selection of the number of transmit antennas for \ac{AA} is made once for the entire frame.

For power modeling of single antenna and dual antenna transmission, the affine power models from Chapter~\ref{ch:device}, \eqref{eq:affinemodel1x} and \eqref{eq:affinemodel2x} are employed. It is assumed that the \ac{BS} is equipped with two antennas out of which one can be deactivated. Since the switch-off process is idealized to be instantaneous, a \ac{DTX} power consumption of $P_{\mathrm{S}} = 107$~W is assumed for both single and dual antenna transmission.

\subsection{Problem Formulation}
The global problem statement is formulated as follows. Given the channel state matrices $\B{H}_{n,t,k}$ on each channel and the vector of target rates per user $\B{r}=(R_1,\dots,R_K)$, sought are:
\begin{itemize}
 \item the set of resources allocated to each user $\mathcal{U}_{k}$, 
 \item the power level $P_{a,e}$ per resource block $a = 1,\dots, |\mathcal{U}_{k}|$ and the spatial channel eigenvalue $\epsilon_e$ with $e =1,\dots,|\mathcal{E}_{a,k}|$,
 \item and the number of active transmit antennas $D$, 
\end{itemize}
such that the supply power consumption is minimized while fulfilling the transmission power constraint, $P_{\mathrm{max}}$.
\nomenclature{$\B{r}$}{Vector of target rates, length $K$}%
\nomenclature{$P_{a,e}$}{Transmission power per resource in W}%

Combining~\eqref{eq:Shannon}, \eqref{eq:overlineR} and \eqref{eq:ergCDITCSIRMIMOcapacity}, the sum capacity of user~$k$ over one transmission frame of duration~$\tau_{\mathrm{frame}}$ is given by
\begin{equation}
\label{eq:R_k}
 R_k = \frac{w \tau}{\tau_{\mathrm{frame}}} \displaystyle\sum_{a=1}^{|\mathcal{U}_{k}|} \sum_{e = 1}^{|\mathcal{E}_{a,k}|} \log_2 \left( 1 + \frac{P_{a,e} \mathcal{E}_{a,k}(e)}{P_{\mathrm{N}}} \right),
\end{equation} 
with subcarrier bandwidth $w$ in Hz.
\nomenclature{$w$}{Subcarrier bandwidth in Hz}%

The \ac{BS} \ac{RF} transmission power in time slot~$t$ is
\begin{equation}
\label{eq:Pt}
 P_{\mathrm{Tx}} = \displaystyle \sum_{a=1}^{|\mathcal{A}_{t}|} \sum_{e=1}^{|\mathcal{E}_{a}|} P_{a,e},
\end{equation}
where $\mathcal{A}_{t}$ is the set of $N$ resources in time slot $t$ and $\mathcal{E}_{a}$ is the vector of channel eigenvalues on resource~$a$.

Given \eqref{eq:affinemodel}, \eqref{eq:R_k} and \eqref{eq:Pt}, the optimization problem that minimizes the supply power consumption is of the form
\begin{subequations}
\begin{align}
  &\underset{   \mathcal{U}_{k}\forall\,k, P_{a,e}\forall\,(a,e), D    }    {\text{minimize}} 	&P_{\mathrm{supply}} (\B{r}) 	&=	\frac{1}{T} \left( \displaystyle \sum_{t=1}^{T_{\mathrm{Active}}} \left( P_{0,D} + \Delta_{\mathrm{P}} P_{\mathrm{Tx}} \right) + \sum_{t=1}^{T_{\mathrm{Sleep}}} P_{\mathrm{S}} \right) \\
  &\text{subject to} 										&P_{\mathrm{Tx}} 					&\leq P_{\mathrm{max}}, \\
  & 												&T_{\mathrm{Active}} + T_{\mathrm{Sleep}} &= T,\\
  & 												&R_k 					&\leq \frac{w \tau}{\tau_{\mathrm{frame}}} \displaystyle\sum_{a=1}^{|\mathcal{U}_{k}|} \sum_{e=1}^{|\mathcal{E}_{a,k}|} \log_2 \left( 1 + \frac{P_{a,e} \mathcal{E}_{a,k}(e)}{P_{\mathrm{N}}} \right)
\end{align}%
\label{eq:OPJSAC}%
\end{subequations}%
with the number of active transmission time slots $T_{\mathrm{Active}}$, and \ac{DTX} time slots $T_{\mathrm{Sleep}}$. This is a set selection problem over the sets $\mathcal{U}_{k}$, and $\mathcal{E}_{a,k}$, as well as a minimization problem in~$P_{a,e}$. 
\nomenclature{$T_{\mathrm{Active}}$}{Number of transmission time slots}%
\nomenclature{$T_{\mathrm{Sleep}}$}{Number of \ac{DTX} time slots}%

\subsubsection{Complexity}
Dynamic subcarrier allocation is known to be a complex problem for a single time slot in frequency-selective fading channels that can only be solved by suboptimal or computationally expensive algorithms~\cite{wclm9901,kll0301,jl0301}. In this study, two additional degrees of freedom are added by considering \ac{AA} and \ac{DTX}, increasing the complexity further. Consequently the problem is considered intractable and divided into two steps:
\begin{enumerate}
 \item Real-valued estimates of the resource share per user, \ac{DTX} duration, and number of active \ac{RF} chains $D$, are derived based on simplified system assumptions. 
 \item The power-minimizing resource allocation over the integer set $\mathcal{U}_{k}$ and power allocation are performed.
\end{enumerate}

%

\subsection{Step~1: \ac{AA}, \ac{DTX} and Resource Allocation}
\label{step1}
The first step to solving the global problem is based on a simplification of the system assumptions. This allows defining a convex subproblem with an optimal solution, which is later (sub-optimally) mapped to the solution of the global problem in step~2, described in Section~\ref{step2}. 

Instead of time and frequency-selective fading it is assumed that channel gains on all resources are equal to the center resource block,
\begin{equation}\label{eq:chnCenter}
 \B{H}_k = \B{H}_{n^{\mathrm{c}}, t^{\mathrm{c}},k},
\end{equation}
where the superscript ${\mathrm{c}}$ signifies the center-most subcarrier and time slot. Step~1 thus assumes a block fading channel per user over $W=N w$ and $\tau_{\mathrm{frame}}$. 
\nomenclature{$\B{H}_{n^{\mathrm{c}}, t^{\mathrm{c}},k}$}{Matrix of channel states of the \ac{OFDMA} frame's central resource}%

The center resource block is selected due to the highest correlation with all other resources. Alternative methods to construct a representative channel state matrix are to take the mean or median of $\B{H}_{n,t,k}$ over the \ac{OFDMA} frame. However, application of the mean or median over a set of \ac{MIMO} channels were found to result in a channel with lower capacity. 
See Section~\ref{results} for a comparison plot between different channel selection methods.

The link capacity is calculated using equal-power precoding and assuming uncorrelated antennas. In contrast to water-filling precoding, equal-power precoding provides a direct relationship between total transmit power and target rate. On block fading channels with real-valued resource sharing, OFDMA and \ac{TDMA} are equivalent. Without loss of generality, resource allocation via \ac{TDMA} is selected. These simplifications allow a convex optimization problem to be established that can be efficiently solved.


In a block fading multi-user downlink with equal power precoding, the average rate for user~$k$ is given by
\begin{equation}
  \overline{R_k} = \mu_k W \sum_{e=1}^{|\mathcal{E}_k|} \log_2 \left( 1 + \frac{P_k}{D} \frac{\mathcal{E}_k(e)}{P_{\mathrm{N}}} \right),
\label{eq:mimocapacity}
\end{equation}
with the vector of channel eigenvalues per user $\mathcal{E}_{k}$, and transmission power on this link, $P_k$.

The transmission power is a function of the target rate, depending on the number of transmit and receive antennas. For the following configurations, \eqref{eq:mimocapacity} reduces to:

1x2 \ac{SIMO},
\begin{equation}
\begin{aligned}
& \quad P_k(\overline{R_k}) = \frac{P_{\mathrm{N}}}{\epsilon_1} \left( 2^{\left(\frac{\overline{R_k}}{W \mu_k}\right)} - 1 \right),
\end{aligned}
\label{eq:1x2SIMO}
\end{equation}

2x2 \ac{MIMO},
\begin{equation}
  \overline{R_k} = \mu_k W \log_2 \left( 1 + \frac{P_k}{2 P_{\mathrm{N}}}(\epsilon_1 + \epsilon_2) + \left(\frac{P_k}{2 P_{\mathrm{N}}}\right)^2 \epsilon_1 \epsilon_2 \right),
\end{equation}
\begin{equation}
 P_k(\overline{R_k}) = P_{\mathrm{N}} \frac{ - (\epsilon_1 + \epsilon_2) + \sqrt{(\epsilon_1 + \epsilon_2)^2 + 4 \epsilon_1 \epsilon_2 ( 2^{\frac{\overline{R_k}}{W \mu_k}} - 1 ) }}{\epsilon_1 \epsilon_2},
\label{eq:2x2MIMO}
\end{equation}
where $\mathcal{E}_k = (\epsilon_1, \epsilon_2)$. These equations can be extended in similar fashion to combinations with up to four transmit or receive antennas. Note that a higher number of antennas would require the general algebraic solution of polynomial equations with degree five or higher, which cannot be found in line with the Abel-Ruffini theorem~\cite{r9501a}.
\nomenclature{$\mathcal{E}_k$}{Vector of spatial channel eigenvalues scheduled to user $k$}%

Like on the \ac{BS} side, the number of \ac{RF} chains used for reception at the mobile could be adapted for power saving. However, the power-saving benefit of receive \ac{AA} is much smaller than in transmit \ac{AA}, where a \ac{PA} is present in each \ac{RF} chain. Moreover, multiple receive antennas boost the useful signal power and provide a diversity gain. Therefore, $M_{\mathrm{Rx}}$ is assumed to be fixed and set to $M_{\mathrm{R}} = 2$ in the following.

Given the transmission power \eqref{eq:1x2SIMO}, \eqref{eq:2x2MIMO}, and the power model \eqref{eq:psupply}, the supply power consumption for \ac{TDMA} is derived, similar to~\eqref{eq:OP},
\begin{equation}
 P_{\mathrm{supply}}(\B{r}) = \sum_{k=1}^{K} \mu_k \left( P_{0,D} + \Delta_{\mathrm{P}} P_k(R_k) \right) + \nu P_{\mathrm{S}}.
\end{equation}


Consequently, the optimization problem is defined as:
\begin{subequations}
\begin{align}
& \underset{(\mu_1,\dots,\mu_{K+1})}{\text{minimize}} 	& P_{\mathrm{supply}}(\B{r}) 		&= \sum_{k=1}^{K} \mu_k \left( P_{0,D} + \Delta_{\mathrm{P}} P_k(R_k) \right) + \mu_{K+1} P_{\mathrm{S}} \\
& \text{subject to} 					&	\sum_{k=1}^{K} \mu_k + \nu 		&= 1\\
&							&			\nu		&\geq 0 \\
&							&			 \mu_k 		&\geq 0\ \forall\,k \\
&							&		0 			& \leq  						P_k(R_k) \leq P_{\mathrm{max}} \forall\,k.
\end{align}
\label{eq:OPRAPS}
\end{subequations}%
The first constraint ensures that all resources are accounted for and upper bounds $\mu_k$. The second and third constraint guarantee positive durations. The fourth encompasses the transmit power budget of the \ac{BS} and acts as a lower bound on $\mu_k$. Note that due to the block fading assumption and \ac{TDMA}, the transmission power per user, $P_k$, is equivalent to $P_{\mathrm{Tx}}$ in~\eqref{eq:affinemodel} at that point in time. 

This problem is convex in its cost function and constraints (as proven in Appendix~\ref{appendix:proof2}). It can therefore be solved with available tools like the interior point method~\cite{book:bv0401}. As part of the \ac{RAPS} algorithm, \eqref{eq:OPRAPS} is solved once for each possible number of transmit antennas. The number of transmit antennas is then selected according to which solution results in lowest supply power consumption.

The solution of this first step yields an estimate for the supply power consumption, the number of transmit antennas, $D$, the \ac{DTX} time share, $\nu$, and the resource share per user $\mu_k$. If a solution for step~1 cannot be found, step~2 is not performed and outage occurs. Outage handling is left to a higher system layer mechanism which could, \eg~prioritize users and reiterate with reduced system load. If a solution to step~1 can be found, step~2 is performed as described in the following section.

\subsection{Step~2: Subcarrier and Power Allocation}
\label{step2}
This section describes the second step of the \ac{RAPS} algorithm. In step~2, the results of step~1 are mapped back to the global problem to find the resource allocation for each user, $\mathcal{U}_{k}$, the power level per resource and spatial channel, $P_{a,e}$, and the number of \ac{DTX} slots, $T_{\mathrm{Sleep}}$. 

First, the real-valued resource share, $\mu_k$, is mapped to the \ac{OFDMA} resource count per user, $m_k \in \mathbb{N}$, such that
\nomenclature{$m_k$}{Number of resources scheduled to user $k$}%
\begin{equation}
 m_k = \lceil \mu_k N T  \rceil \quad \forall\,k.
\label{eq:mk}
\end{equation}
Possible rounding effects through the ceiling operation in~\eqref{eq:mk} are compensated for by adjusting the number of \ac{DTX} time slots, with
\begin{equation}
 T_{\mathrm{Sleep}} = \left\lfloor \frac{T N \mu_{K+1} - K}{N} \right\rfloor = \lfloor T \mu_{K+1} - K/N \rfloor.
\label{eq:tsleep}
\end{equation}
The remaining time slots are available for transmission,
\begin{equation}
 T_{\mathrm{Active}} = T - T_{\mathrm{Sleep}}.
\end{equation}
The remaining unassigned resources,
\begin{equation}
 m_{\mathrm{rem}} = N T - \sum_{k=1}^{K} m_k - N T_{\mathrm{Sleep}},
\end{equation}
are assigned to $m_k$ in a round-robin fashion. After this allocation, it holds that $m_k = |\mathcal{U}_{k}|$.
\nomenclature{$m_{\mathrm{rem}}$}{Remaining resources during quantization}%

Next, the number of assigned resource blocks per user, $m_k$, is equally subdivided into the number of resources per user and time slot, $m_{k, t}$, with
\begin{equation}
 m_{k, t} = \left \lfloor \frac{m_k}{\sum_{l=1}^{K} m_l} N \right\rfloor.
\end{equation}
The remaining unassigned resources per time slot,
\begin{equation}
 m_{t,\mathrm{rem}} = N - \sum_{k=1}^{K} m_{k,t},
\end{equation}
are allocated to different $m_{k,t}$ in a round-robin fashion.
\nomenclature{$m_{k,t}$}{Number of resources in time slot $t$ scheduled to user $k$}%
\nomenclature{$m_{t,\mathrm{rem}} $}{Remaining resources in time slot $t$ during quantization}%

Time slots considered for \ac{DTX} are assigned statically, starting from the back of the frame. Note that the dynamic selection of \ac{DTX} slots creates additional opportunities for capacity gains or power savings as they could be assigned to time slots with poor channel states, \eg~time slots experiencing deep fades. This opportunity is revisited in Chapter~\ref{ch:network}.

A corner case exists when $T N \mu_{K+1} < K$ and \eqref{eq:tsleep} becomes negative. This occurs when the \ac{DTX} time share, $\nu$, is very small and thus traffic load is high. Due to the high traffic load, it is possible that the target rates cannot be fulfilled within the transmit power constraint, leading to outage for at least one user. Accordingly, if $T N \mu_{K+1} < K$, $T_{\mathrm{Active}} = T$ and the resource mapping strategy in \eqref{eq:mk} is adapted such that 
\begin{equation}
 m_k = \lfloor \mu_k N T \rfloor,
\label{mkhighload}
\end{equation}
which guarantees that $m_k < NT$. The remaining resources are allocated to users as outlined above.

A subcarrier allocation algorithm from Kivanc \emph{et al.}~\cite{kll0301} is adopted, which has been shown to work effectively with low complexity. The general idea of the algorithm is as follows: First assign each subcarrier to the user with the best channel. Then start trading subcarriers from users with too many subcarriers to users with too few subcarriers based on a nearest-neighbour evaluation of the channel state. The algorithm is outlined in Algorithm~\ref{alg1} and applied to each time slot~$t$ consecutively. The original algorithm from the literature is adapted as indicated in the caption.

\begin{algorithm}
\caption{Adapted \ac{RCG} algorithm performed on each time slot~$t$. As compared to~\cite{kll0301} the cost parameter $h_{n,t,k}$ has been adapted and the absolute value has been added to the search of the nearest neighbour. $\mathcal{A}_{k,t}$ holds the set of subcarriers assigned to user~$k$.}
\label{alg1}
\begin{algorithmic}[1]
\ENSURE $m_{k,t}$ is the target number of subcarriers allocated to each user $k$, $h_{n,t,k} = |\overline{\B{H}_{n,t,k}}|^2$ and $\mathcal{A}_{k,t} \leftarrow\{\}$ for $k=1,\dots,K$.
\FORALL{subcarriers $n$}
  \STATE $k^* \leftarrow \displaystyle\argmax_{1\leq k \leq K} h_{n,t,k}$
  \STATE $\mathcal{A}_{k^*,t} \leftarrow \mathcal{A}_{k^*,t} \cup \{n\}$
\ENDFOR
\FORALL{users $k$ such that $|\mathcal{A}_{k,t}| > m_{k,t}$}
  \WHILE{$|\mathcal{A}_{k,t}| > m_{k,t}$}
    \STATE $l^* \leftarrow \displaystyle\argmin_{\{l:|\mathcal{A}_{l,t}|<m_{l,t}\}} \displaystyle\min_{1\leq n \leq N} |-h_{n,t,k} + h_{n,t,l}|$
    \STATE $n^* \leftarrow \displaystyle\argmin_{1\leq n \leq N} |-h_{n,t,k} + h_{n,t,l^*}|$
    \STATE $\mathcal{A}_{k,t} \leftarrow \mathcal{A}_{k,t} / \{n^*\}, \mathcal{A}_{l^*,t} \leftarrow \mathcal{A}_{l^*,t} \cup \{n^*\}$
  \ENDWHILE
\ENDFOR
\end{algorithmic}
\end{algorithm}
\nomenclature{$\mathcal{A}_{k,t}$}{Set of subcarriers on time slot $t$ assigned to user $k$}%

At this stage, the \ac{MIMO} configuration, the number of \ac{DTX} time slots, and the subcarrier assignment are determined. Transmit powers are assigned in both spatial and time-frequency domains via an algorithm termed margin-adaptive power allocation~\cite{ygc0201}. The notion of this algorithm is as follows: First, channels are sorted by quality and a `water-level' is initialized on the best channel, such that the rate target is fulfilled. Then, in each iteration of the algorithm, the next best channel is added to the set of used channels, thus reducing the water-level in each step. The search is finished once the water-level is lower than the next channel metric to be added. Refer to Appendix~\ref{appendix:iwf} for the derivation. 

Let the number of bits to be transmitted to each user be
\begin{equation}
\label{eq:btarget}
 B_{\mathrm{target}, k} = R_k \tau_{\mathrm{frame}}.
\end{equation}
To fulfill $B_{\mathrm{target}, k}$, the following constraint must be met
\begin{equation}
 B_{\mathrm{target}, k} - \displaystyle\sum_{a=1}^{|\mathcal{U}_{k}|} \sum_{e=1}^{|\mathcal{E}_{a,k}|} w \tau \log_2 \left( 1 + \frac{P_{a,e} \mathcal{E}_{a,k}(e)}{P_{\mathrm{N}}} \right) = 0,
\end{equation}
which assesses the sum capacity according to~\eqref{eq:R_k}.
\nomenclature{$B_{\mathrm{target}, k}$}{Target rate for user $k$ in bps}%

The water-level~$\upsilon$ can be found via an iterative search over the set of channels that contribute a positive power,~$\Omega_k$, with
\begin{equation}
   \log_2(\upsilon) = \frac{1}{|\Omega_{k} |}\!\left(\frac{B_{\mathrm{target}, k}}{w \tau} - \sum_{e=1}^{|\Omega_k|} \log_2 \!\left( \frac{\tau\, \mathcal{E}_{a,k}(e)w}{P_{\mathrm{N}} \log(2)} \right)\!\right).
\end{equation}
The water-level is largest on the first iteration and decreases on each iteration until it can no longer be decreased. 
\nomenclature{$\Omega_k$}{Set of channels which contribute positive power in margin adaptive power allocation}%
\nomenclature{$\upsilon$}{Water-level in margin-adaptive power allocation}%

Using the Lagrangian method detailed in Appendix~\ref{appendix:iwf}, one arrives at a power-level per spatial channel of
\begin{equation}
 P_{a,e} = \frac{\upsilon w \tau }{\log(2)} - \frac{P_{\mathrm{N}}}{\mathcal{E}_{a,k}(e)}.
\end{equation}

\begin{figure}
\centering
\includegraphics[width=0.6\textwidth]{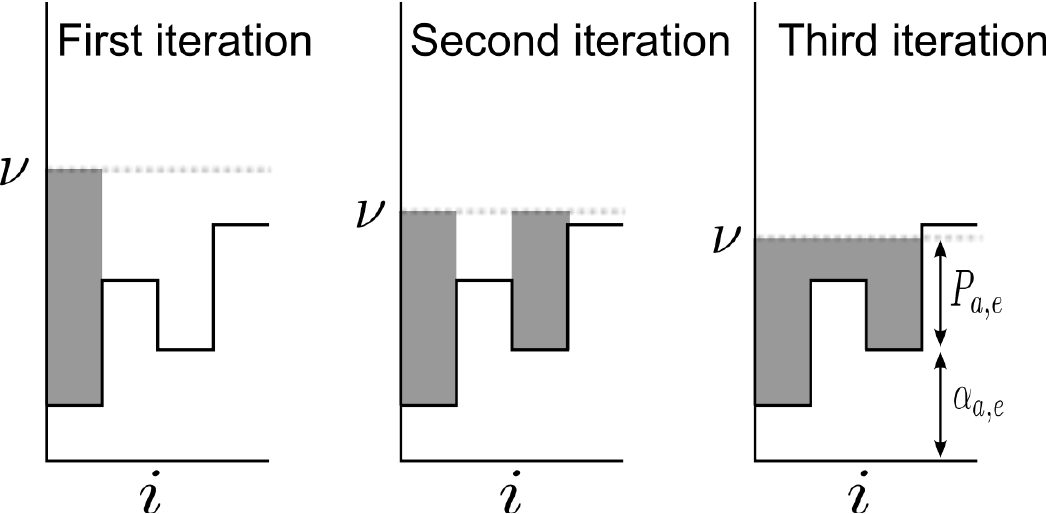}
\caption[Illustration of margin-adaptive power allocation over three steps]{Illustration of margin-adaptive power allocation with three steps. The height of each patch~$i$ is given by $\alpha_{a,e} = P_{\mathrm{N}}/\mathcal{E}_{a,k}(e)$. The water-level is denoted by~$\upsilon$. The height of the water above each patch is $P_{a,e}$. The first step sets the water-level such that the rate target is fulfilled on the best patch. The second and third step add a patch, thus reducing the water-level. After the third step, the water-level is below the fourth patch level, thus terminating the algorithm.}
\label{fig:IWF}
\end{figure}

\begin{figure}
\centering
\includegraphics[width=0.6\textwidth]{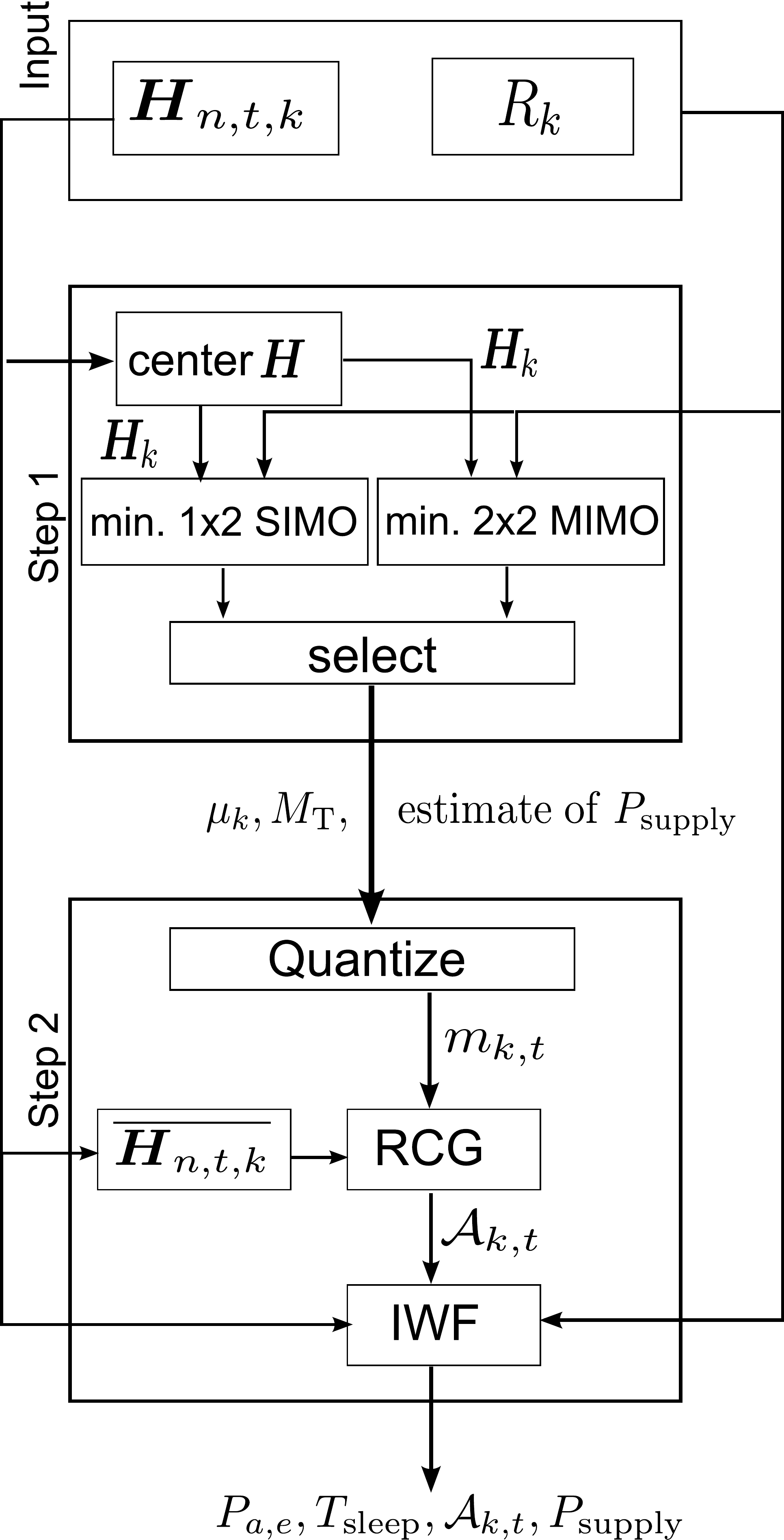}
\caption{Outline of the \ac{RAPS} algorithm.}
\label{fig:algflowchart}
\end{figure}

The margin-adaptive algorithm is illustrated for four channels in Fig.~\ref{fig:IWF}. The outcome of margin-adaptive power allocation as part of \ac{RAPS} are the transmission power levels $P_{a,e}$ for each resource block. The supply power consumption after application of \ac{RAPS} can be found by summation of transmission powers in each time slot and application of the affine power model. The entire \ac{RAPS} algorithm is outlined in Fig.~\ref{fig:algflowchart}.

\subsection{Results}
\label{results}

In order to assess the performance of the \ac{RAPS} algorithm Monte Carlo simulations are conducted. The simulations are configured as follows: mobiles are uniformly distributed around the \ac{BS} on a circle with radius 250\,m and a minimum distance of 40\,m to the \ac{BS} to avoid peak \acp{SNR}. Fading is computed according to the NLOS model described in \ac{3GPP}  TR 25.814~\cite{std:3gpp-faeutrapla} with 8\,dB shadowing standard deviation, and the frequency-selective channel model B5 described in~\cite{std:ist-chanmod} with 3\,m/s mobile velocity. All transmit and receive antennas are assumed to be mutually uncorrelated. 
Further system parameters are listed in Table~\ref{tab:notationsummary}. 
\begin{table}
\centering
\begin{tabular}{l|l|l}
      Variable 	&  & Value\\
\hline
      $K$ 				& Number of users 							& 10\\
      $N$ 				& Number of subcarriers 						& 50 \\
      $T$ 				& Number of time slots in an \ac{OFDMA} frame 				& 10\\
      $D $ 		& Number of transmit antennas 						& [1,2]\\
      $M_{\mathrm{R}}$ 			& Number of receive antennas 						& 2\\
      $P_{\mathrm{max}} $ 		& Maximum transmission power 						& 46~dBm\\
      $\tau_{\mathrm{frame}}$/$\tau$	& Duration of frame/time slot				& 10\,ms/1\,ms\\
      $W$/$w$ 				& System/subcarrier bandwidth 						& 10\,MHz/200\,kHz\\
      $P_{\mathrm{N}}$			& Thermal noise power per subcarrier					& $4w\times 10^{-21}$\,W\\
\end{tabular}%
\caption{System parameters}%
\label{tab:notationsummary}%
\end{table}%

\vfill 
\subsubsection{Benchmarks}
The following transmission strategies are evaluated for comparison purposes:
\begin{itemize}
 \item The theoretical limit of the \ac{BS} power consumption is obtained by constant transmission at maximum power, $P_1$. 
 \item \ac{BA}, which finds the minimum number of subcarriers that achieves the rate target. No sleep modes are utilized and all scheduled subcarriers transmit with a transmission power spectral density of $P_{\mathrm{max}}/N$. This benchmark represents the power consumption of \ac{SOTA} \acp{BS} which are neither capable of \ac{DTX} nor \ac{AA}. 
 \item \ac{DTX}, where the \ac{BS} transmits with full power, $P_1$, and switches to \ac{DTX} once the rate requirements are fulfilled.  This benchmark assesses the attainable savings when only \ac{DTX} is applied.
\end{itemize}

\subsubsection{Performance Analysis}
\begin{figure}
\centering
 \includegraphics[width=1\textwidth]{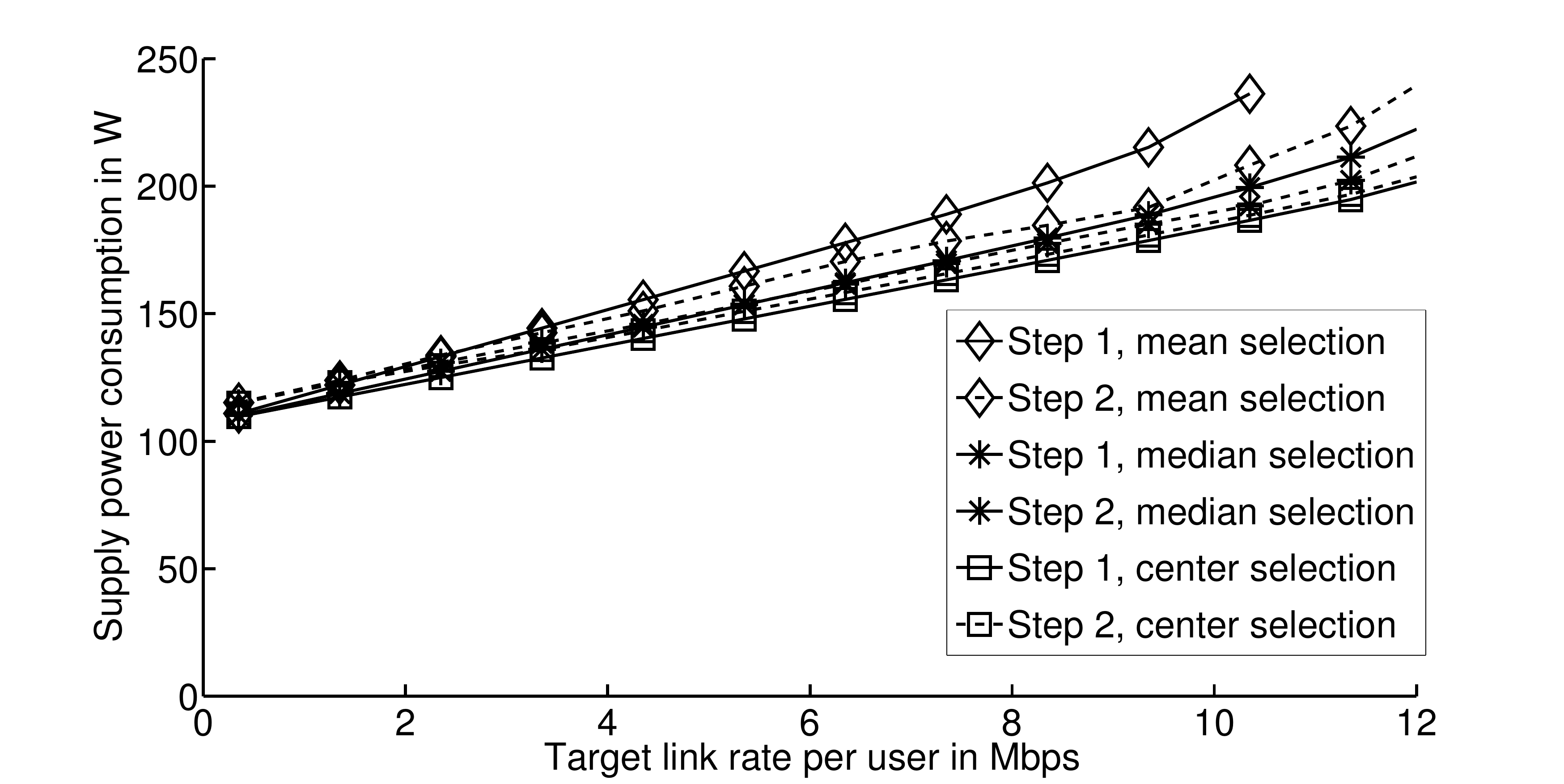}
\caption[Performance comparison in step~1]{Performance comparison of different channel state selection alternatives.}
\label{fig:resource_selection_comparison_S1}
\end{figure}

The channel value selection in~\eqref{eq:chnCenter} serves as the channel gain for the block fading assumption of step~1. How the channel value selection of three possible alternatives (mean, median center) affects the supply power consumption is examined in Fig.~\ref{fig:resource_selection_comparison_S1}. It can be seen that the selection of the mean channel gain results in the highest supply power consumption estimate after step~1. While the estimate can be improved in step~2, it is still inferior to the other alternatives. Use of the mean channel states causes step~1 to underestimate the channel quality. Consequently, too few time slots are selected for \ac{DTX} in step~2. Choosing the median provides a better estimate than the mean and after step~2 the solution matches `step~2, center selection'. However, in center channel state selection the step~1 estimate and the step~2 solution have both the best match and the lowest supply power consumption. Therefore, as mentioned in Section~\ref{step1}, center channel state selection is chosen in the \ac{RAPS} algorithm and all further analyses.

\begin{figure}
\centering
 \includegraphics[width=0.8\textwidth]{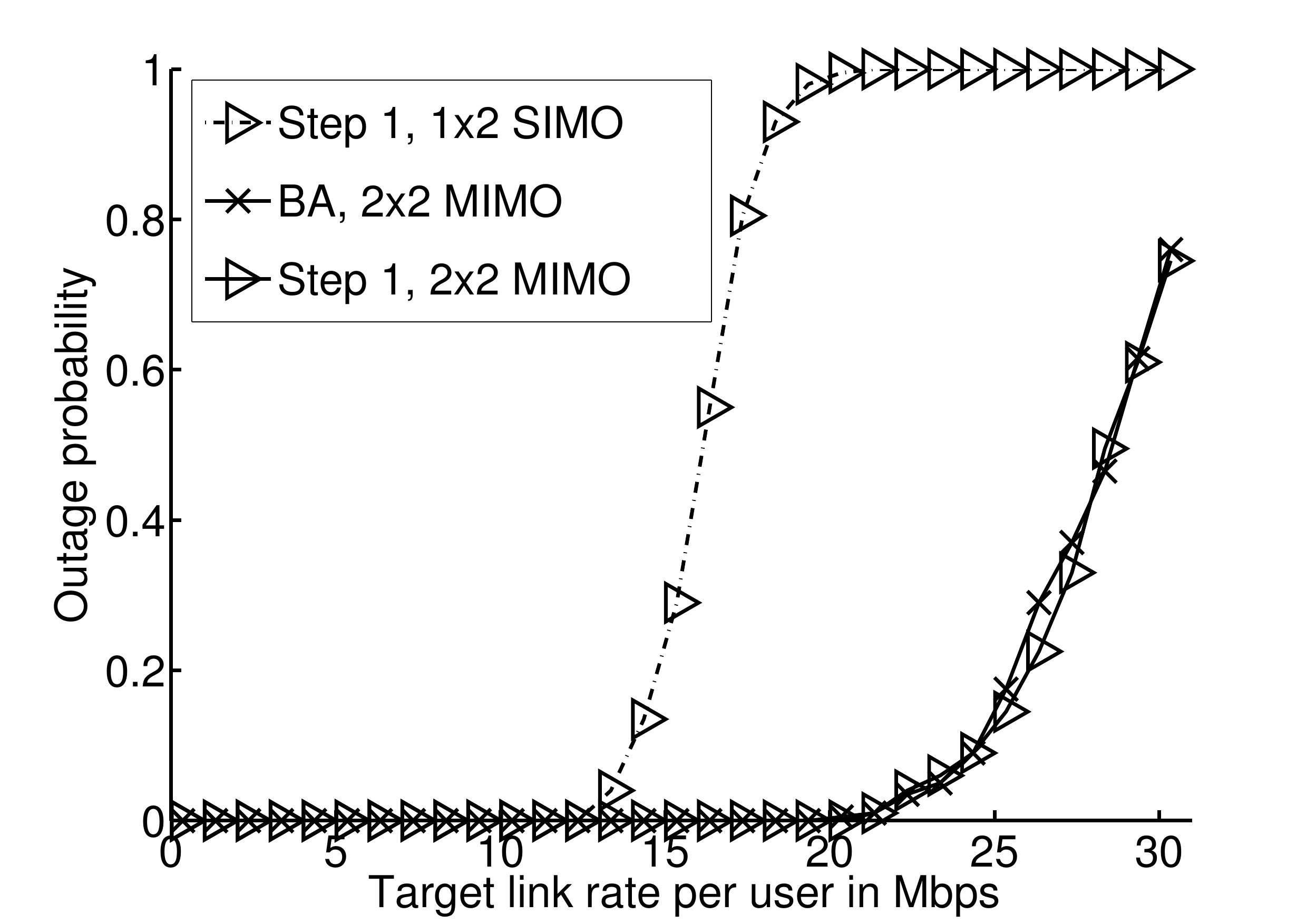}
\caption[Outage probability in step~1 and the \acs{BA} benchmark]{Outage probability in step~1 and the \ac{BA} benchmark.}
\label{fig:outage}
\end{figure}

Fig.~\ref{fig:outage} compares the outage probabilities of step~1 with that of the \ac{BA} benchmark. Outage refers to the lack of a solution to step~1; when the user target data rates are too high for the given channel conditions, the convex subproblem has no solution, which causes the algorithm to fail. A reduction of user target data rates based on, \eg~priorities, latency or fairness, is specifically not covered by \ac{RAPS}. One suggested alternative is to introduce admission control, in the way that some users are denied access, so that the remaining users achieve their target rates. Note that step~2 has no effect on the outage; if a solution exists after step~1, then step~2 can be completed. If a solution does not exist after step~1, then step~2 is not performed. Fig.~\ref{fig:outage} illustrates that \ac{RAPS} selects two transmit antennas for target link rates above 14\,Mbps with high probability, as target link rates cannot be achieved with a single transmit antenna. The \ac{BA} benchmark and the step~1 \ac{MIMO} solution have similar outage behavior.

\begin{figure}
\centering
 \includegraphics[width=0.8\textwidth]{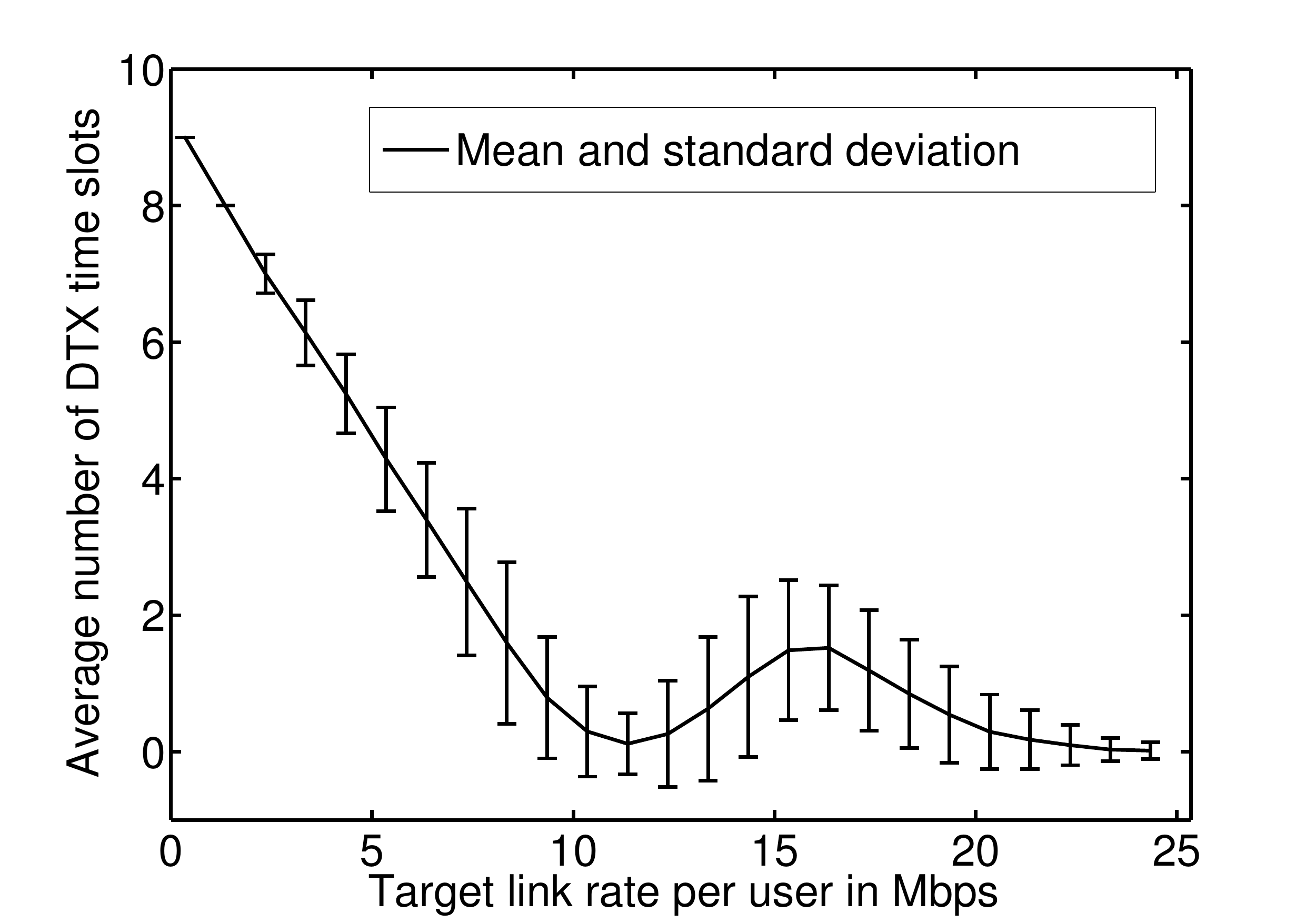}
\caption[Average number of \acs{DTX} time slots over increasing target link rates]{Average number of \ac{DTX} time slots over increasing target link rates. Error bars portray the standard deviation. Total number of time slots $T = 10$.}
\label{fig:avgsleepslots}
\end{figure}

Fig.~\ref{fig:avgsleepslots} depicts the average number of \ac{DTX} time slots over increasing target link rates. The effect of \ac{AA} (\ie~switching between \ac{SIMO} and \ac{MIMO} transmission) can clearly be seen. For low target link rates, a large proportion of all time slots is selected for \ac{DTX}. For higher target link rates, the number of \ac{DTX} time slots must be reduced when operating in \ac{SIMO} mode. For target link rates above 12\,Mbps, which approach the \ac{SIMO} capacity (as described in the previous paragraph), the system switches to \ac{MIMO} transmission. The added \ac{MIMO} capacity allows the \ac{RAPS} scheduler to allocate more time slots to \ac{DTX}. In the medium load region (around 15~Mbps) the standard deviation is highest, indicating that here the \ac{RAPS} algorithm strongly varies the number of \ac{DTX} time slots depending on channel conditions and whether \ac{SIMO} or \ac{MIMO} transmission is selected. These variations contribute strongly to the additional savings provided by \ac{RAPS} over the benchmarks. Another observation from Fig.~\ref{fig:avgsleepslots} is that it is unlikely that more than two out of ten \ac{DTX} time slots are scheduled at target link rates above 8\,Mbps. This means that for a large range of target rates, no more than two \ac{DTX} time slots are required to minimize power consumption. This is an important finding for applications of \ac{RAPS} in established systems like \ac{LTE}, where the number of \ac{DTX} time slots may be limited due to constraints imposed by the standard.

\begin{figure}
\centering
 \includegraphics[width=1\textwidth]{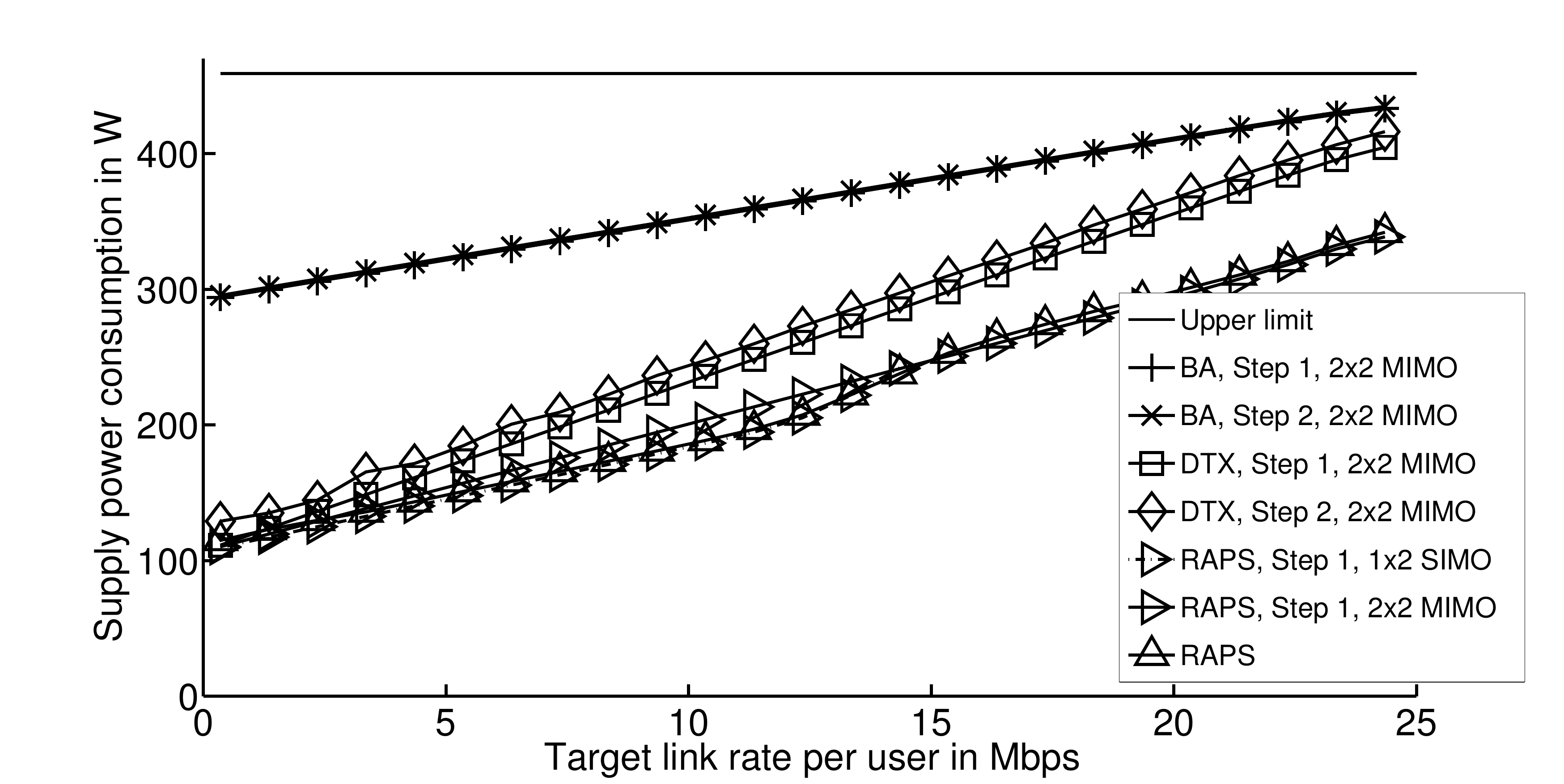}
\caption[Supply power consumption for different \acs{RRM} schemes overall]{Supply power consumption for different \ac{RRM} schemes on block fading and frequency-selective fading channels (comparison of step~1 and step~2) for ten users. Overlaps indicate the match between the step~1 estimate and the step~2 solution.}
\label{fig:fullsim_general}
\end{figure}

A comparison of the supply power consumption estimates of step~1 and step~2 in Fig.~\ref{fig:fullsim_general} verifies that the estimate taken in step~1 as input for step~2 are sufficiently accurate. Although step~1 is greatly simplified with the assumption of block fading and its output parameters cannot be applied readily to an \ac{OFDMA} system, it precisely estimates power consumption which is the optimization cost function. The slight difference between the step~1 estimate and power consumption after step~2 is caused by quantization loss and resource scheduling. Note that while step~1 supplies a good estimate of the power consumption, it is not a solution to the original \ac{OFDMA} scheduling problem, since it does not consider the frequency-selectivity of the channel and does not yield the resource and power allocation.

\begin{figure}
\centering
 \includegraphics[width=1\textwidth]{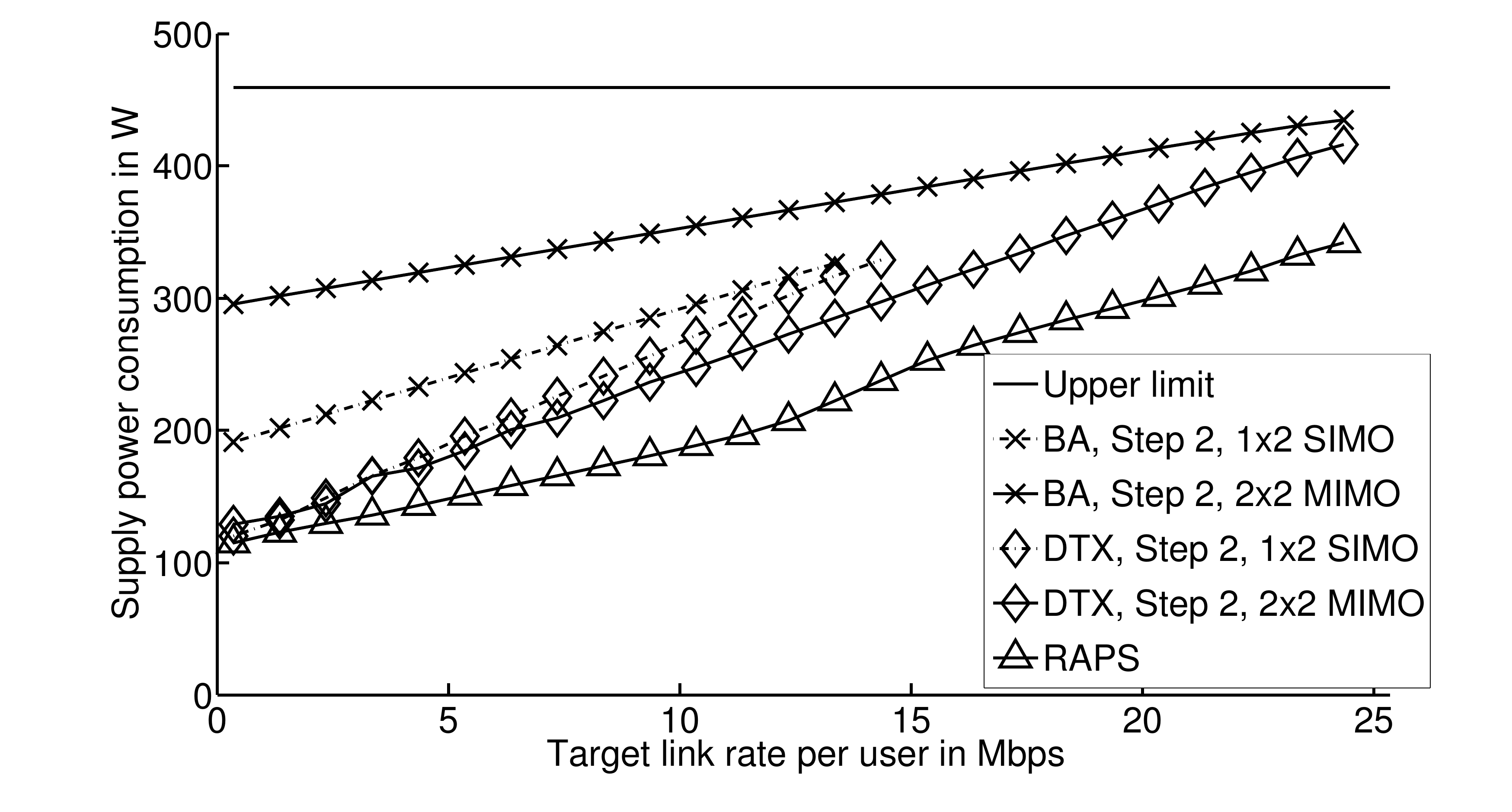}
\caption[Supply power consumption for different \acs{RRM} schemes in step~2]{Supply power consumption on frequency-selective fading channels for different \ac{RRM} schemes and \ac{RAPS} for ten users. For a bandwidth-adapting \ac{BS}, power consumption is always reduced by \ac{AA}, while \ac{AA} is never beneficial for a sleep mode capable \ac{BS}. For \ac{RAPS} energy consumption is reduced by \ac{AA} at low rates. In general, \ac{RAPS} achieves substantial power savings at all \ac{BS} loads.}
\label{fig:fullsim_step2}
\end{figure}

The performance of \ac{RAPS} in comparison to each benchmark is separately analyzed in Fig.~\ref{fig:fullsim_step2}. Here, a first observation is that even the supply power consumption of \ac{BA} is significantly lower than the theoretical maximum for most target link rates. \ac{BA} with a single transmit antenna always consumes less power than with two antennas, as long as the rate targets with one antenna can be met. \ac{AA} is thus a valid power-saving mechanism for \ac{BA}. 

Second, the \ac{DTX} power consumption is significantly lower than for \ac{BA}, especially at low target rates, because lower target rates allow the \ac{BS} to enter \ac{DTX} for longer periods of time. As the \ac{BS} load increases, the opportunities of the \ac{DTX} benchmark to enter sleep mode are reduced, so that \ac{DTX} power consumption approaches that of \ac{BA}. Unlike in \ac{BA}, it is never beneficial for a \ac{BS} capable of \ac{DTX} to switch operation to a single antenna, because transmitting for a longer time with a single antenna always consumes more power than a short two-antenna transmission, which allows for a longer \ac{DTX} duration. (Note that this finding may not apply to other power models.)

Third, \ac{RAPS} reduces power consumption further than \ac{DTX} by employing \ac{AA} at low rates and \ac{PC} at high rates. Through \ac{PC}, the slope of the supply power is kept low between 15 and  20\,Mbps. At higher rates an upward trend becomes apparent, since link rates only grow logarithmically with the transmission power. In theory, when the \ac{BS} is at full load, the supply power of all energy saving mechanism will approach maximum power consumption. However, when the \ac{BS} load is very high, not all users may achieve their target rate and outage occurs (see Fig.~\ref{fig:outage}). In other words, operating the \ac{BS} with load margins controls outage and allows for large transmission power savings. Power savings of \ac{RAPS} compared to the \ac{SOTA} \ac{BA} range from 102.7\,W (24.5\%) to 136.9\,W (41.4\%) depending on the target link rate per user.

\begin{figure}
\centering
 \includegraphics[width=1\textwidth]{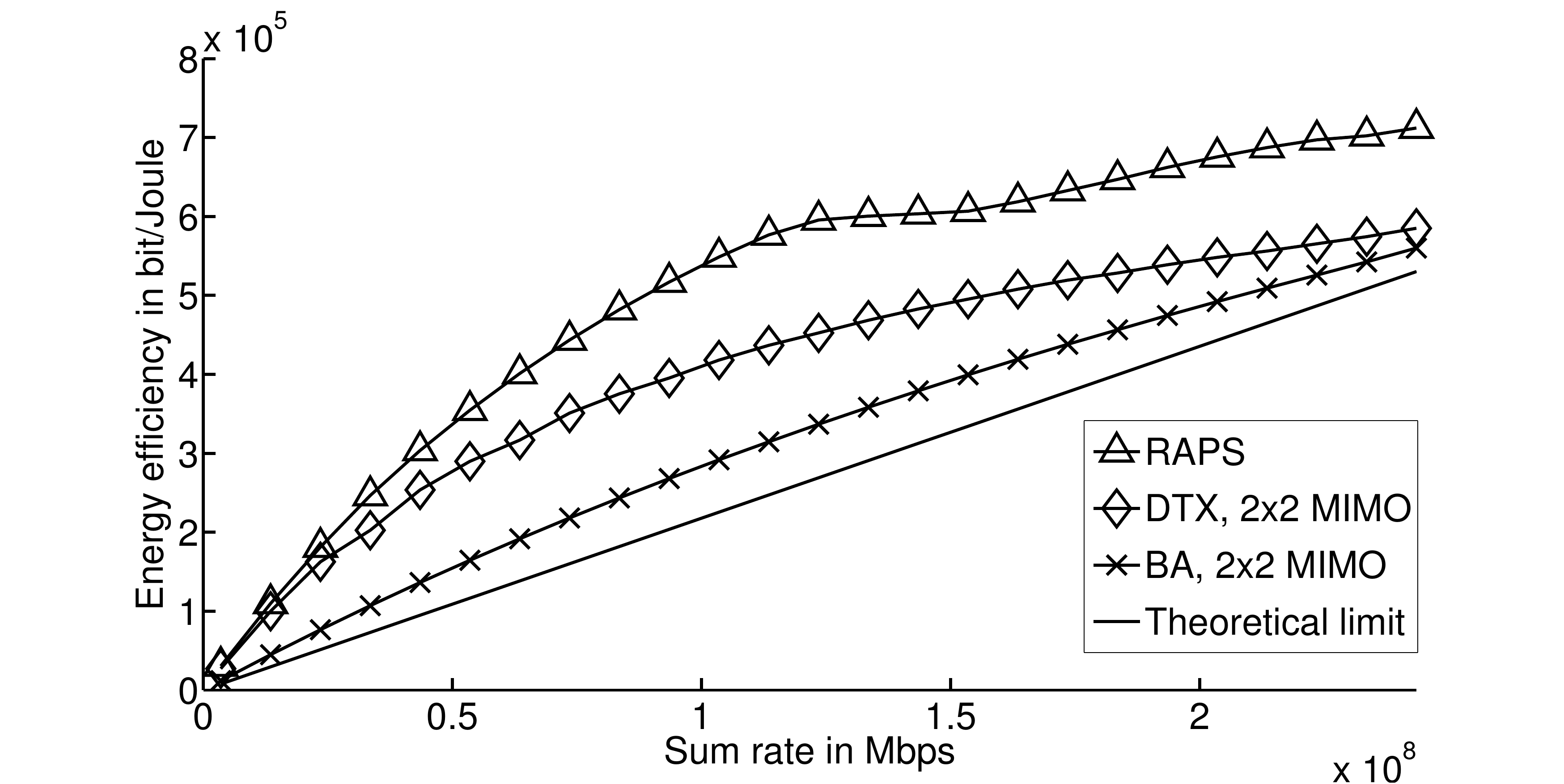}
\caption[Energy efficiency as a function of sum rate]{Energy efficiency as a function of sum rate. Energy efficiency can be increased by the \ac{RAPS} algorithm. The \ac{BS} remains most efficient at peak rate.}
\label{fig:step2_EE}
\end{figure}

In addition to absolute consumption, the energy efficiency of \ac{RAPS} is inspected in Fig.~\ref{fig:step2_EE}. Energy efficiency is defined in bit/Joule as
\begin{equation}
 E = P_{\mathrm{supply}}^{-1} \sum_{k=1}^{K} R_k.
\end{equation}
Observe that all data series are monotonically increasing. Therefore, \ac{RAPS}  does not change the paradigm that a \ac{BS} is most efficiently operated at peak rates. However, \ac{RAPS} will always offer the most efficient operation at any requested rate.
\nomenclature{$E$}{Energy efficiency in bit/Joule}%

\section{Summary}
\label{ch4:summary}
Starting with the Shannon limit, this chapter derived supply power optimal resource allocation in an \ac{LTE} cell.

First, a literature review described the \ac{SOTA} of energy efficient \ac{RRM} for cellular networks.

Next, the general interrelationships between bandwidth, power and time in transmissions were studied. As the \ac{PA} consumption can be reduced by \ac{PC}, a multi-user downlink \ac{PC} allocation mechanism was proposed. Consideration of single links was found to be insufficient for power-saving, as a \ac{BS} schedules multiple users on the same resources. It was shown that the power model, due to its affine mapping, does not affect the \ac{PC} solution. Thus, it is also the solution to supply power minimization, when sleep modes are not considered. 

As the power consumption reduction through \ac{PC} is lower bounded by the idle power consumption of a transmitter, short sleep modes, \ie~\ac{DTX}, have also been taken into consideration. It was shown that the joint application of \ac{PC} and \ac{DTX} brings additional savings over the application of each individual scheme. Using convex optimization, a resource allocation mechanism called \ac{PRAIS} was proposed. Using \ac{PRAIS}, fundamental limits of power-saving resource allocation were identified on several power models. Power savings estimated on the basis of the affine model presented in Chapter~\ref{ch:device} are in the range of 50\%. 

When adding the integer value constraint, which is present in practical systems where resources are only available in chunks and each has an individual channel state, the resource allocation problem becomes a mixed integer program. As this problem is highly complex, it was split into two steps. A convex subproblem was derived and solved in similar fashion as in the \ac{PRAIS} algorithm. The solution to the subproblem yields an approximation of quantization, providing a configuration for the mixed integer problem. The overall algorithm is labelled \ac{RAPS}. Application of the \ac{RAPS} algorithm on this stricter constraint set achieved savings in the order of 25\% to 40\%. 

The \ac{RAPS} algorithm could be readily applied in current \ac{LTE} systems. It adds complexity to scheduling, but offers significant power savings over all loads. Limitations lie in the fact that switching is not instantaneous in practice and \acp{PA} have limits on the \ac{PC} range. If switching times are slower than the fraction of a time slot, power savings will not be as high as predicted. Due to linearity requirements, some \acp{PA} have a lower limit on transmission powers. Thus, \ac{PC} may not be as effective as predicted in this chapter. Overall, these limitations only affect the achievable savings, not the functionality of the power-saving \ac{RRM} algorithms presented in the chapter.

\acresetall 

\renewcommand{\dir}{chapter5/chapter/}
\chapter{Power Saving on the Network Level (Multi-cell)}
\label{ch:network}

\section{Overview}
\begin{figure}
\centering
\includegraphics[width=0.8\textwidth]{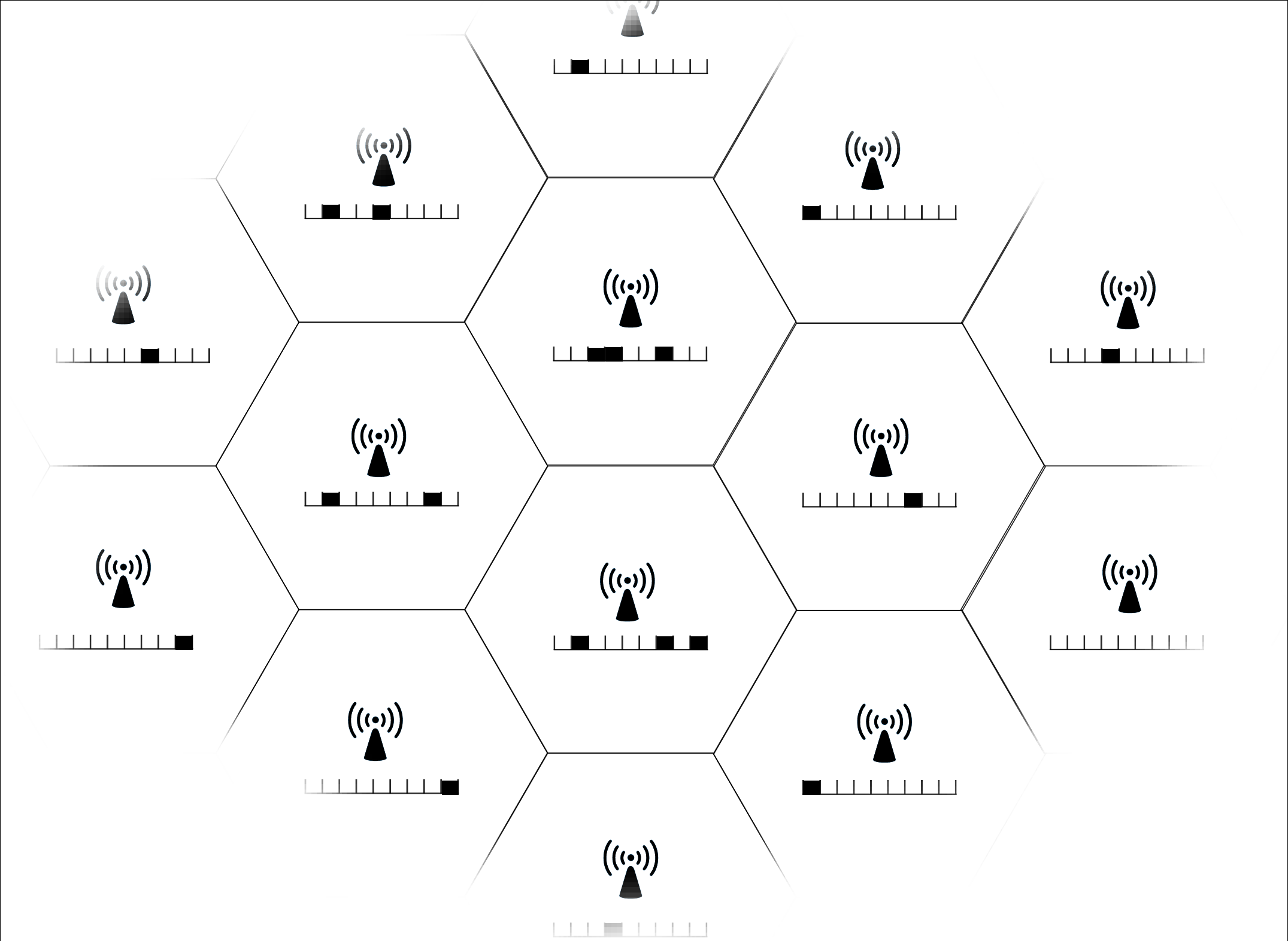}
\caption[Illustration of \ac{DTX} alignment in a network]{Each \ac{BS} in a network select time slots for transmission (black) or \ac{DTX} (blank).}
\label{fig:channelalignment}
\end{figure}

Chapter~\ref{ch:cell} proposed how the power consumption of a \ac{BS} could be minimized regardless of interference. This chapter expands the power-saving perspective to the network level, thus taking inter-cell interference into account. In particular, it studies how the \ac{DTX} time slots scheduled by each \ac{BS} in a network can be aligned constructively to avoid mutual interference. See Fig.~\ref{fig:channelalignment} for an illustration of multiple cells, each choosing to transmit or schedule \ac{DTX} in the available time slots.

In this chapter, the findings on channel allocation from literature are combined and adapted for power-saving. Four \ac{DTX} alignment strategies for the cellular downlink are derived, one of which is an original contribution in which each \ac{BS} independently and without coordination prioritizes time slots for transmission while maintaining system stability.

The remainder of the chapter is structured as follows. Section~\ref{ch5:problem} formulates the system model and the problem at hand. In Section~\ref{ch5:solutions}, three \ac{SOTA} channel alignment solutions are described and the novel alignment method is introduced. Findings obtained from simulation are presented in Section~\ref{ch5:results}. The chapter is summarized in Section~\ref{ch5:summary}.

The contents of this chapter have been previously accepted for publication in~\cite{hh1301} and submitted for publication in~\cite{hdh1301}.

\section{Channel Allocation in Literature}
\label{ch5:literature}
Generally, in channel allocation, the challenge is to assign frequency channels to different cells or links of a network such that mutual interference is avoided. Several centralized approaches to interference avoidance have been proposed~\cite{ll0601,uab1101,ry1001,lcyzchk1201}. However, to be applicable for future small-cell cellular networks, such channel allocation needs to be distributed, uncoordinated and dynamic. Distributed operation prevents the delay and backhaul requirements imposed by a central controller. If the allocation is also uncoordinated, it does not require message exchange through a backhaul of limited capacity. Furthermore, as networks, channels and traffic change rapidly, the allocation must be highly dynamic and updated often. Problems of this type are known to be NP-hard~\cite{knun01}. Therefore, different suboptimal approaches have been proposed. The simplest approach to distributing channels over a network is a fixed assignment in which it is predefined which links will use which channels, \eg~see~\cite{book:p0001}. However, this is inflexible to changing or asymmetric traffic loads. For flexibility, dynamic channel allocation methods are required. A simple dynamic channel allocation method, Sequential Channel Search~\cite{sg9301}, assigns channels with sufficient quality in a predefined order. This technique allows adjusting the number of channels flexibly, but is suboptimal due to its strong channel overlap between neighbors. Taking the channel quality into consideration by measurement is proposed in the Minimum \ac{SINR} method~\cite{sg9301}. This technique offers increased spectral efficiency, but can lead to instabilities when allocations happen synchronously. Ellenbeck~\ea~\cite{ehb0801} introduce methods of game theory and respond to the instabilities by adding p-persistence to avoid simultaneous bad player decisions. However, the proposed method only applies to single user systems and does not address target rates. Dynamic Channel Segregation~\cite{aa9301} introduces the notion of memory of channel availability, allowing a transmitter to track which channels tend to be favorable for transmission. However, the algorithm only applies to sequential channel decisions based on an idle or busy state in circuit switched networks and cannot be applied to the concurrent alignment of channels as is required in \ac{OFDMA} networks.

\section{System Model and Problem Formulation}
\label{ch5:problem}
The network is considered as follows. \acp{BS} in a reuse-one \ac{OFDMA} cellular network schedule a target rate, $B_k$, per \ac{OFDMA} frame to be transmitted to each mobile $k=\{1,\dots,K\}$. All time slots are available for scheduling in all cells. Mobiles report perceived \ac{SINR} from the previous \ac{OFDMA} frame, $s_{n,t,k}$, on subcarrier $n=\{1,\dots,N\}$, time slot $t=\{1,\dots,T\}$ to their associated \ac{BS}. \acp{BS} have a \ac{DTX} mode available for each time slot during which transmission and reception are disabled and power consumption is significantly reduced to $P_{\mathrm{S}}$ compared to power consumption during transmission, which is a function of the power allocated to each resource block, $\rho_{\mathrm{Tx}}$. \ac{DTX} is available fast enough to enable it within individual time slots of an \ac{OFDMA} transmission such that transmission time slots are not required to be consecutive. A \ac{BS} can schedule no transmission in one or more time slots and go to \ac{DTX} mode instead. Scheduling a \ac{DTX} time slot in one \ac{BS} reduces the interference on that time slot to all other \acp{BS}. The \ac{OFDMA} frames of all \acp{BS} are assumed to be aligned such that the interference over one time slot and subcarrier is flat and that all \acp{BS} can perform alignment operations in synchrony. The \ac{OFDMA} frame consists of $N T$ resource blocks. The channel is subject to block fading as described in Section~\ref{sect:multicarrier}. Interference is treated as noise.

To describe the resource allocation problem formally, the function 
\begin{equation}
\begin{aligned}
\Pi: \{1,\dots,T\}\times\{1,\dots,N\} &\rightarrow \{0,1,\dots,K\}\\ 
(n,t)&\mapsto k
\end{aligned}
\end{equation}
is defined, which maps to each resource block a user $k$ or 0, where 0 indicates that the resource block is not scheduled.
\nomenclature{$\Pi$}{Function mapping resource blocks to users}

 With $r_{n,t,k}$ the capacity of resource $(n,t)$ if it is scheduled to $k$, the optimization problem for the total \ac{BS} power consumption, $P_{\mathrm{total}}$, is as follows.
\begin{subequations}
\begin{align}
 \underset{\Pi}{\text{minimize}} \quad P_{\mathrm{total}} &= P_{\mathrm{S}} \frac{T_{\mathrm{S}}}{T} + \rho_{\mathrm{Tx}} N_{\mathrm{Tx}} \left(\frac{T-T_{\mathrm{S}}}{T}\right) \label{eq:ptotal}&\\
\text{subject to}&&\notag\\ 
\quad N_{\mathrm{Tx}} &= &\label{eq:nt}\\
 |\{ n \in \{1,\dots,N\} &| \left( \Pi(n,t) \neq 0 \quad\forall t \in \{1,\dots,T\}\right)\}|&\notag\\
\quad T_{\mathrm{S}} &= &\label{eq:ts}\\
 |\{ t \in \{1,\dots,T\} &| \left( \Pi(n,t) = 0 \quad\forall n \in \{1,\dots,N\}\right)\}|&\notag\\
  B_k &\leq \displaystyle \sum_{\{(n,t)|\Pi(n,t)=k\}} r_{n,t,k} \quad \forall k.&\label{eq:bk}
\end{align}
\end{subequations}
With $|\cdot|$ the cardinality operator, $T_{\mathrm{S}}$ is the number of time slots in which for all $n$, no user is allocated and the \ac{BS} can enable \ac{DTX} and $N_{\mathrm{Tx}}$ is the number of resource blocks that are scheduled for transmission. The constraint~\eqref{eq:bk} provides the rate guarantee for each user.
\nomenclature{$r_{n,t,k}$}{Capacity of subcarrier $n$, time slot $t$ when scheduled to user $k$}
\nomenclature{$T_{\mathrm{S}}$}{Number of time slots in which for all subcarriers no user is allocated}
\nomenclature{$N_{\mathrm{Tx}}$}{Number of resource blocks scheduled for transmission}

The difficulty lies in finding the mapping $\Pi$. Under a brute force approach, there exist $(K+1)^{NT}$ possible combinations. Due to the large number resource blocks present in typical \ac{OFDMA} systems like \ac{LTE}, the computation of the solution is infeasible. Consequently, the next section resorts to heuristic methods.

Some of the methods compared in the following make use of a time slot ranking. In other works, \eg~\cite{aa9301}, ranking is proposed to be based on \ac{SINR}. However, in a multi-user \ac{OFDMA} system each subcarrier and mobile terminal have a different \ac{SINR}, thus generating a problem of comparability between time slots. Consequently, it is proposed to compare time slots by their hypothetical sum capacity, $B_t$, over all mobiles and subcarriers with
\begin{equation}
\label{eq:bt}
\begin{aligned}
R_t &= \sum_{k=1}^K \sum_{n=1}^N \log_2 \left( 1 + s_{n,t,k} \right).
\end{aligned}
\end{equation}
\nomenclature{$R_t$}{Virtual sum capacity of time slot $t$ in bits}

\section{\ac{DTX} Alignment Strategies}
\label{ch5:solutions}
In this section, four strategies to tackle the problem at hand are identified which differ in performance and complexity.

\subsection{Sequential Alignment}
\begin{figure}
 \centering
\includegraphics[width=0.4\textwidth]{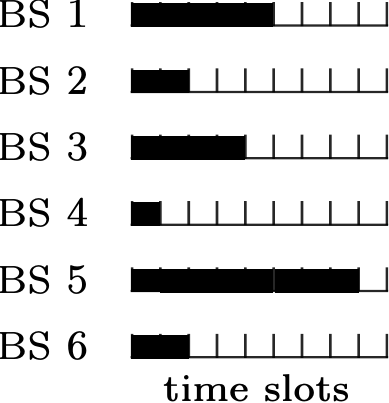}
\caption[Illustration of sequential alignment]{Illustration of sequential alignment of time slots for transmission (black) or \ac{DTX} (blank) in six \acp{BS}.}
\label{fig:seqalign}
\end{figure}

In \emph{sequential alignment}, the strategy is to always allocate as many time slots for transmission as required in sequential order as illustrated in Fig.~\ref{fig:seqalign} and set the remainder to \ac{DTX}. This leads to strongest overlap and consequently highest interference on the first time slot and lowest overlap and possibly no transmissions at all on the last time slot. This strategy does not make use of the available channel quality information per subcarrier and provides valuable insights into questions of stability, reliability and convergence.

\subsection{Random Alignment}
\emph{Random alignment} refers to a random selection of transmission time slots after every \ac{OFDMA} frame and setting the remainder to \ac{DTX}. This results in chaotic interference patterns. This strategy provides a reference for achievable gains, allows to assess the worst cast effects of randomness and represents the \ac{SOTA} in today's unsynchronized, unaligned networks.

\subsection{P-persistent Ranking}
The synchronous alignment of uncoordinated \acp{BS} can lead to instabilities, when neighboring \acp{BS}--perceiving similar interference information--schedule the same time slots for transmission. This leads to oscillating scheduling which never reaches the desired system state~\cite{ehb0801}. To address this problem, p-persistence is introduced to break the unwanted synchrony by only changing established \ac{DTX} schedules with a probability $p=0.3$. In initial tests, the value of $p$ did not have a significant effect on the power consumption; exact tuning is presented in~\cite{ehb0801}. P-persistent ranking first ranks time slots by their sum capacity as in~\eqref{eq:bt}, schedules time slots in that order, but only applies this new selection with probability $p$. Otherwise, the schedule from the last iteration remains active. After ranking, time slots are selected for transmission in order of $B_t$. The remainder of time slots is set to \ac{DTX}.

\subsection{Distributed \ac{DTX} Alignment with Memory}
To counter oscillation and achieve a convergent network state memory of past schedules is introduced into the alignment process. The notion of the \emph{distributed \ac{DTX} alignment with memory} algorithm is as follows. Taking into account current time slot capacities, $R_t$, past scores and slot allocations, each \ac{BS} first updates the internal score of each time slot and then returns the priority of time slots by score. In case of equal scores, time slots are further sub-sorted by $R_t$. The score is updated in integers. All time slots which were used for transmission in the previous \ac{OFDMA} frame receive a score increment of one. Furthermore, the time slot with highest $R_t$, $Q_0$, receives an increment of one. Time slots which were not used for transmission in the previous \ac{OFDMA} frame receive a decrement of one, except for $Q_0$. Scores have an upper limit, $\psi_{\mathrm{ul}}$, beyond which there is no increment and a lower limit, $\psi_{\mathrm{ll}}$, below which there is no decrement. The difference between $\psi_{\mathrm{ul}}$ and $\psi_{\mathrm{ll}}$ can be interpreted as the depth of the memory buffer.

\begin{algorithm}
\caption{Distributed DTX alignment with memory}
\label{ch5alg1}
\begin{algorithmic}[1]
\ENSURE {$\psi, \Upsilon_{\mathrm{u}}$}
\STATE $Q \leftarrow $sort-desc-by-capacity$(\Upsilon)$
\FORALL{$\upsilon$ in $\Upsilon_{\mathrm{u}}$}
  \IF {$\psi (\upsilon) < \psi_{\mathrm{ul}}$}
    \STATE $\psi(\upsilon) \leftarrow \psi(\upsilon) + 1$
  \ENDIF
\ENDFOR
\FORALL{$\upsilon$ in $\Upsilon_{\mathrm{uu}} \backslash \{Q_0\}$}
  \IF {$\psi (\upsilon) > \psi_{\mathrm{ll}}$}
    \STATE $\psi(\upsilon) \leftarrow \psi(\upsilon) - 1$
  \ENDIF
\ENDFOR
\STATE $\psi (Q_0) \leftarrow \psi(Q_0) + 1$
\STATE $V \leftarrow $ sort-desc-by-score$(Q, \psi, \Upsilon)$
\RETURN $\psi, V, \Upsilon_{\mathrm{u}}$
\end{algorithmic}
\end{algorithm}
\nomenclature{$\psi$}{Scoring map, assigning scores to users}
\nomenclature{$\Upsilon_{\mathrm{u}}, \Upsilon_{\mathrm{uu}}$}{Set of used/unused time slots, respectively}
\nomenclature{$Q$}{Ranking tuple, time slots in order of capacity}
\nomenclature{$\psi_{\mathrm{ul}}, \psi_{\mathrm{ll}}$}{Upper/lower limit of score, respectively}
\nomenclature{$Q_0$}{Time slot with highest virtual capacity}
\nomenclature{$V$}{Priority tuple, time slots in order of score}

The algorithm is shown in Algorithm~\ref{ch5alg1} with scoring map $\psi$, ranking tuple $Q$, and priority tuple $V$. The set of used time slots, $\Upsilon_{\mathrm{u}}$, and unused time slots, $\Upsilon_{\mathrm{uu}}$, make up the set of time slots, $\Upsilon$, the cardinality of which is $T$. The function `sort-desc-by-capacity($\Upsilon$)' returns a list of time slots ordered descending by $R_t$. The function `sort-desc-by-score$(Q, \psi, \Upsilon)$' returns a list of time slots ordered primarily by descending score and secondarily by descending $R_t$. After the execution of Algorithm~\ref{ch5alg1}, the first $T_{\mathrm{Tx}}$ time slots in $V$ are scheduled for transmission.

In the following, the iterations over three \ac{OFDMA} frames of Algorithm~\ref{ch5alg1} for a system with three time slots are illustrated. In the example, the algorithm delays the change of $\Upsilon_{\mathrm{u}}$ from $c$ to $b$ to buffer scheduling changes. The iteration begins with arbitrarily chosen $\Upsilon=\{a,b,c\}$, $\psi=\{a:0,b:2,c:5\}$:
\begin{enumerate}
 \item $\Upsilon_{\mathrm{u}}=\{c\}$, $Q = ( b,c,a ) $\\
$\rightarrow \psi = \{a:0,b:3,c:5\}$, $V = (c,b,a)$
 \item $\Upsilon_{\mathrm{u}}=\{b,c\}$, $Q = ( b,c,a ) $\\
$\rightarrow \psi = \{a:0,b:5,c:5\}$, $V = (b,c,a)$
 \item $\Upsilon_{\mathrm{u}}=\{b\}$, $Q = ( b,a,c ) $\\
$\rightarrow \psi = \{a:0,b:5,c:4\}$, $V = (b,c,a)$
\end{enumerate}

When applied, Algorithm~\ref{ch5alg1} strongly benefits time slots which were used for transmission in the past ($b,c$ in step 1 of the example). These time slots tend to repeatedly receive score increments until they all have maximum score $\psi_{\mathrm{ul}}$ ($b,c$ in step 2 of the example). When the score is equal for some time slots ($b,c$ in step 2 of the example), the ranking is based on $R_t$. A time slot which was used for transmission and has highest  $R_t$, $Q_0$, receives the highest increment (slot $b$ in step 2 of the example). When a time slot has highest  $R_t$, but was not selected for transmission in the previous \ac{OFDMA} frame, it receives a score increment, but is not guaranteed to be used for transmission (slot $b$ in step 1 of the example). When a time slot repeatedly has highest $R_t$, it reaches $\psi_{\mathrm{ul}}$ ($b$ in the example). The algorithm thus buffers short term changes in the channel quality setting in favor of long term time slot selection.

\section{Results}
\label{ch5:results}
This section analyzes the four alignment strategies in simulation with regard to power consumption, convergence, reliability of delivered rates and algorithmic complexity after the introduction of the simulation environment and the resource block scheduling scheme.

\subsubsection{Simulation Environment and Resource Block Scheduling}
The four strategies were tested in a network simulation with 19-cell hexagonal arrangement with uniformly distributed mobiles and fixed target rates per mobile. Data was collected only from the center cell which is thus surrounded by two tiers of interfering cells. Power consumption of a cell is modelled by the affine model in~\eqref{eq:affinemodel1x}. Table~\ref{tab:params} lists additional parameters used which approximate an \ac{LTE} system. The simulation is started with the assumption of full transmission power on all resources with a power consumption of 355~W per cell as a worst-case initial configuration.

\begin{table}
\centering
      \begin{tabular}{c|c}
	Parameter          			& Value\\
	\hline
	Carrier frequency  			& 2\,GHz\\
	Intersite distance 			& 500\,m\\
	Pathloss model  			& 3GPP UMa, NLOS, \\
						& shadowing~\cite{std:3gpp-faeutrapla}\\
	Shadowing standard deviation		& 8\,dB\\
	Bandwidth				& 10\,MHz \\
	Transmission power per resource block	& 0.8~W \\
	Thermal noise temperature		& 290\,K \\
	Interference tiers			& 2 (19 cells)\\
	Mobile target rate			& 1~Mbps to 3~Mbps\\
	\ac{OFDMA} subframes (time slots)	& 10\\
	Subcarriers				& 50\\
	Mobile terminals			& 10\\
	$\psi_{\mathrm{ul}}$			& 5\\
	$\psi_{\mathrm{ll}}$			& 0\\
	Power model parameters, \eqref{eq:affinemodel1x}\\ (idle; load factor; DTX)& 186~W; 4.2; 107~W\\
	Subcarrier bandwidth 		& 200\,kHz\\
	\end{tabular}
\caption[Simulation parameters]{Simulation parameters used.}
\label{tab:params}
\end{table}

In order to assess the performance of the four strategies, it is necessary to make assumptions about how the individual resource blocks are scheduled within a time slot. Resource block scheduling is itself challenging and can have a strong effect on power consumption. In order to avoid masking effects of the time slot scheduling algorithms under test, only a very simple sequential resource block allocation is applied rather than sophisticated solutions from the literature. Sequential resource allocation is performed after the \ac{DTX} alignment step is completed and allocates as many bits to an \ac{OFDMA} resource as possible according to the Shannon capacity, followed in order by the next resource in the same time slot (sequentially), until the target rate has been scheduled to each mobile. Time slots are scheduled in the order provided by each of the four strategies. This sequential resource scheduler deliberately omits the benefits of multi-user diversity. This leads to underestimating achievable rates in simulation compared to a system which exploits multi-user diversity, in favor of allowing a fair comparison of the quality of different \ac{DTX} alignment algorithms.

\subsubsection{Power Consumption}
\begin{figure}
\centering
\includegraphics[width=0.8\textwidth]{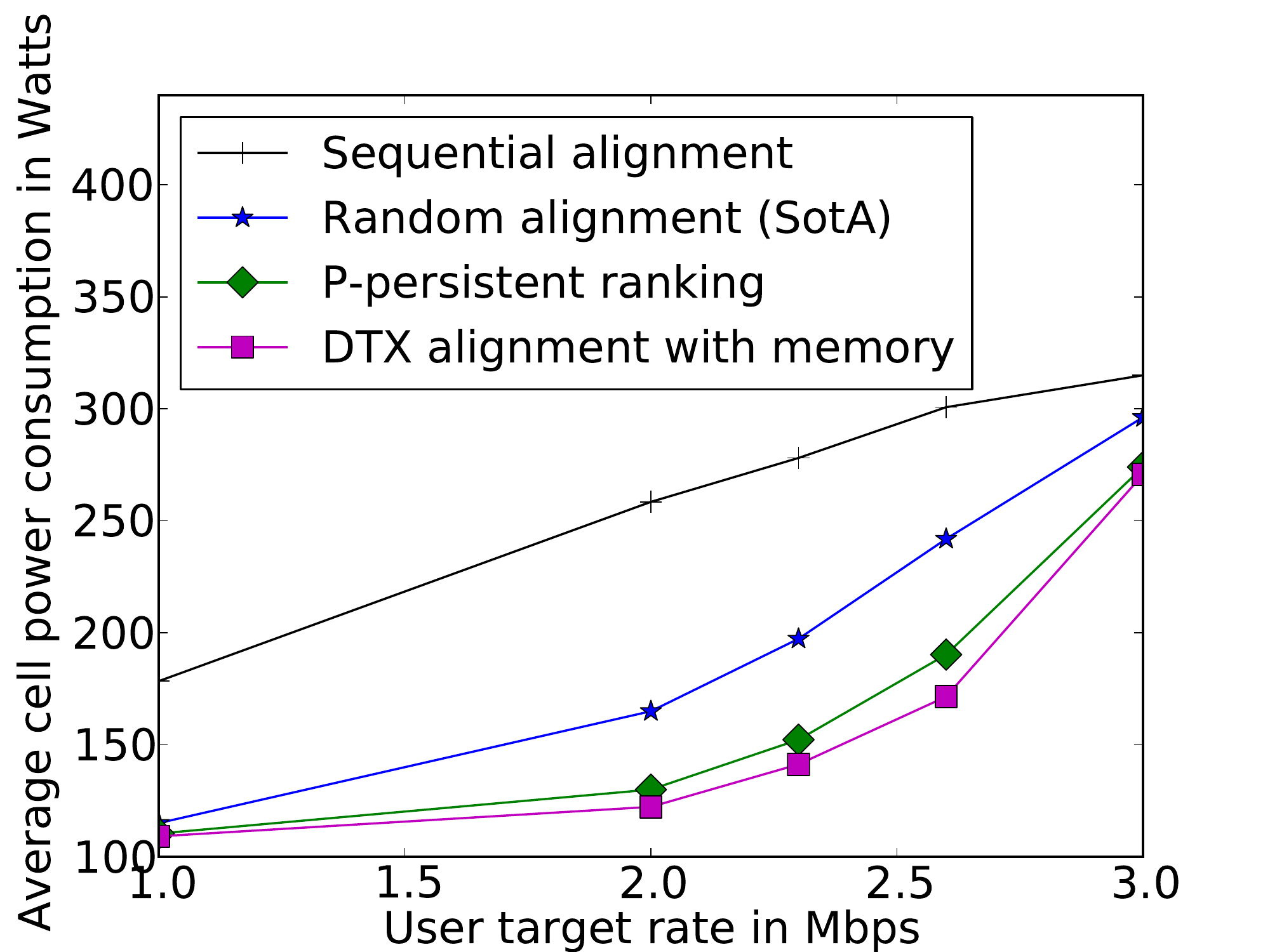}
 \caption[\acs{BS} power consumption over different cell sum rates]{\ac{BS} power consumption over different cell sum rates.}
\label{fig:sumrate}
\end{figure}

To assess achievable power savings and the dynamic adaptivity over a large range of cell loads, the cell total power consumption is shown in Fig.~\ref{fig:sumrate}. 
At low load very few resource blocks are required to deliver the target rate and more time slots can be scheduled for \ac{DTX} than at high traffic loads, leading to monotonously rising power consumption over increasing target rates for all alignment methods. 

Sequential alignment causes the highest power consumption over any target rate with an almost linear relationship between user target rates and power consumption. Sequential alignment consumes high power, as it schedules many resource blocks to achieve the target rate due to the high interference level present. 

This power consumption is significantly lower for the \ac{SOTA}, random alignment. The randomness of time slot alignment creates a much lower average interference than with sequential alignment allowing more data to be transmitted in each resource block. As fewer resource blocks are required, less power is consumed.

P-persistent ranking and \ac{DTX} with memory both achieve similarly low power consumption of up to 40\% less than random alignment. The relationship between power consumption and target rate is noticeably non-linear, as it is flat at low target rates and grows more steeply at high target rates. This behavior is caused by the low interference level these strategies manage to create. Only at high rates, when the number of sleep time slots has to be significantly decreased, does the interference increase, leading to higher power consumption.

Also noteworthy is the fact that at 1~Mbps and 3~Mbps, random alignment performs nearly as good as p-persistent ranking. At these extreme points the network is almost unloaded and almost fully loaded, respectively. Consequently, either most time slots are scheduled for \ac{DTX} or none, leaving very little room for optimization compared to randomness. The largest potential for time slot alignment for power saving is in networks which are medium loaded. Under medium load, the number of transmission and \ac{DTX} time slots is similar, causing the effects of alignment to be most pronounced.

\subsubsection{Convergence}
\begin{figure}
\centering
\includegraphics[width=0.8\textwidth]{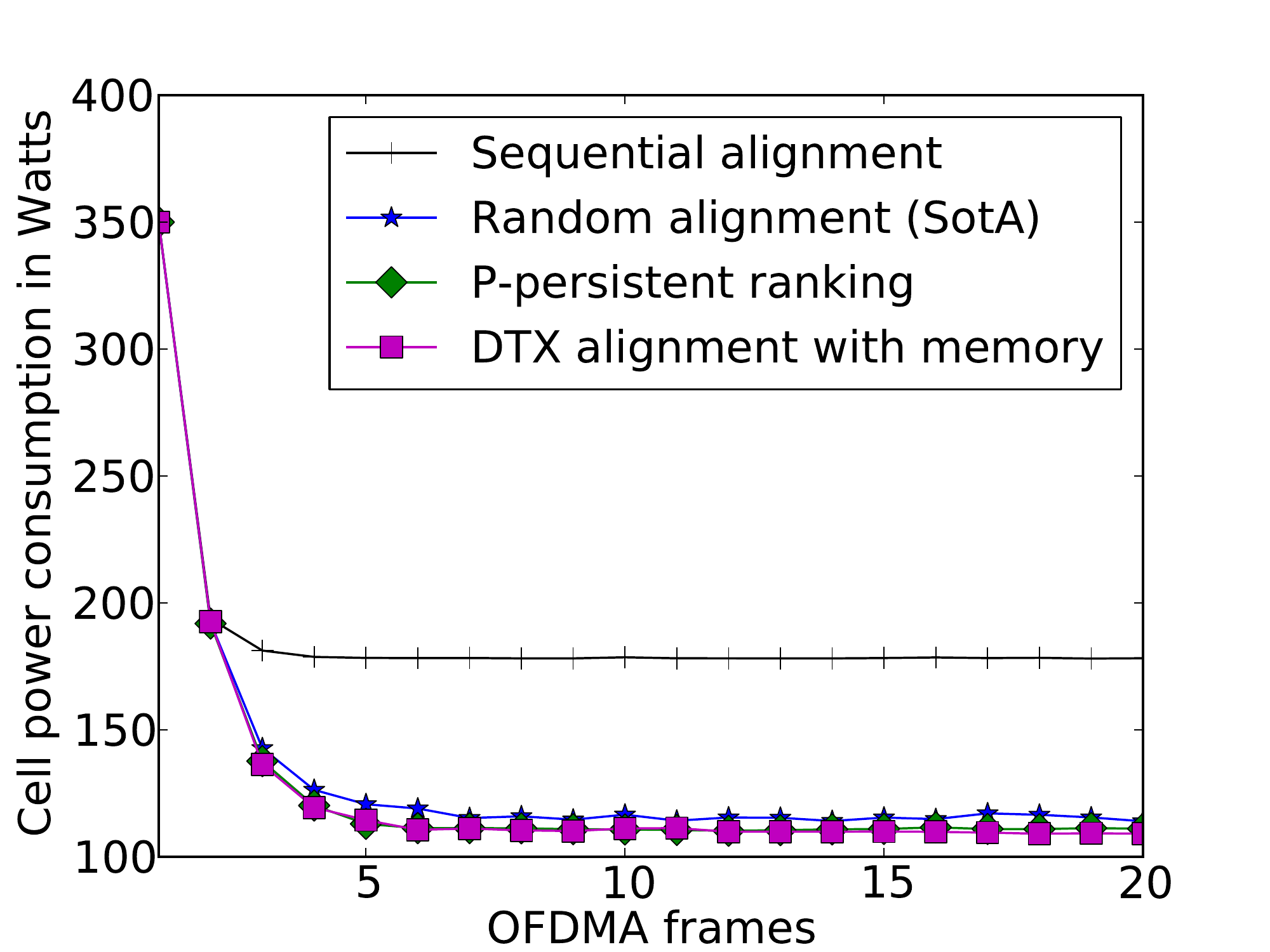}
 \caption[\acs{BS} power consumption over \acs{OFDMA} frames at 1~Mbps per mobile]{\ac{BS} power consumption over \ac{OFDMA} frames at 1~Mbps per mobile.}
\label{fig:power_consumption_1000000}
\end{figure}

Another relevant aspect is the convergence of the network to a stable state. As each \ac{BS} makes iterative adjustments to its selection of time slots for transmission, the speed of convergence as well as the convergence to a stable point of operation are relevant criteria. The effect of the iterative execution of the four strategies on the cell power consumption is illustrated in Fig.~\ref{fig:power_consumption_1000000}. Power consumption is found to converge to a stable value within within six \ac{OFDMA} frames (alignment iterations). All strategies converge to the average power consumption values shown in Fig.~\ref{fig:sumrate}. The simulation starts from a worst-case schedule of transmission on all resource blocks and then iteratively schedules time slots for transmission and \ac{DTX}. In the case of 1~Mbps per user, one transmission time slot is sufficient to schedule the target rate. With transmissions only taking place during one time slot, there is very little difference between a random alignment and p-persistent ranking or \ac{DTX} with memory. 

\begin{figure}
\centering
\includegraphics[width=0.8\textwidth]{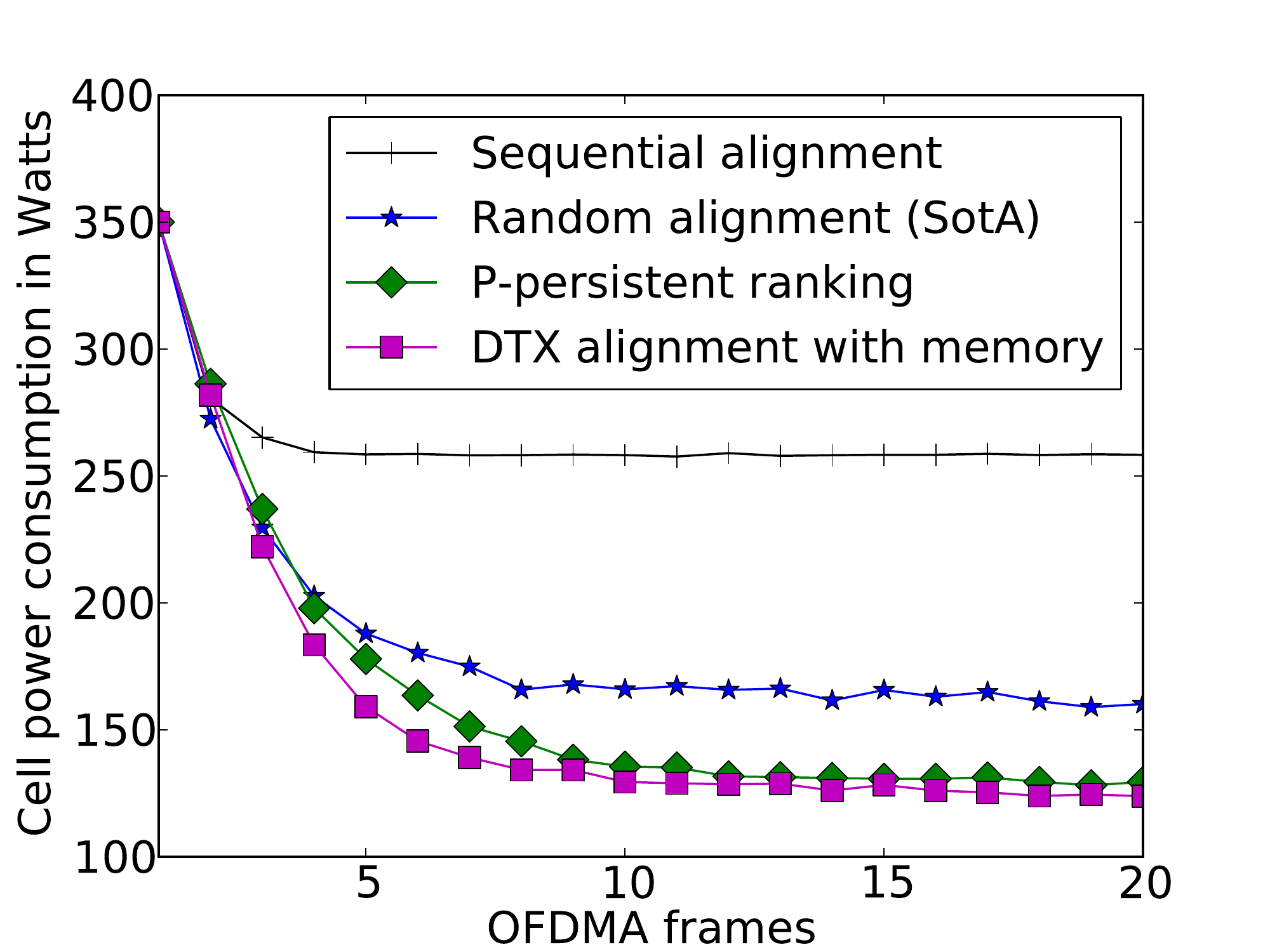}
 \caption[\acs{BS} power consumption over \acs{OFDMA} frames at 2~Mbps per mobile]{\ac{BS} power consumption over \ac{OFDMA} frames at 2~Mbps per mobile.}
\label{fig:power_consumption_2000000}
\end{figure}

At 2~Mbps per user, see Fig.~\ref{fig:power_consumption_2000000}, random alignment occasionally causes higher interference than p-persistent ranking or \ac{DTX} with memory, leading to higher power consumption. Also, p-persistent ranking converges more slowly than \ac{DTX} alignment with memory.

\subsubsection{Reliability}
\begin{figure}
\centering
\includegraphics[width=0.8\textwidth]{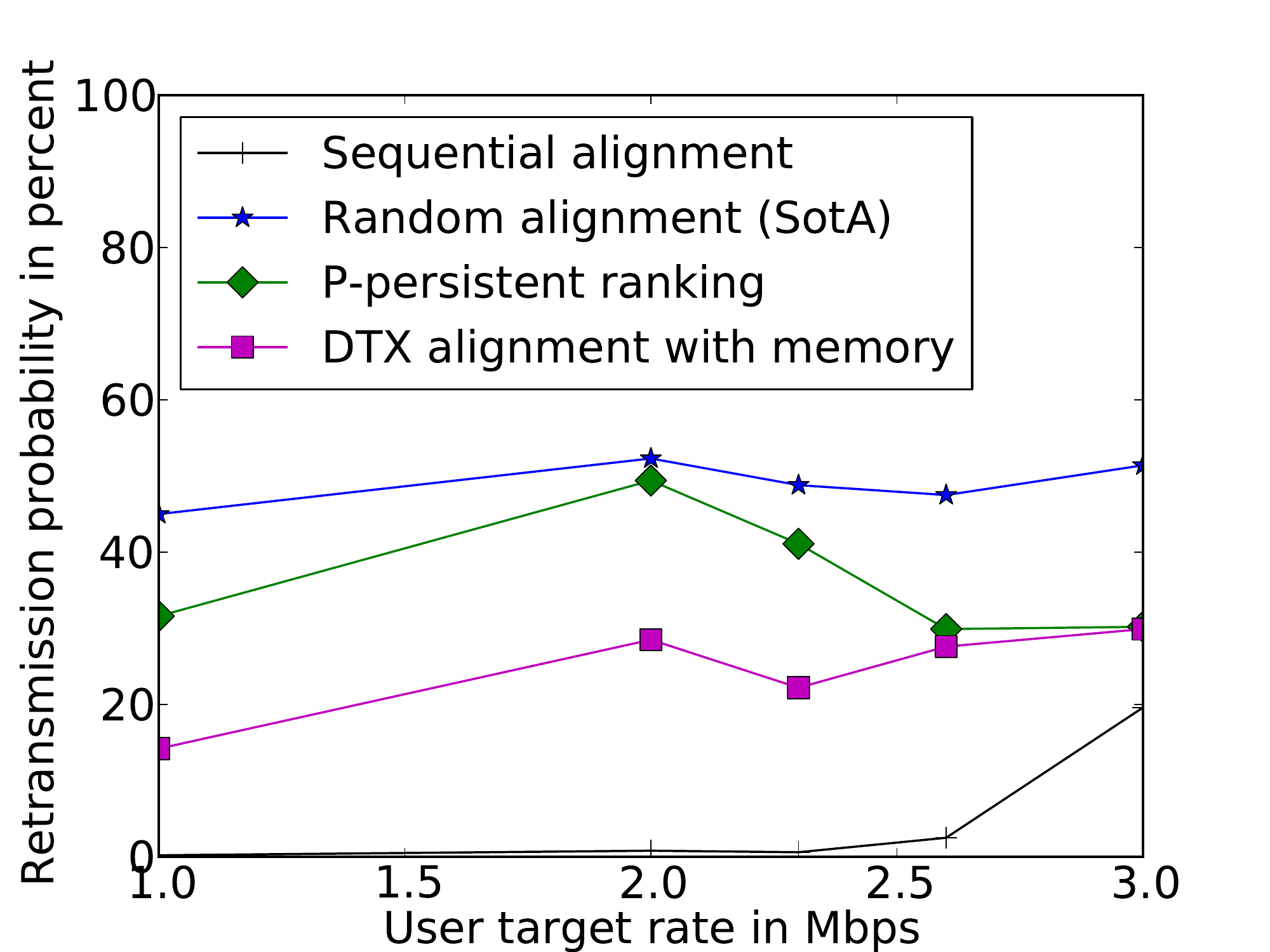}
 \caption{Retransmission probability over targeted rate.}
\label{fig:percentage_satisfied_over_target_rate}
\end{figure}

An important aspect in dynamical systems with target rates is that scheduled target rates cannot always be fulfilled. As resource block scheduling is based on channel quality information which was collected in the previous \ac{OFDMA} frame, the actual channel quality during transmission may differ, leading to lower than expected rates. Thus, although certain target rates are scheduled and although the system is not fully loaded, a \ac{BS} may fail to deliver the targeted rate and require retransmission of some resource blocks. This metric is assessed by considering the retransmission probability for each strategy. The results are shown in Fig.~\ref{fig:percentage_satisfied_over_target_rate} against a range of user target rates. 

Easiest to interpret is sequential alignment which does not require retransmission for target rates up to 2.3~Mbps due to its determinism. The increase in the retransmission probability at high rates is not caused by a failure of the alignment, but by system overload. When high rates are combined with bad channel conditions, the system may be unable to deliver the target data rate, independent of the alignment strategy. This increase of the retransmission probability at high rates is present for all alignment strategies and constitutes outage.

The retransmission probability is highest for random alignment. This is caused by the strong fluctuation of interference under randomized scheduling. Channel quality measurements used for resource block scheduling are of very little reliability, as the interference changes quickly due to random time slot alignment.

P-persistent ranking performs slightly better than random alignment at 1~Mbps target rate. At 2~Mbps per user, where the alignment potential is highest, oscillation causes the highest retransmission probability.

\ac{DTX} alignment with memory achieves a much lower retransmission probability in the range of 15\% to 20\%, due to the reduction in interference fluctuation introduced by the memory score. 

\subsubsection{Complexity}
With regard to complexity, sequential and random alignment are of minimal complexity. These strategies involve no algorithmic decision-making on the set of transmission time slots. P-persistent ranking requires one ranking and time slot selection with probability $p$ per iteration. The highest complexity is present in \ac{DTX} with memory, which requires two executions of the time slot sort, one for the generation of the score and one for the output of the ranking. Although \ac{DTX} with memory comprises the highest complexity in comparison, these operations pose a small burden on modern hardware as the number of time slots is typically small. For example, typical \ac{LTE} systems are designed with 10~subframes (time slots). 

\subsubsection{Interpretation}
\ac{DTX} with memory was found to provide the best results. Under the present assumptions, it provides both a lower power consumption and lower retransmission probability than the \ac{SOTA} and p-persistent ranking. Future \ac{DTX} capable networks can and should exploit this alignment potential.

\section{Summary}
\label{ch5:summary}
In this chapter, the constructive uncoordinated and distributed alignment of \ac{DTX} time slots between interfering \ac{OFDMA} \acp{BS} for power-saving under rate constraints was studied. The open problem was first formally established. Due to its complexity, it was approached with four alternative heuristic strategies: Sequential alignment, random alignment, p-persistent ranking and distributed \ac{DTX} with memory. Distributed \ac{DTX} with memory is an original contribution of this thesis. It addresses shortcomings of the other strategies by introducing memory to overcome short-term fluctuations and networks oscillations. The performance of the four strategies was tested in simulation. It was found that power consumption can be reduced significantly, especially at medium cell loads. All strategies converge within six \ac{OFDMA} frames. \ac{DTX} with memory reduces power consumption up to 40\% compared to the \ac{SOTA}, combined with around 20\% reduction in retransmission probability.

\acresetall 

\renewcommand{\dir}{chapter6/chapter/}
\chapter{Conclusions, Limitations and Future Work}
\label{ch:conclusion}


\section{Summary and Conclusions}
In this thesis, the power consumption of \ac{LTE} \ac{BS} devices was studied and modelled. The findings from the model were applied to reduce power consumption through energy efficient \ac{RRM} mechanisms while upholding user rate constraints. \ac{RRM} mechanisms were applied to control links on the cell level and interference on the multi-cell level.

In Chapter~\ref{ch:introduction}, the importance of energy efficiency for current mobile networks was emphasized. As mobile traffic, the number of devices and the size of the infrastructure are constantly increasing, the \coo~emission of mobile networks is expected to double between 2007 and 2020. Green Radio research efforts, such as this thesis, are directed to revert that trend. 

Chapter~\ref{ch:motivation} first inspected the power consumption life cycle of a mobile network. While the production of mobile equipment causes significant emissions of \coo, the operation is power efficient. The situation was found to be almost the opposite for cellular \acp{BS} which cause the largest share of emissions during operation. This is caused by the significant number of stations in operation combined with their long life times. This situation motivated the reduction of operation power consumption of \acp{BS} as the topic of this thesis. Next, the \ac{SOTA} of Green Radio was identified through a detailed literature survey. Finally, the chapter also discussed fundamental technical concepts in preparation for the original contributions in Chapters~\ref{ch:device}, \ref{ch:cell}, and \ref{ch:network}.

Chapter~\ref{ch:device} explored the sources of power consumption in an \ac{LTE} \ac{BS}. First,  the difficulty of accurate power modeling and other existing power models were described. Next, the typical architecture of a \ac{BS} was introduced. Each of its components was discussed and individually modelled. Components were found to be either load adaptive, like the \ac{PA} and \ac{BB} unit, have constant power consumption, like the \ac{RF} transceiver, or have a power consumption which scales with other components' consumption, like power conversion and cooling. The power consumptions of the subcomponents were combined to find the power consumption of the entire \ac{BS}. Since the component model is too elaborate for many applications, a parameterized model was proposed which abstracts many parameters. The parameterized model only considers those parameters which are typically modified during operation such as bandwidth, transmission power, the number of transmission antennas, and sleep mode. Parameters were provided. The parameterized model was once more simplified into an affine model which can be applied in optimization. Overall, it was found that an increase of the energy efficiency in \acp{BS} must be based on reducing the \ac{PA} power consumption through \ac{PC} and setting the entire \ac{BS} to a sleep mode. Hardware research should focus on enabling and amplifying these two mechanisms. Other manipulation such as the adaptation of coding, modulation or bandwidth are less promising. Simple models are found to capture the most relevant aspects sufficiently. 

With the power model and saving techniques identified in the prior chapter, Chapter~\ref{ch:cell} explored three \ac{RRM} mechanisms to reduce \ac{BS} power consumption. The first technique is \ac{PC}, through which power was adapted to match only the requested capacity per \ac{UE} instead of over-achieving. The second technique is \ac{DTX}, a very short sleep mode. Of the two, \ac{PC} was found to provide strong savings in high load when transmission power is generally high, whereas \ac{DTX} was most effective at low load when there is ample opportunity for sleep mode. It was shown that the joint application of \ac{PC} and \ac{DTX} achieves lower power consumption than either individual technique over all cell traffic loads in the \ac{PRAIS} scheme. This joint problem was shown to be convex and thus efficiently solvable. This finding was combined with the third technique, \ac{AA}, and applied by using it as the core of a power-saving \ac{MIMO} \ac{OFDMA} resource scheduler, called \ac{RAPS}. \ac{RAPS} first solves a simplified convex real-valued subproblem. The solution of the subproblem is then quantized to the integer-valued \ac{OFDMA} scheduling problem and joined with subcarrier allocation and power allocation. \ac{AA} was found to provide little added value on top of the other two techniques. Application of \ac{RAPS} was shown to reduce power consumption of \ac{LTE} \acp{BS} by 25-40\% in single-cell simulation. \ac{RAPS} could be readily built into current \ac{LTE} \acp{BS}. 

Going beyond the cell level, Chapter~\ref{ch:network} studied the exploration of intercell interference for power saving. It is beneficial to align the \ac{DTX} mode of \acp{BS} constructively, to the effect that interference is reduced. Four alternative alignment techniques have been tested in simulation: sequential alignment, random alignment, p-persistent ranking, and distributed \ac{DTX} with memory. The latter is an original contribution by the author. Sequential alignment is a conservative technique which maximizes interference for the benefit of predictability. Random alignment is assumed was the \ac{SOTA}. P-persistent ranking is a popular technique for the suppression of oscillation. Distributed \ac{DTX} with memory goes further and keeps track of past allocations and channel quality to enable convergence to a stable network state. The important finding was that---unlike for the single cell in Chapter~\ref{ch:cell}---the power consumption could not be reduced through these mechanisms without affecting the \ac{QoS}. The interference dynamics cause occasional violations of the rate constraints, thus necessitating retransmissions. Random alignment caused regular rate constraint violations at medium power consumption. P-persistent ranking was slightly more reliable with lower retransmission probability at low power consumption. However, distributed \ac{DTX} with memory provided a much lower retransmission probability than p-persistent ranking at lower power consumption. It thus provided the best best retransmission probability and power saving. The power saving of distributed \ac{DTX} with memory over the \ac{SOTA} was found to be up to 40\%, depending on the cell loads.

In conclusion, this thesis provides three readily applicable supply power models for \ac{LTE} \acp{BS}, including the model derivations. From these models, \ac{DTX} and \ac{PC} were identified as most effective and applied to reduce \ac{BS} operating power consumption. \ac{PC} and \ac{DTX} were optimized jointly and integrated into a comprehensive \ac{MIMO} \ac{OFDMA} scheduler. The inherent \ac{DTX} time slots of such a scheduler were then aligned constructively between neighbouring \acp{BS}. Reducing \ac{BS} power consumption by 50\% through software is thus well within reach.

\section{Limitations and Future Work}
The most important limitation to the techniques proposed in this thesis lies in the restrictions imposed on the scheduler by a transmission standard like \ac{LTE}. In this research, it was assumed that a scheduler can select any resource block for transmission or sleep. In practice, however, this selection will be limited as the standard reserves some resources for signaling, channel sounding, or pilot transmission. Since the scheduler will be less potent due to this limited selection, power consumption is expected to be higher under these constraints.

Next, the power model in Chapter~\ref{ch:device} as well as all later chapters assume that a \ac{BS} can be switched on and off in an instant. Yet, although switching times are assumed to be in the order of tens of microseconds~\cite{fmmjg1101}, they are not fully instantaneous. This has to be considered in \ac{DTX} and wake--up scheduling. If sleep modes are deeper, switching times will become longer. Following this line of argument, there may actually be more than one sleep mode available; for example, an OFF state with zero power consumption, a deep sleep with very low consumption, and a light sleep for \ac{DTX}. In that case, a scheduler would have to consider the wake--up times in its selection of sleep modes.

Finally, the validity of the results presented over the coming years is strongly dependent on the assumptions made in the power model. Drastic changes in the architecture of \ac{BS}, for example through the proliferation of massive \ac{MIMO}~\cite{ltem1301}, may change the way \acp{BS} consume power. In such a case, the effectiveness of different saving techniques will have to be re-evaluated. 

For future research, the most immediate goal is to combine the \ac{RAPS} scheduler with distributed \ac{DTX} with memory and assess the joint power savings. As the \ac{RAPS} scheduler makes no assumptions about the time slots allocated for \ac{DTX}, it can provide instructions for the \ac{DTX} alignment scheme. A further planned extension to the \ac{DTX} alignment scheme is an decrease of the retransmission probability. It is expected that a rate margin imposed on the rate constraint, \ie~requesting a higher rate target, will be beneficial.

Another important addition is the consideration of standards limitations. The strength of the impact of standards limitations has to be assessed. If the standard in its current form severely caps power savings, future standards could be designed to be more open to power saving.

The power-saving \ac{RRM} mechanisms established herein can also be applied to \acp{BS} which are smaller than macro. In smaller \ac{BS} types, the load dependence is smaller. Thus, \ac{PC} is less effective compared to \ac{DTX}. Consideration of smaller \ac{BS} types is also related to the field of HetNets~\cite{hqkg1101}. Since coverage is generally assumed to overlap in HetNets, new options for longer sleep modes arise as \acp{UE} will have guaranteed coverage even when some \acp{BS} are asleep.

All analyses in this work assumed a uniform distribution of \acp{BS}. However, with the advance of HetNets, it may also be relevant to explore distributions such as Poisson Point Processes~\cite{habdf0901}. This will have an effect particularly on the multi-cell \ac{DTX} alignment work.

\renewcommand{\chaptermark}[1]{%
 \markboth{\MakeUppercase{%
 \appendixname} \ \thechapter.%
 \ #1}{}}
\renewcommand{\dir}{appendix/appendix}
\begin{appendices}


\chapter{Appendix} 

\section{Proof of Convexity for Problem~\eqref{eq:OP}}
\label{appendix:proof1}
This proof shows that \eqref{eq:OP} is convex. 
\begin{proof}
For $P_{\mathrm{S}}=0$, the first derivative of the cost function for link~$k$ is
\begin{equation}
\begin{aligned}
 f_{0,k}^\prime = \frac{\partial P_{\mathrm{supply}}(\overline{R}_k)}{\partial \mu_k} & = P_0 + \Delta_{\mathrm{p}}\frac{P_{\mathrm{N}}}{G_k} \left[ - 1 + 2^\frac{\overline{R_k}}{W \mu_k} \left( 1 - \frac{\overline{R_k}}{W \mu_k} \ln(2) \right) \right].
\end{aligned}
\label{eq:f0iprimePC}
\end{equation}
The second derivative of~$f_{0,k}$ is given by
\begin{equation}
 f_{0,k}^{\prime\prime} = \frac{\partial^2 P_{\mathrm{supply}}(\overline{R}_k)}{\partial^2 \mu_k} = \Delta_{\mathrm{p}}\frac{P_{\mathrm{N}}}{G_k} \left(\frac{\overline{R_k}}{W} \right)^2 \ln^2(2) \frac{ 2^\frac{\overline{R_k}}{W \mu_k}}{\mu_k^3}.
\label{eq:f0iprimeprimePC}
\end{equation}
All variables in the second derivative are positive, thus $f_{0,k}^{\prime\prime} \geq 0$ within the parameter bounds. Therefore, each $f_{0,k}$ is convex within the bounds. 

The non-negative sum preserves convexity. Thus, $f_{0}$ is convex. \end{proof}

\section{Proof of Convexity for Problem~\eqref{eq:OPJSAC}}
\label{appendix:proof2}
This proof shows that \eqref{eq:OPJSAC} is convex.
\begin{proof}
The partial second derivative of the \ac{MIMO} cost function in \eqref{eq:OPJSAC} with respect to $\mu_k$ is
\begin{equation}
\begin{aligned}
 & \frac{\partial^2 P_{\mathrm{supply},k}(\B{r})}{\partial^2 \mu_k} = \\
 & \Delta_{\mathrm{p}}  \log^2(2) \frac{R_k}{W} \frac{ 2^{\frac{R_k}{W \mu_k}} \left( (\epsilon_1 + \epsilon_2)^2 + 2 \epsilon_1 \epsilon_2 \left( 2^{\frac{R_k}{W \mu_k}}-2 \right) \right)} {\mu_k^3 \left( (\epsilon_1 + \epsilon_2)^2 + 4 \epsilon_1 \epsilon_2 \left( 2^{\frac{R_k}{W \mu_k}}-1 \right) \right)^{\frac{3}{2}}},
\end{aligned}
\end{equation}
which is non-negative since
\begin{equation}
 2 \epsilon_1 \epsilon_2 2^{\frac{R_k}{W \mu_k}} \geq 2 \epsilon_1 \epsilon_2
\end{equation}
within the parameter bounds. Thus the cost function is convex in $\mu_k$ within the bounds. 
\end{proof}
Convexity of the \ac{SIMO} cost function and the constraint function can be shown similarly and is omitted here for brevity.

\section{Margin-adaptive Resource Allocation}
\label{appendix:iwf}
Margin-adaptive power allocation over a set of resource blocks, $\mathcal{U}_k$, and a vector of channel eigenvalues, $\mathcal{E}_{a,k}$, minimizes power consumption while fulfilling a target rate~\cite{ygc0201}. 

Assuming block-diagonalization precoding, the problem is to
\begin{subequations}
\begin{align}
& \underset{P_{a,e}}{\text{minimize}} 		& \displaystyle\sum_{a=1}^{|\mathcal{U}_k|} \sum_{e=1}^{|\mathcal{E}_{a,k}|} P_{a,e}  	& \\
& {\text{subject to} }	& B_{\mathrm{target}, k} &= \displaystyle\sum_{a=1}^{|\mathcal{U}_k|} \sum_{e=1}^{|\mathcal{E}_{a,k}|} w \tau \log_2 \left( 1 + \frac{P_{a,e} \mathcal{E}_{a,k}(e)}{P_{\mathrm{N}}} \right), \\
& 			&P_{a,e} 										&\geq 0 \quad \forall a,e, \\
& 			&\displaystyle\sum_{a=1}^{|\mathcal{U}_k|} \sum_{e=1}^{|\mathcal{E}_{a,k}|} P_{a,e} 	&\leq P_{\mathrm{max}}.
\end{align}
\label{eq:iwfop}
\end{subequations}
With Lagrange multipliers $\lambda, \beta, \upsilon$, the Karush-Kuhn-Tucker optimality conditions for \eqref{eq:iwfop} are
\begin{subequations}
 \begin{align}
 -P_{a,e} &\leq 0 \quad\forall a,e \label{eq:pf1}\\
 \displaystyle\sum_{a=1}^{|\mathcal{U}_k|} \sum_{e=1}^{|\mathcal{E}_{a,k}|} P_{a,e} 	- P_{\mathrm{max}} &\leq 0 \label{eq:pf2}\\
 B_{\mathrm{target}, k} - \displaystyle\sum_{a=1}^{|\mathcal{U}_k|} \sum_{e=1}^{|\mathcal{E}_{a,k}|} w \tau \log_2 \left( 1 + \frac{P_{a,e} \mathcal{E}_{a,k}(e)}{P_{\mathrm{N}}} \right) &= 0 \label{eq:ct}\\
 \lambda_{a,e}^\star \geq 0 \quad\forall a,e, \quad\beta &\geq 0 \label{eq:df1}\\
 \lambda_{a,e}^\star (-P_{a,e}^\star) &= 0 \quad\forall a,e \label{eq:lambdastarf}\\
 \beta \left( \displaystyle\sum_{a=1}^{|\mathcal{U}_k|} \sum_{e=1}^{|\mathcal{E}_{a,k}|} P_{a,e} 	- P_{\mathrm{max}} \right) &= 0 \label{eq:betaf}\\
 \frac{\partial \mathcal{L}}{\partial P_{a,e}} &= 0,
 \end{align}
\end{subequations}
where \eqref{eq:pf1}, \eqref{eq:pf2}, \eqref{eq:ct} ensure the primal feasibility, \eqref{eq:df1} the dual feasibility and \eqref{eq:lambdastarf}, \eqref{eq:betaf} the complementary slackness.

The derivative of the Lagrangian at the minimum is
\begin{equation}
 \frac{\partial \mathcal{L}}{\partial P_{a,e}} = 1 - \lambda + \beta - \frac{\upsilon\, w \tau\; \mathcal{E}_{a,k}(e)}{\log(2)\left( P_{\mathrm{N}} + \mathcal{E}_{a,k}(e) P_{a,e}  \right)} = 0,
\label{eq:lagrdev}
\end{equation}
with $\lambda$ the multiplier for the first inequality constraint, $\beta$ for the second (power-limit) constraint and $\upsilon$ the equality constraint multiplier. When reducing the conditions to $P_{a,e} > 0$, $\lambda=0$. When considering that the power constraint is always loosely bound with $\left( \displaystyle\sum_{a=1}^{|\mathcal{U}_k|} \sum_{e=1}^{|\mathcal{E}_{a,k}|} P_{a,e} - P_{\mathrm{max}} \right) < 0$, then by \eqref{eq:betaf}, $\beta = 0$.

Solving the remainder for $P_{a,e}$,
\begin{equation}
 P_{a,e} = \frac{\upsilon w \tau }{\log(2)} - \frac{P_{\mathrm{N}}}{\mathcal{E}_{a,k}(e)},
\end{equation}
which can be inserted into the equality constraint of \eqref{eq:iwfop} to yield the water-level,~$\upsilon$, by
\begin{equation}
   \log_2(\upsilon) = \frac{1}{|\Omega_k|} \left( \frac{B_{\mathrm{target}, k}}{w \tau} - \sum_{e=1}^{|\Omega_k|} \log_2 \!\left( \frac{\tau\,\mathcal{E}_{a,k}(e) w}{P_{\mathrm{N}} \log(2)} \right)\right).
\end{equation}

The water-level can be found via an iterative search over the vector $\Omega_k$ of channels that contribute a positive power. Since the sum power is reduced on each iteration, the power constraint (accounted for by the multiplier~$\beta$) only needs to be tested after the search is finished.

\chapter{List of Publications}
This chapter contains a list of publications related to this thesis ordered by academic publication status (published, accepted, submitted) and other type (project reports and book contributions).

\section{Published}
\begin{itemize}
 \item Gunther Auer and Istv\'an G\'odor and L\'aszl\'o H\'evizi and Muhammad Ali Imran and Jens Malmodin and P\'eter Fasekas and Gergely Bicz\'ok and \textbf{Hauke Holtkamp} and Dietrich Zeller and Oliver Blume and Rahim Tafazolli, {``Enablers for Energy Efficient Wireless Networks''}, \emph{Proceedings of the IEEE Vehicular Technology Conference (VTC) Fall-2010}, 2010~\cite{aghimfbhz1001}
 \item \textbf{Hauke Holtkamp} and Gunther Auer, {``Fundamental Limits of Energy-Efficient Resource Sharing, Power Control and Discontinuous Transmission''}, \emph{Proceedings of the Future Network \& Mobile Summit 2011}, 2011~\cite{ha1101}
 \item \textbf{Hauke Holtkamp} and Gunther Auer and Harald Haas, {``{Minimal Average Consumption Downlink Base Station Power Control Strategy}''}, \emph{Proceedings of the 2011 IEEE 22nd International Symposium on Personal Indoor and Mobile Radio Communications (PIMRC)}, 2011~\cite{hah1101a}
 \item \textbf{Hauke Holtkamp} and Gunther Auer and Harald Haas, {``{On Minimizing Base Station Power Consumption}''}, \emph{Proceedings of the Vehicular Technology Conference (VTC) Fall-2011}, 2011~\cite{hah1101}
 \item Claude Desset and Bj\"{o}rn Debaillie and Vito Giannini and Albrecht Fehske and Gunther Auer and \textbf{Hauke Holtkamp} and Wieslawa Wajda and Dario Sabella and Fred Richter and Manuel Gonzalez and Henrik Klessig and Istv\'an G\'odor and Per Skillermark and Magnus Olsson and Muhammad Ali Imran and Anton Ambrosy and Oliver Blume, {``{Flexible Power Modeling of LTE Base Stations}''}, \emph{Proceedings of the 2012 IEEE Wireless Communications and Networking Conference:}, 2012~\cite{ddgfahwsr1201}
 \item \textbf{Hauke Holtkamp} and Gunther Auer and Samer Bazzi and Harald Haas, {``{Minimizing Base Station Power Consumption}''}, \emph{IEEE Journal on Selected Areas in Communications}, 2014~\cite{habh1301}
\end{itemize}

\section{Accepted}
\begin{itemize}
  \item \textbf{Hauke Holtkamp} and Gunther Auer and Vito Giannini and Harald Haas, {``{A Parameterized Base Station Power Model}''}, \emph{IEEE Communications Letters}, 2013~\cite{hag1301}
 \item \textbf{Hauke Holtkamp} and Harald Haas, {``{OFDMA Base Station Power-saving Via Joint Power Control and DTX in Cellular Systems}''}, \emph{Proceedings of the Vehicular Technology Conference (VTC) Fall-2013}, 2013~\cite{hh1301}
\end{itemize}

\section{Submitted}
\begin{itemize}
 \item \textbf{Hauke Holtkamp} and Guido Dietl and Harald Haas, {``{Distributed DTX Alignment with Memory}''}, \emph{Proceedings of the 2014 IEEE International Conference on Communications (ICC)}, 2014~\cite{hdh1301}
\end{itemize}

\section{Project Reports}
\begin{itemize}
\item {EARTH Project Work Package 2}, {``{Deliverable D2.2: Reference Systems and Scenarios}''}, 2012
 \item {EARTH Project Work Package 3}, {``{Deliverable D3.3: Green Network Technologies}''}, 2012~\cite{std:earthd33}
\end{itemize}

\section{Contributions}
\begin{itemize}
 \item Jinsong Wu and Sundeep Rangan and Honggang Zhang, {``{Green Communications: Theoretical Fundamentals, Algorithms and Applications}''}, \emph{Taylor \& Francis Group}, 2012~\cite{wrz1201}
\end{itemize}

\chapter{Attached Publications}
This chapter contains all work either published or submitted for publication to academic conferences or journals. For brevity, \ac{EARTH} project deliverables~\cite{std:earthd33} and a book chapter~\cite{wrz1201} are not listed here.

\end{appendices}

\ifthenelse{\boolean{edthesis}}
 {
 \begin{singlespace}
}
{
  \backmatter
}

\def\bibname{Literature References}

\bibliography{general,DOCOMO}
\ifthenelse{\boolean{edthesis}}
 {
  \end{singlespace}
 }{}

\end{document}